\documentclass[prb,twocolumn,amsmath,amssymb]{revtex4-1}

\usepackage{graphicx}
\usepackage{dcolumn}
\usepackage{bm}
\usepackage{comment}
\usepackage{color}

\begin{document}

\title{Wilson loop approach to fragile topology of split elementary band representations and topological crystalline insulators with time reversal symmetry}
\author{Adrien Bouhon}
\email{adrien.bouhon@physics.uu.se}

\author{Annica M. Black-Schaffer}
\affiliation{Department of Physics and Astronomy, Uppsala University, Box 516, SE-751 21 Uppsala, Sweden }

\author{Robert-Jan Slager}
\affiliation{Max-Planck-Institut f\"ur Physik komplexer Systeme, 01187 Dresden, Germany}
\affiliation{Department of Physics, Harvard University, Cambridge, Massachusetts 02138, USA}

\date{\today}

\begin{abstract}

We present a general methodology towards the systematic characterization of crystalline topological insulating phases with time reversal symmetry (TRS).~In particular, taking the two-dimensional spinful hexagonal lattice as a proof of principle we study windings of Wilson loop spectra over cuts in the Brillouin zone that are dictated by the underlying lattice symmetries.~Our approach finds a prominent use in elucidating and quantifying the recently proposed ``topological quantum chemistry" (TQC) concept.~Namely, we prove that the split of an elementary band representation (EBR) by a band gap must lead to a topological phase.~For this we first show that in addition to the Fu-Kane-Mele $\mathbb{Z}_2$ classification, there is $C_2\mathcal{T}$-symmetry protected $\mathbb{Z}$ classification of two-band subspaces that is obstructed by the other crystalline symmetries, i.e.~forbidding the trivial phase. This accounts for all nontrivial Wilson loop windings of split EBRs \textit{that are independent of the parameterization of the flow of Wilson loops}.~Then, by systematically embedding all combinatorial four-band phases into six-band phases, we find a refined topological feature of split EBRs. Namely, we show that while Wilson loop winding of split EBRs can unwind when embedded in higher-dimensional band space, two-band subspaces that remain separated by a band gap from the other bands conserve their Wilson loop winding, hence revealing that split EBRs are at least ``stably trivial'', i.e. necessarily non-trivial in the non-stable (few-band) limit but possibly trivial in the stable (many-band) limit.~This clarifies the nature of \textit{fragile} topology that has appeared very recently.~We then argue that in the many-band limit the stable Wilson loop winding is only determined by the Fu-Kane-Mele $\mathbb{Z}_2$ invariant implying that further stable topological phases must belong to the class of higher-order topological insulators.
  
  

\end{abstract}

\pacs{Valid PACS appear here}
\maketitle

\section{Introduction}

With the prediction and experimental verification of topological insulators\cite{KM1,KaneMeleZ2,BernevigHughesZhang,FuKane3D, Rmp1,Rmp2} interest in band structures has been heavily reinvigorated over the past decade. 
In particular, it has been found that nontrivial topological structures may emerge due to the presence of symmetries leading to robust physical signatures such as helical edge states and defect modes,\cite{FuKaneZ2, FuKaneInversion,Schnyder08, Mode1, Mode2, Mode3, regdef, Mode4}  which have significantly extended the seminal founding of the quantum Hall effect (QHE). More recently, further interest has been focused on the interplay between topology and the underlying crystal symmetries, leading to the uncovering of an enriched landscape of topological phases.\cite{Clas1, InvTIBernevig, InvTIVish, Clas2, probes_2D, PointGroupsTI,  MorimotoFurusaki_clifford, Shiozaki14, SchnyderClass, Wi1,  Wi3,Alex_BerryPhase, hourglass,Vishwanath_filling_gapless, Nodal_chains, AlexAdiabaticSOC,Clas3, ShiozakiSatoGomiK,bouhon2019nonAbraid,Wi2, Clas4,Clas5,BbcWeyl,NodalLines1}
These classification pursuits have revealed the existence of new invariants, effectively conveying obstructions to deform the band structure to a trivial insulating state in the atomic limit, and accordingly new types of edge and surface states. Still, it is of large practical and theoretical interest to understand these phases from a concrete perspective and consider their response to well established probes. 
A prime example in this regard are Wilson loops.\cite{WindingKMZ2, InvTIBernevig, PointGroupsTI, Wi1, Wi3, hourglass, Alex_BerryPhase, Wi2, AlexAdiabaticSOC, BBS_nodal_lines, HolAlex_Bloch_Oscillations,EBR_2} 


In this work we study the relation between crystalline symmetries and the structure of Wilson loop spectra with the aim of using them as clear indicators of nontrivial topologies. In order to do so we systematically account for all symmetries that span the little groups at high symmetry points (HSP) of the Brillouin zone (BZ) and the compatibility relations between them. This is effectively done through the characterization of the band structures in terms of the irreducible representations (IRREPs) of the underlying space groups, which can be shown to match the descriptive equivariant K-theory in the absence of spins and TRS,\cite{Clas3, ShiozakiSatoGomiK} and has been extended heuristically to the spinful case including TRS.\cite{Clas4,2017arXiv171104769K,Khalaf_sum_indicators} In particular, we derive the symmetry protected windings of the Wilson loop spectra over patches of the BZ that have been chosen such as to take advantage of all available crystalline symmetries. In the process we map out the different topological sectors and also quantitatively evaluate whether insulating band structures, which split an elementary band representation\cite{Zak_EBR1, Zak_EBR2, Zak_EBR3, Zak_EBR4, Zak_EBR5, Clas5, HolAlex_Bloch_Oscillations, EBR_1, EBR_2} (EBR), must be topological. This general idea was postulated in Ref.~\onlinecite{Clas5, EBR_1} and lies at the core of the ``topological quantum chemistry'' (TQC) concept. 

Preceding our work, Ref.~\onlinecite{HolAlex_Bloch_Oscillations} has rigorously related the topology of split EBRs with the symmetry protected quantization of Wilson loop spectra over special base loops. Furthermore, Ref.~\onlinecite{Wi1} had early revealed the existence of a Wilson loop winding in spinful topological insulators with inversion symmetry. An example of Wilson loop winding in a split spinless EBR has also been shown in Ref.~\onlinecite{EBR_2}, however without a proof that it cannot be trivialized. 

In this work we rigorously characterize the topology of split (spinful) EBRs in terms of an obstructed $\mathbb{Z}$ classification of the winding of Wilson loop spectra over the BZ combining the homotopy of the flow of two-band Wilson loops and the symmetry quantization of Wilson loops over special base loops. Furthermore, we reveal a refined topological feature of split EBRs under the addition of extra bands. Namely, we show that even when multi-band Wilson loops can be unwinded, two-band subspaces that remain separated by a band gap from the other bands conserve their Wilson loop winding. This in particular allows us to elucidate the recently coined concept of ``fragile topology''\cite{Ft1} from a quantitative as well as clear conceptual perspective. Finally, we argue that only the stable Fu-Kane-Mele $\mathbb{Z}_2$ invariant characterizes the winding of the Wilson loop in the many-band limit implying that the further stable topological phases fall in the class of the higher-order topological insulators.\cite{SOT1, SOT2} 

Altogether, this sets the basis towards a systematic classification of topological crystalline phases with TRS, where we rigorously position fragile topology of split EBRs as the bridge in crystalline systems between unstable topology, i.e.~pertaining to few-band models (e.g.~Hopf insulators), and stable topology, i.e.~unaffected by the addition of trivial bands (e.g.~Chern insulators). 

\begin{figure}[t]
\centering
	\includegraphics[width=1.\linewidth]{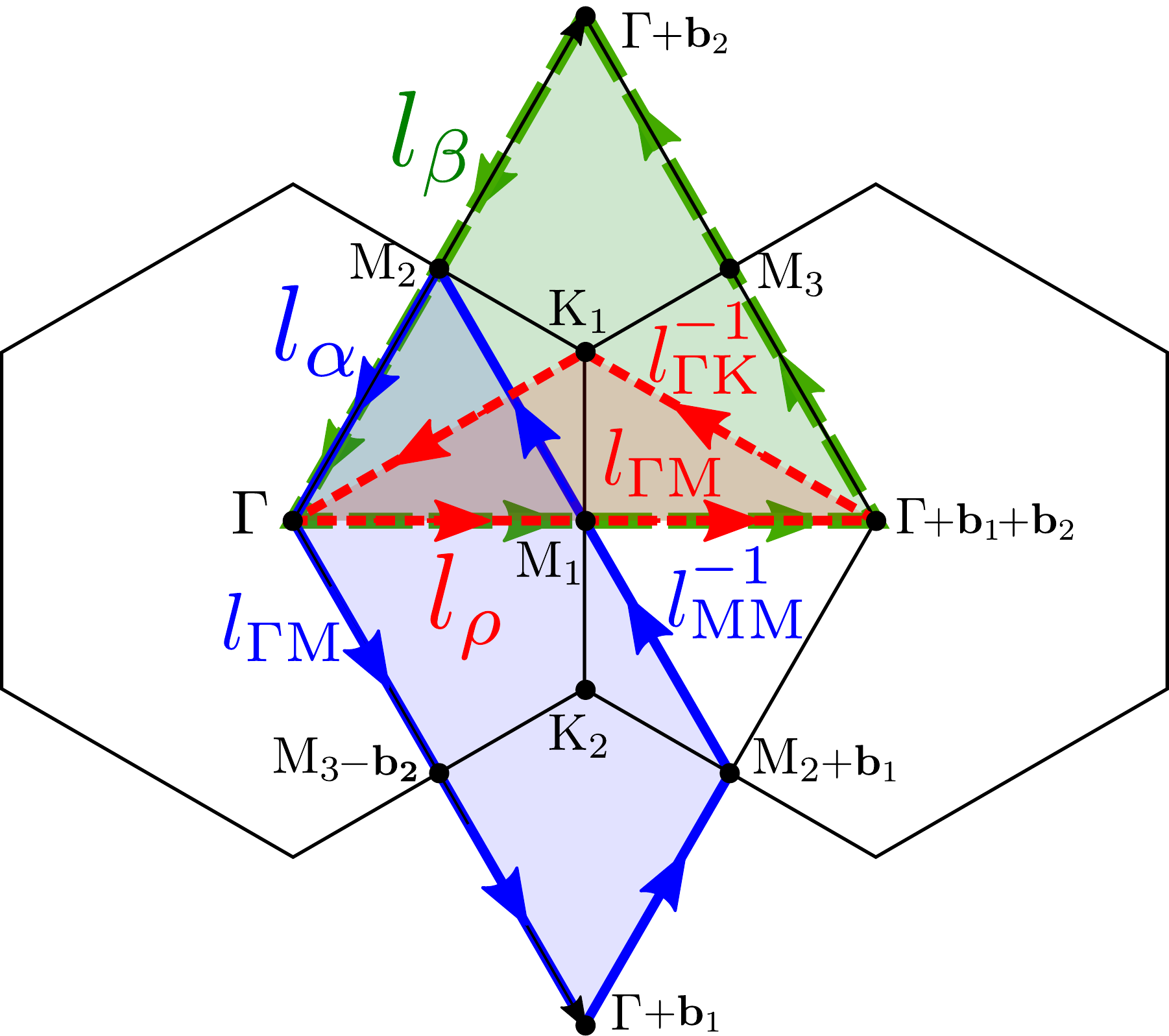} 
\caption{\label{fig_BZ} First BZ of the honeycomb lattice spanned by the primitive reciprocal vectors $\boldsymbol{b}_1$ and $\boldsymbol{b}_2$. High-symmetry points (HSP) $\Gamma$, K and M. Inequivalent high-symmetry points are written with different subscripts, i.e.~K$_1$, K$_2$, M$_1$, M$_2$, M$_3$. Half Brillouin zone (HBZ) patches $S_{\alpha}$ (blue) and $S_{\beta}$ (green), with their respective oriented boundaries $l_{\alpha}\sim l^{-1}_{\text{M}\text{M}} \circ l_{\Gamma\text{M}}$ (blue) and $l_{\beta} $ (dashed green), and the patch $S_{\rho}$ (red) bounded by $l_{\rho}= l^{-1}_{\Gamma\text{K}} \circ l_{\Gamma\text{M}} $ (dotted red) covering one sixth of the BZ. $l_{P_2P_1}$ is the oriented open loop that threads the BZ starting at point $P_1$, crossing $P_2$ and ending at the shifted point $P_1+\boldsymbol{K}$ with the reciprocal lattice vector $\boldsymbol{K} = \pm \boldsymbol{b}_{1(2)}$. We mean by $\sim$ equal up to segments that cancel each other through a translation by a reciprocal lattice vector. Note that the different segments of the curves directly correspond to the underlying crystal symmetries. 
}
\end{figure} 
To concretize the discussion we consider the honeycomb lattice and focus on the specific symmorphic layer groups No.~80 (L80) and No.~77 (L77).\cite{ITCA,ITCD} For simplicity, we set one orbital per site. This nonetheless does not compromise the generality of our results as they illustrate an entirely generic and systematic method that can be extended to any other context as it reveals a direct relation between crystalline invariants and the flow of Wilson loop spectra over patches of the Brillouin zone that are dictated by the lattice symmetries. L80 contains inversion symmetry ($\mathcal{I}$) and is characterized by the point group $D_{6h}$, which incidentally also contains mirror symmetry with respect to the basal plane ($\sigma_h$). L77 on the contrary does not include inversion nor basal mirror symmetries and is characterized by the point group $C_{6v}$. For both layer groups the honeycomb lattice corresponds to Wyckoff's position $2b$ such that the unit cell contains only two sub-lattice sites.\footnote{All the crystallographic and group theoretic data can be found in the International Tables for Crystallography\cite{ITCA,ITCD} and the Bilbao Crystallographic Server.\cite{Bilbao,Double_IRREPs} The tables of IRREPs used in the work were retrieved from the Bilbao Crystallographic Server using the space groups SG183 (L77), SG191 (L80), and SG175 (L75).} Later in the work we also consider the effect of coupling the honeycomb lattice with the triangular lattice made of the sub-lattice sites located at the center of the honeycombs, i.e. they realize the Wyckoff's position $1a$ of L80 and L77. We assume time reversal symmetry (TRS) throughout the work.

Having set the stage, the work is organized in the following way. In Section~\ref{spinless} we derive the exhaustive classification of two-band spinless topological semi-metallic phases for the honeycomb lattice. We show that on top of the well known essential nodal point at K (``Dirac cone'') characterized by a `$\pi$'-Berry phase, an infinity of distinct topological phases exists corresponding to the accumulation of symmetry protected nodal points which cannot be removed as long as the band gap remains open at the high-symmetry points ($\Gamma$, M). For this we use a global approach that goes beyond local $\boldsymbol{k}\cdot\boldsymbol{p}$-type analysis. 

Once the spinless classification is known, we move on to answering the question of the effect of spin-orbit coupling (SOC). We start with the inversion symmetric case for which SOC always gaps the spinless band structure. In Section~\ref{spinful_inversion} we derive the exhaustive classification of four-band spinful topological insulating phases for the honeycomb lattice when inversion symmetry is conserved. The classification of the spinful inversion symmetric case is done in terms of the flow of the spin-polarized $U(1)$-Berry phase (i.e.~providing the spin-polarized Chern number). When the spin components cannot be separated, i.e. by breaking inversion symmetry, the classification must be done in terms of the flow of $U(2)$-Wilson loop over the BZ. For this we introduce in Section \ref{WLW} the Wilson loop winding and relate it to the symmetry protected quantization of Wilsonian phases over special base loops of the BZ. In particular, we show that while $C_{2z}\mathcal{T}$ symmetry (rotation combined with TRS) protects a $\mathbb{Z}$ Wilson loop winding for two-band subspaces, the $C_3$ symmetry of the hexagonal lattice acts as an obstruction that forbids the zero winding. As a consequence a split EBR must be topologically nontrivial with a finite Wilson loop winding. 

We then go further without inversion symmetry in Section \ref{spinful_ad_inv}, where we show that the classification of spinful topological insulating phases with adiabatically broken inversion symmetry, i.e.~for weak Rashba SOC, agrees with the classification with inversion symmetry. In particular, in this case the spinless classification can be lifted into the spinful context according to a one-to-two correspondence. When inversion symmetry is broken non-adiabatically however, i.e.~when Rashba SOC triggers a topological phase transition, the correspondence between the spinless and the spinful classifications is lost. We overcome this in Section \ref{non_adiabatic} where we present a mechanism based on the spin-locking between $\Gamma$ and K of a set of smooth, periodic and rotation-symmetric Bloch functions that spans the occupied subspace, i.e.~a frame made out of sections of the vector bundle. Since this frame diagonalizes both the Wilson loop and the matrix representation of rotations we obtain the symmetry protected quantization of Wilson loop spectra and the constraints from symmetry on the set of allowed Wilson loop windings over the BZ. This leads to the quantitative characterization of the topology of split EBRs in terms of Wilson loop winding. Furthermore, our approach gives the quantitative as well as conceptual clarification of a new topological phase reported recently by Ref.~\onlinecite{Ft1}, where it was called ``fragile topology'' after showing that it can be adiabatically mapped onto an atomic insulator (i.e.~with a symmetric and localized Wannier representation) when extra trivial bands are included.

Given the finding of Ref.~\onlinecite{Ft1} a natural task is to determine the stability of the topology of split EBRs. For this, we study numerically in Section \ref{six_band} the six-band case by including an extra sub-lattice site at the center of the unit cells, i.e.~coupling the honeycomb lattice (Wyckoff's position $2b$ of L77) with the triangular lattice (Wyckoff's position $1a$ of L77). We find that the Wilson loop windings of two-band subspaces that originate from a split EBR are unstable under the \textit{nonadiabatic} coupling with extra trivial bands (i.e. closing the energy gap), and this independently of the set of IRREPs of the bands under consideration hence greatly generalizing the finding of Ref.~\onlinecite{Ft1}. Based on the results of Section \ref{six_band} we give in Section \ref{discussion} a detailed discussion for all the combinatorial ways of forming two-band and four-band subspaces. We show that while the many-band Wilson loops loose the $\mathbb{Z}$-type winding breaking down to the $\mathbb{Z}_2$ Fu-Kane-Mele classification, the two-band subspaces originating from a split EBR that remain separated from the other bands by a band gap do conserve their nonzero Wilson loop winding. We argue that this fragile topology of split EBRs is related to the existence for two-band subspaces of a smooth (periodic) and rotation symmetric Bloch frame over special base loops of the BZ, while such a frame fails to exist in four-band subspaces. This provides a rigorous characterization of the fragile topology of split EBRs within homotopy theory of vector bundles. Furthermore, we identify the patterns of Wilson loop spectra that survive as the number of bands is increased, thereby identifying the Fu-Kane-Mele $\mathbb{Z}_2$ invariant as the only invariant that characterizes the Wilson loop winding in the many-band limit. Our results thus bridge the gap between unstable topology, as described by homotopy theory for few-band models, and stable topology that is captured by K-theory and is unchanged under the addition of trivial bands. We eventually briefly discuss the breaking of $C_{2z}\mathcal{T}$ and TRS from which insulating phases with a high Chern number can be generated. 

Let us end this introduction with a note on terminologies. We follow for convenience the vocabulary of TQC given very recently in Ref.~\onlinecite{Clas5, EBR_1}. A band representation (BR) is any band structure in reciprocal space that is induced from a basis set of degrees of freedom localized in real space, i.e.~represented by localized Wannier functions, such that the set is closed under all the symmetries of the system, i.e. each symmetry acts as a permutation of the Wannier functions. A BR is composite if it is equivalent to a direct sum of BRs. An EBR is a BR that is not composite. Whenever a BR also satisfies TRS is called a physical band representation (PBR). Therefore, in this work all the BRs are physical. A general band representation, or a quasiband representation (qBR), is any group of bands isolated from the other bands by an energy gap, i.e.~it must satisfy all the compatibility relations that connect the IRREPs of the different regions of the BZ. 

\section{Spinless case}\label{spinless}

Let us start the discussion with the well known characterization of nodal charges in spinless systems in terms of the Berry phase. A special feature of the spinless two-band model is that inversion symmetry ($\mathcal{I}$) and spinless TRS ($\mathcal{T}^2=+1$) are both effectively satisfied. Hence, the results of this section characterize both Altland-Zirnbauer (AZ) classes\cite{Schnyder08} A and AI, and the two layer groups L77 and L80, i.e.~AI+L80($2b$)$\cong$A+L77($2b$) for a two-band model. While we refer to the symmetry class AI+L80($2b$) in this section, we use the fewer IRREPs of L77 for convenience. We give the relevant IRREPs for the little co-groups at $\Gamma$, K and M in Table \ref{T_comp}. 

{\def\arraystretch{1.3}  
\begin{table}[t]
\caption{\label{T_comp} Spinless IRREPs of rotations of the little co-groups\cite{BradCrack} at $\Gamma$, K, and M, for the layer groups L77, L80, and L75. Retrieved from the Bilbao Crystallographic Server.\cite{Bilbao}}  
\begin{tabular}{ c | c cc|cc }
\hline 
\hline 
	$\Gamma $ & $C_{6v}$ & $D_{6h}$ &     $C_{6h}$ & $C_{2z}$ & $C_{3z}$ \\
	\hline 
	& $\Gamma_1$ & $\Gamma_2^-$ & $\Gamma_1^-$ & $1$ & $1$ \\
	& $\Gamma_4$ & $\Gamma_3^+$ & $\Gamma_2^+$ & $-1$ & $1$ \\
	\hline
	$\text{K} $ &  $C_{3v}$ & $D_{3h}$ &      $C_{3h}$ & $C_{3z}$ &  \\
	\hline 
	& $K_3$ & $K_6$ & $\begin{array}{c}K_4 \\ K_6 \end{array}$ & $\begin{array}{c} e^{i 2\pi/3} \\ e^{-i 2\pi/3} \end{array}$ &  \\
	\hline
	$\text{M} $  &  $C_{2v}$ & $D_{2h}$ &      $C_{2h}$ & $C_{2z}$ & \\
	\hline 
	& $M_1$ & $M_2^-$ & $M_1^-$ & $1$ & \\
	& $M_4$ & $M_3^+$ & $M_2^+$ & $-1$ & \\
	\hline
	\hline
\end{tabular}
\end{table} 
}

\begin{figure}[t]
\centering
\begin{tabular}{cc}
	\includegraphics[width=0.5\linewidth]{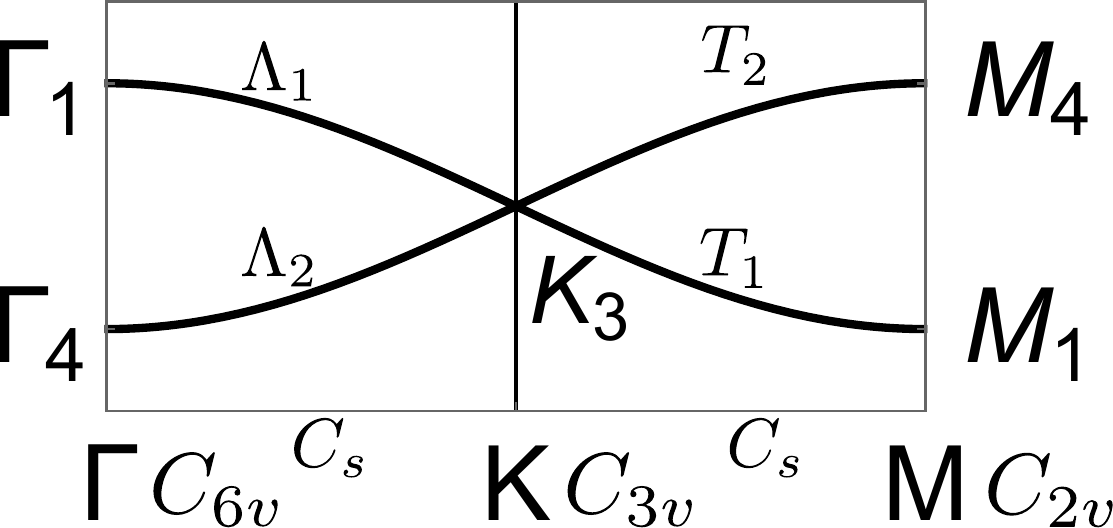} &
	\includegraphics[width=0.5\linewidth]{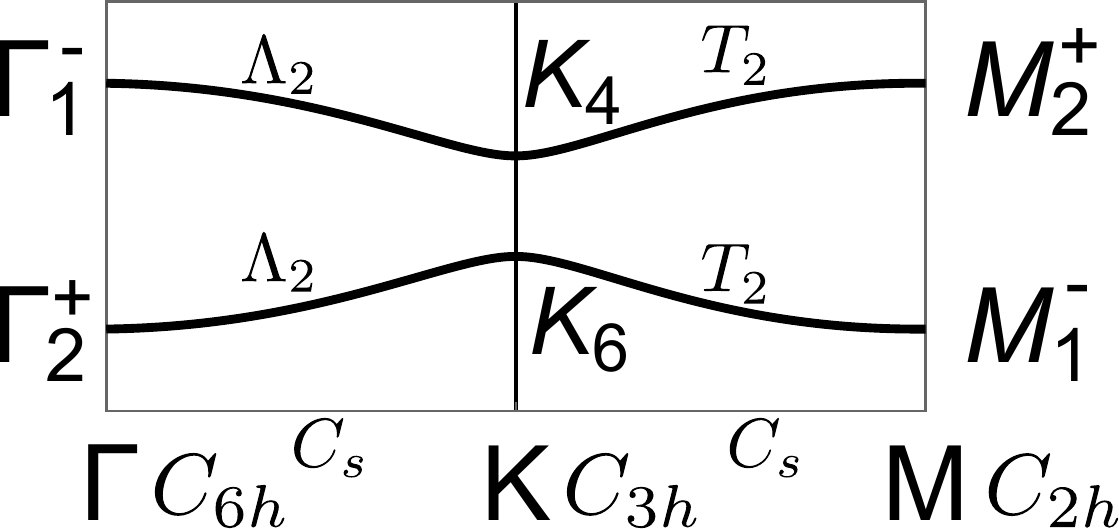} \\
	(a) & (b) 
\end{tabular}
\caption{\label{fig_BS_IRREPs} Examples of band structures for the EBRs of (a) spinless symmetry class A+L77($2b$) $\cong$ AI+L80($2b$), and (b) spinless symmetry class A+L75($2b$), when restricted to Wyckoff's position $2b$. Notations from the Bilbao Crystallographic Server\cite{Bilbao} for the IRREPs of the space groups.}
\end{figure}

The EBR of the spinless honeycomb lattice is two-dimensional as a result of the fact that the two sub-lattice sites are inseparable under L77, e.g.~$C_{6}$-symmetry exchanges $A$ and $B$ sites. The two-band EBR is characterized by the IRREPs at the high-symmetry points (HSPs) and high-symmetry lines (HSLs) of the BZ according to Fig.~\ref{fig_BS_IRREPs}(a). The simple nodal point (NP) at K is an essential degeneracy of AI+L80($2b$) such that the EBR cannot be split, i.e.~no band gap can separate the two bands over the BZ. It follows that the system must be a topological semimetal at half filling, i.e.~there is one (spinless) electron per unit cell. In the following we always consider the half-filled case with a formal Fermi level set such that half of the eigenstates have energy below and half have energy above the Fermi level. 

The stability of the NP is captured by the Berry phase computed over a base loop that encircles the NP and where only the occupied eigenstate is taken into account. Let us first consider the half BZ (HBZ) patch $S_{\alpha}$ that is bounded by the oriented loop $l_{\alpha} \sim  l^{-1}_{\text{M}\text{M}}  \circ l_{\Gamma\text{M}}$ (blue solid line in Fig.~\ref{fig_BZ}, where $\sim$ means that we discard the two segments that cancel each other under a translation by a reciprocal lattice vector), where we write $l_{P_2P_1}$ an oriented open loop in momentum space that threads the BZ starting at $P_1$, crossing $P_2$ and ending at the shifted point $P_1+\boldsymbol{K}$ ($P_{1,2}\in\{\Gamma,\text{K}_{1,2},\text{M}_{1,2,3}\}$, $\boldsymbol{K}=\pm \boldsymbol{b}_{1(2)}$, see Fig.~\ref{fig_BZ}), and $l^{-1}$ is the loop $l$ in the reversed orientation. Folding\cite{PointGroupsTI,hourglass,AlexAdiabaticSOC,Wi2} $l_{\Gamma\text{M}}$ and $l_{\text{M}\text{M}}$ with $C_{2z}$, we find the Berry phase factor ($\gamma[l_{\alpha}]\equiv \gamma_{\alpha}$)
\begin{equation}
\label{gamma_b}
	e^{i \gamma_{\alpha}} = \xi^{\Gamma}_{2} (\xi^{\text{M}}_{2})^3\;,
\end{equation}
where $\xi^{\bar{\boldsymbol{k}}}_{n}$ is the $C_{nz}$-rotation symmetry eigenvalue of the occupied eigenstate at HSP $\bar{\boldsymbol{k}}$, i.e.~given the allowed IRREPs of Fig.~\ref{fig_BS_IRREPs}(a) these are $\xi^{\Gamma}_{2} = \xi^{\text{M}}_{2} = +1$ for the IRREPs $\Gamma_1$ and $M_1$, and $\xi^{\Gamma}_{2} = \xi^{\text{M}}_{2}= -1$ for the IRREPs $\Gamma_4$ and $M_4$. This leads to a $\mathbb{Z}_2$ quantization of the Berry phase. For instance, given the occupied IRREPs of Fig.~\ref{fig_BS_IRREPs}(a) we find that the NP at K is characterized by a `$\pi$' Berry phase, or strictly speaking, by the set $\gamma_{\alpha} \in \{\pi + n 2\pi\}_{n\in \mathbb{Z}}$ which we write as a congruence relation $ \gamma_{\alpha} = \pi ~(\mathrm{mod}~2\pi) $. Performing a band inversion at $\Gamma$, $\Gamma_4 \leftrightarrow \Gamma_1$, (or similarly at M, $M_1 \leftrightarrow M_4$) an extra simple NP must appear on each HSL $\Lambda \equiv \overline{\Gamma \text{K}}$ (respectively $\text{T}\equiv \overline{\text{M} \text{K}}$). Indeed, band crossings over $\Lambda$ (T) cannot be avoided due to the distinct symmetry characters of the bands with respect to the vertical mirror symmetries of the HSL as marked by the IRREPs $\Lambda_{1,2}$ and $T_{1,2}$, see Fig.~\ref{fig_BS_IRREPs}(a). We now have four simple NPs within the HBZ $S_{\alpha}$, but Eq.~(\ref{gamma_b}) gives $\gamma_{\alpha}  = 0 ~(\mathrm{mod}~ 2\pi)$ such that this $C_{2z}$-based computation of the Berry phase looses track of the NPs. 

\begin{figure*}[t]
{\centering
\begin{tabular}{c|c|c|c}
	\includegraphics[width=0.25\linewidth]{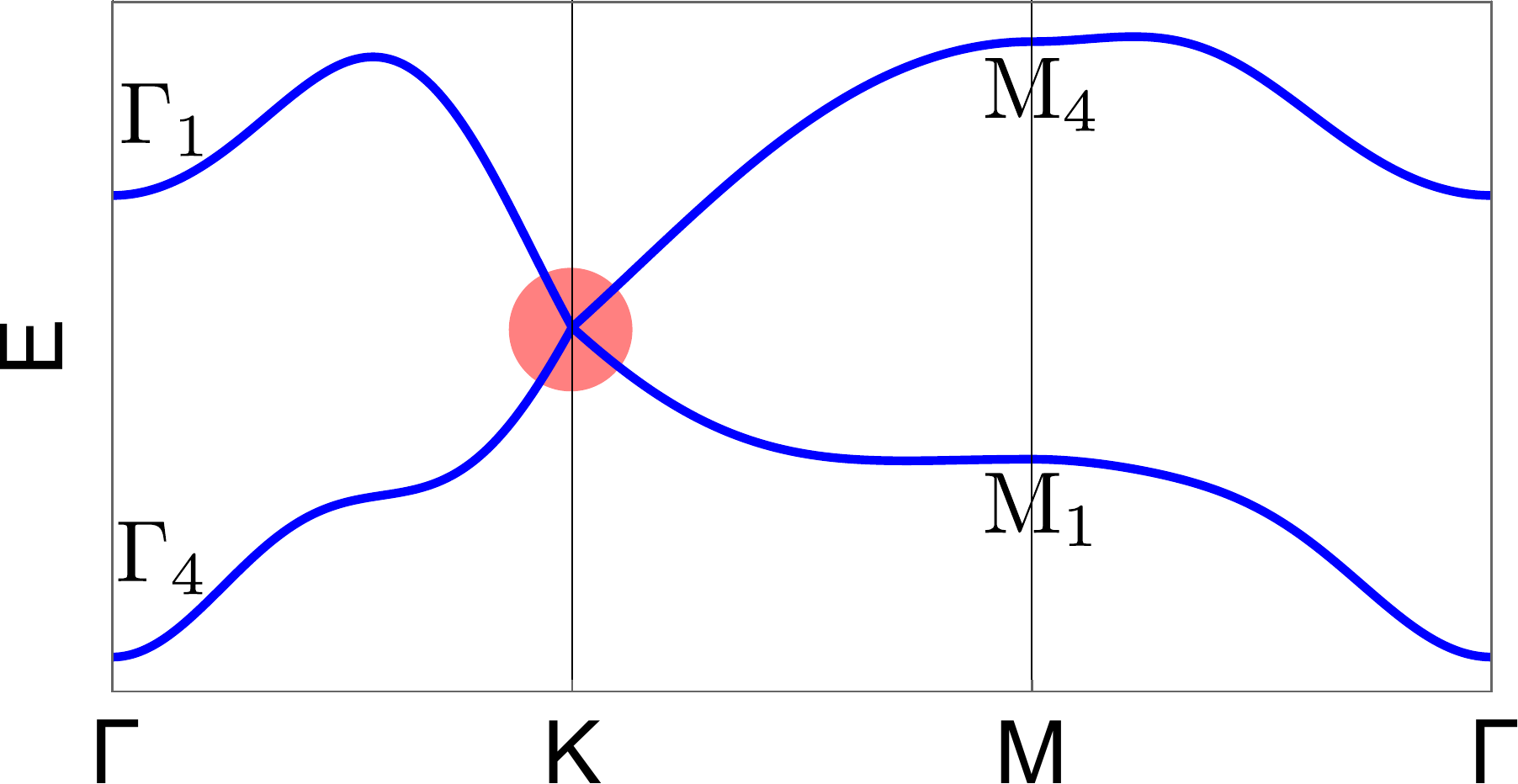} & 
	\includegraphics[width=0.25\linewidth]{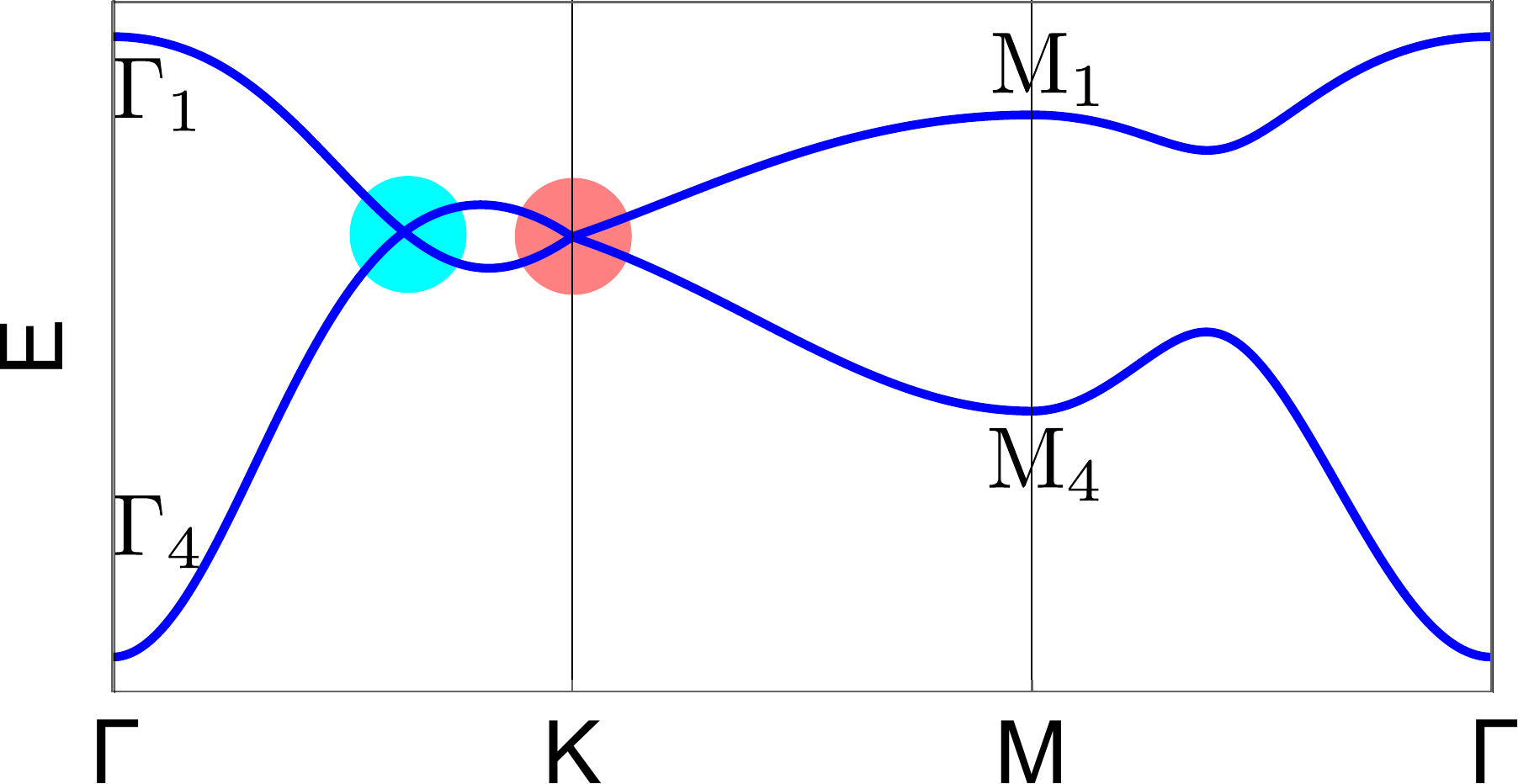} &
	\includegraphics[width=0.25\linewidth]{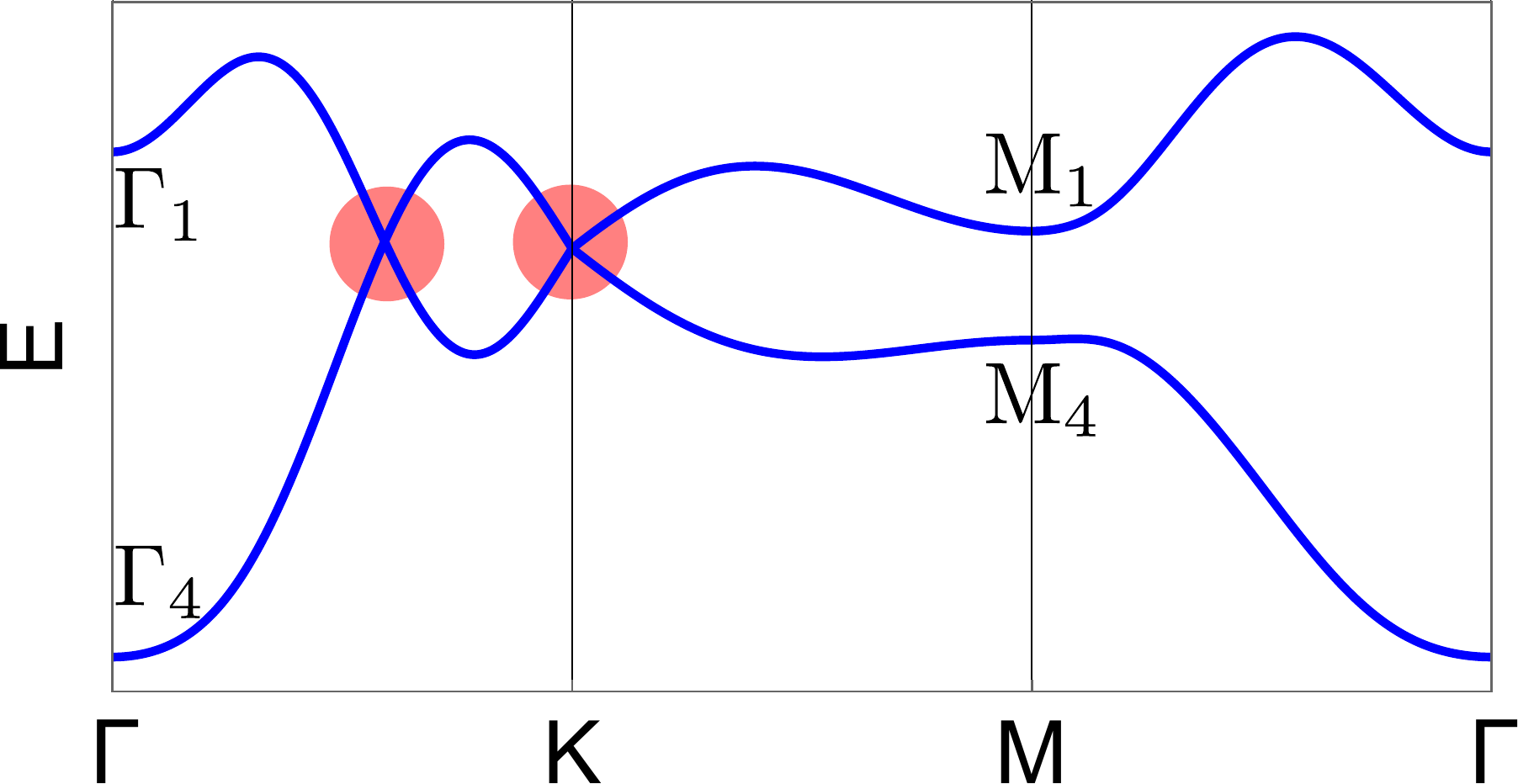} & 
	\includegraphics[width=0.25\linewidth]{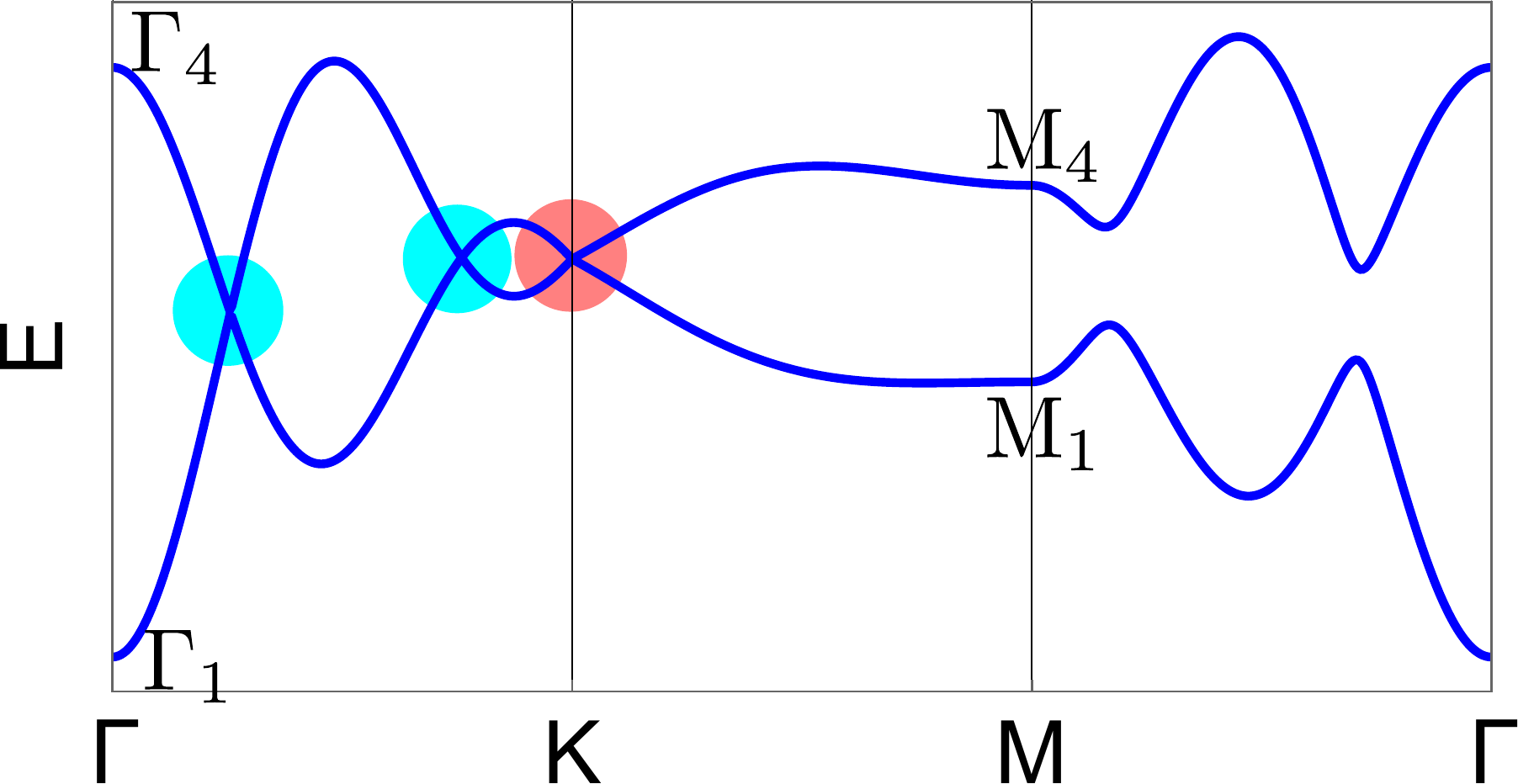}\\
	\includegraphics[width=0.25\linewidth]{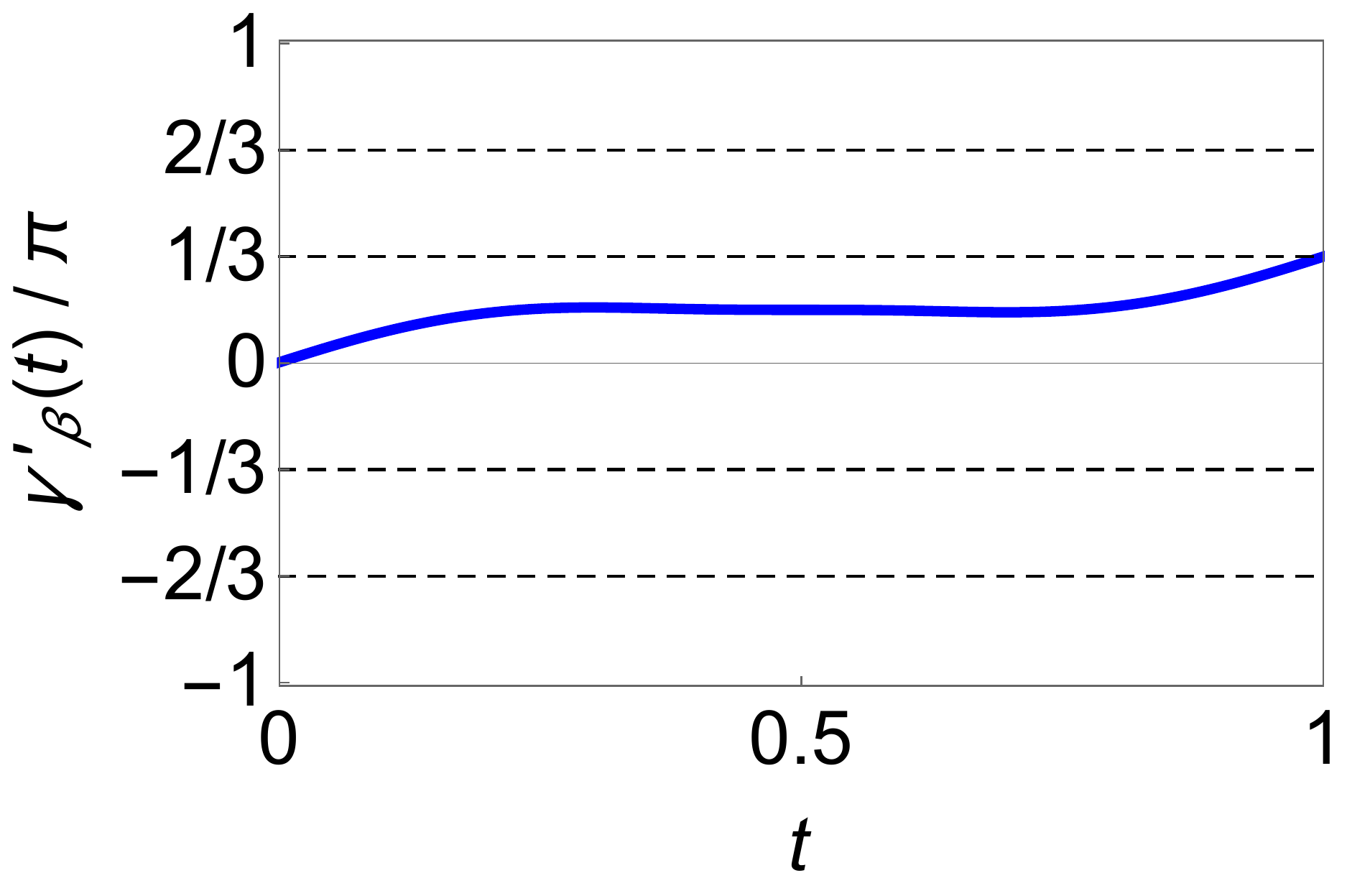} &
	\includegraphics[width=0.25\linewidth]{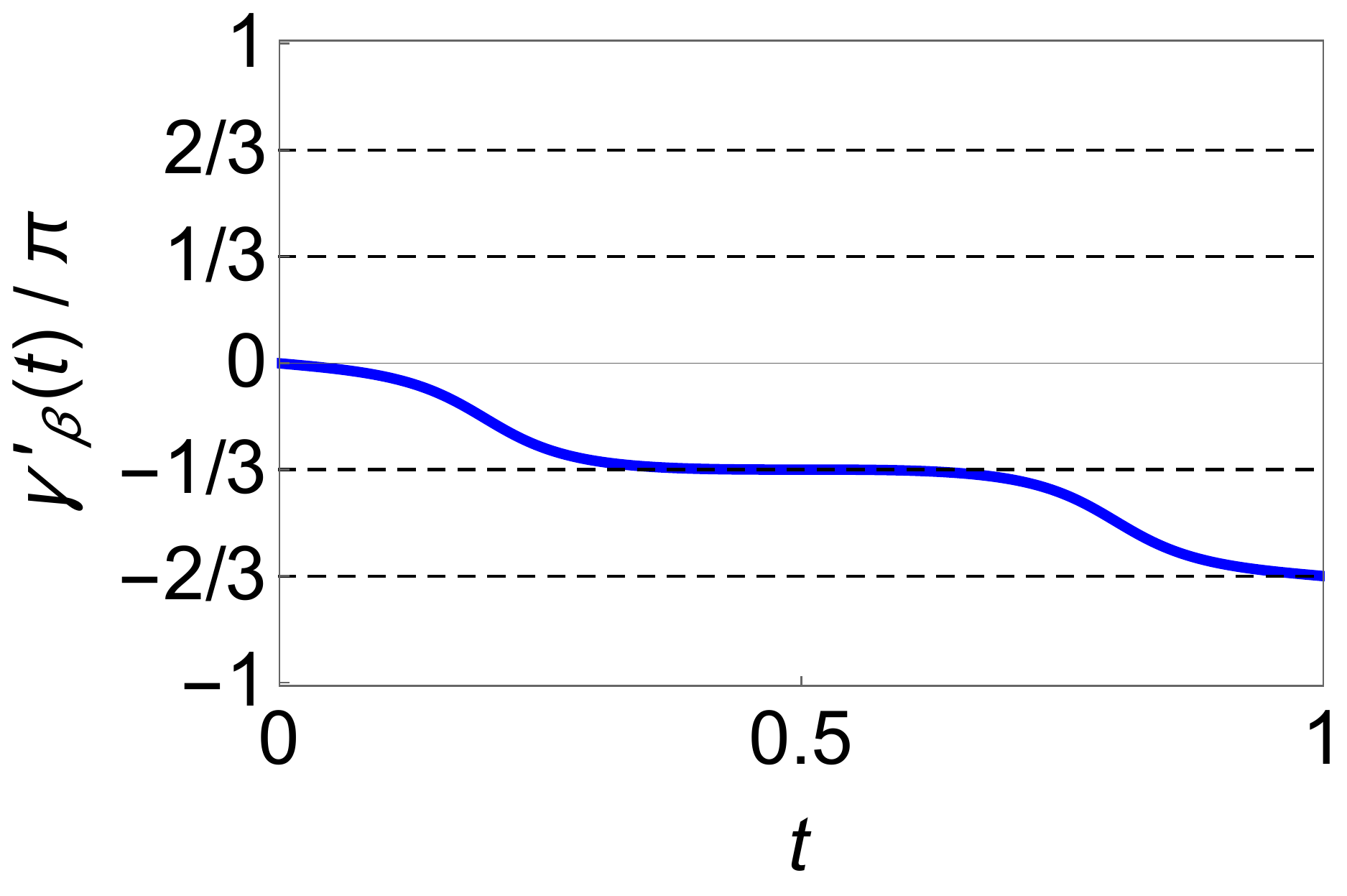} &
	\includegraphics[width=0.25\linewidth]{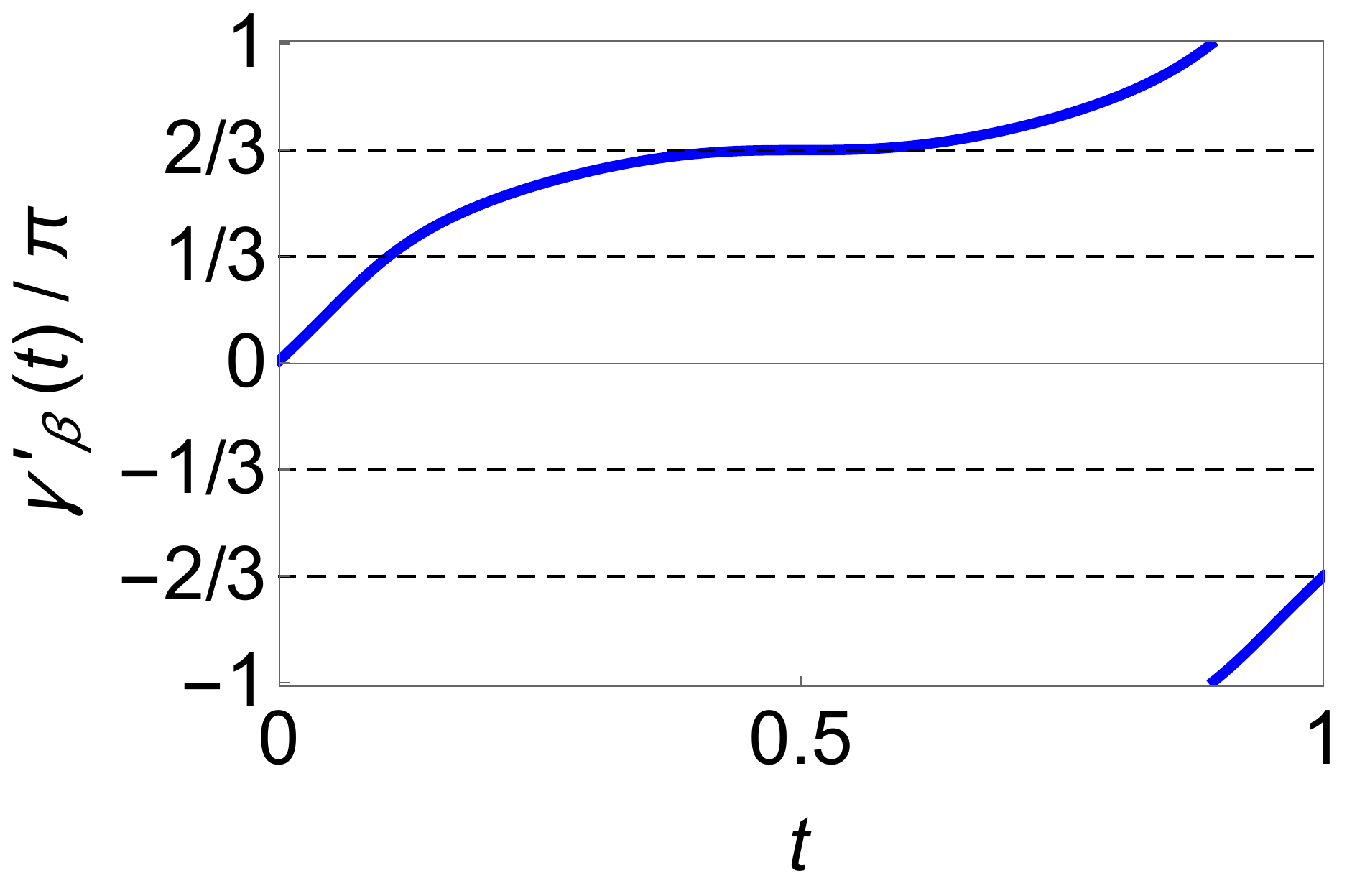} & 
	 \includegraphics[width=0.25\linewidth]{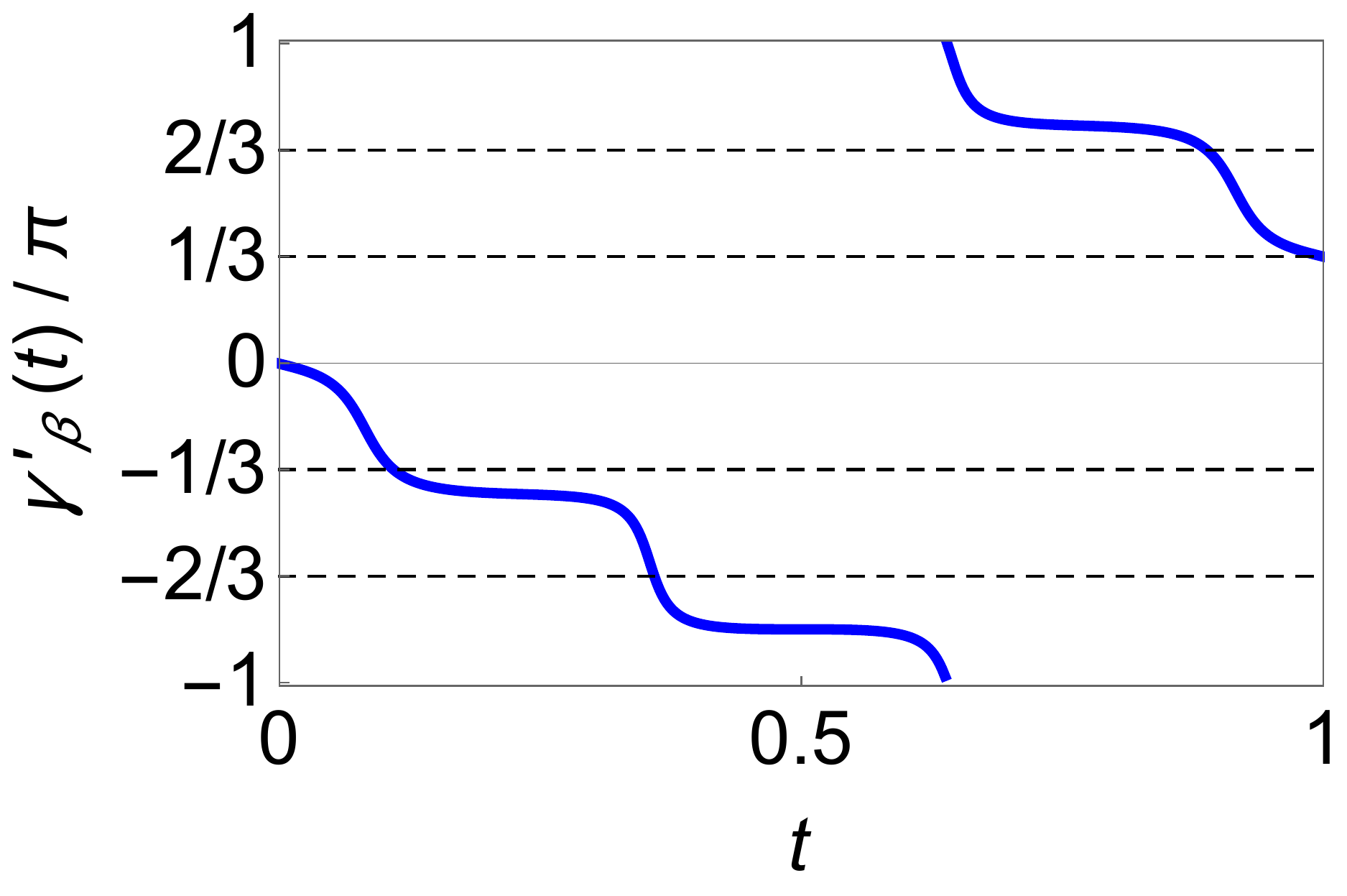} \\
	 $\gamma'_{\beta} = \pi/3$ & $\gamma'_{\beta} = -2\pi/3$ & 
	 $\gamma'_{\beta} = 4\pi/3$ & $\gamma'_{\beta} = -5\pi/3$\\
	 \includegraphics[width=0.2\linewidth]{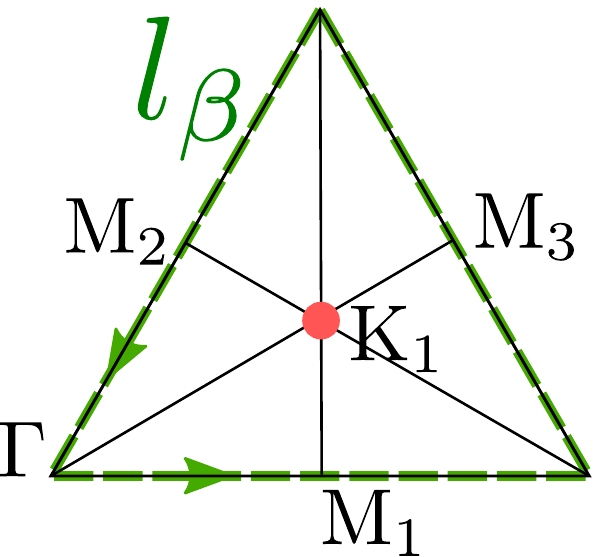} &
	\includegraphics[width=0.2\linewidth]{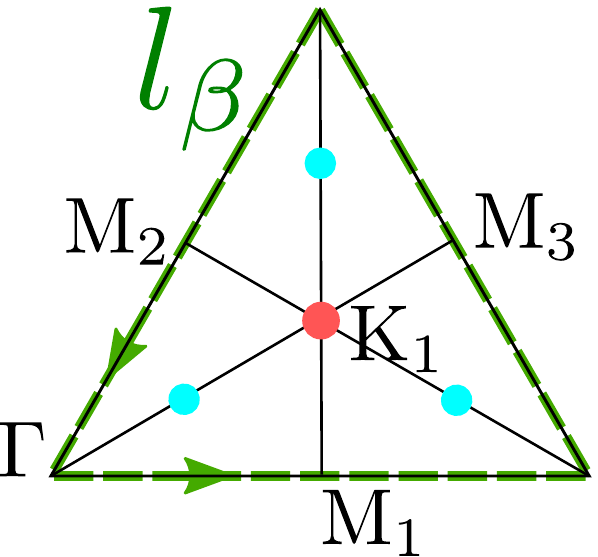} &
	\includegraphics[width=0.2\linewidth]{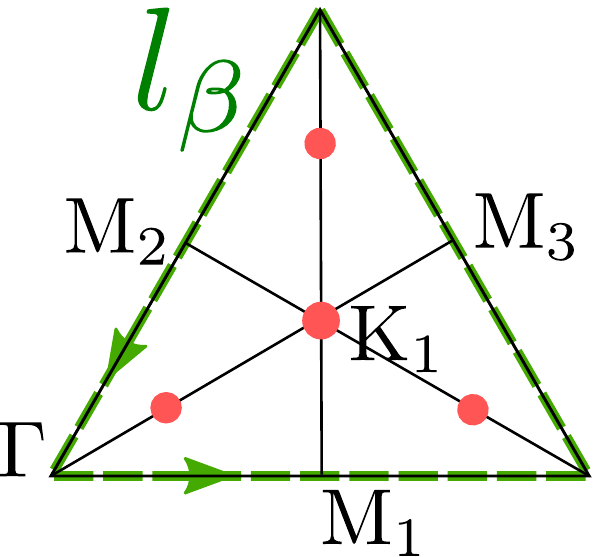} & 
	\includegraphics[width=0.2\linewidth]{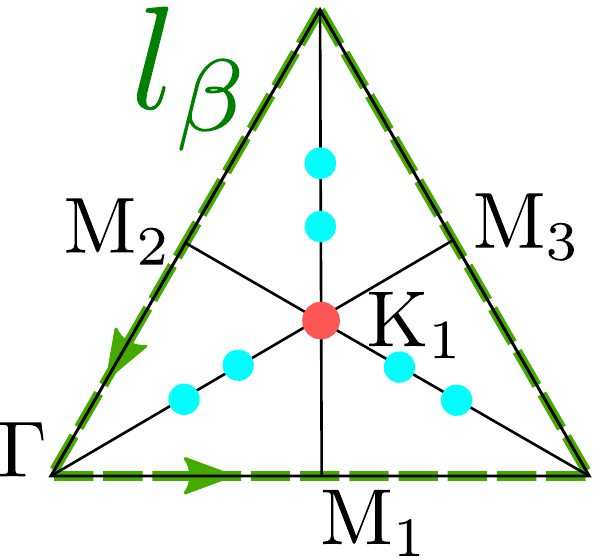} \\
	(a)  $W_I = +1$ & (b)  $W_I = -2$ & 
	(c) $W_I = +4$ & (d)  $W_I = -5$ 
\end{tabular}
}
\caption{\label{fig_nodal_points} Spinless topological semi-metallic phases of the honeycomb lattice reached through band inversions at $\Gamma$ and M triggered by tight-binding parameters of increasing range. Columns (a-d) correspond to an increasing number of simple nodal points protected by crystalline symmetries. First row: band structures along the high symmetry lines of the BZ. Second row: noncyclic Berry phase along one third of the $C_{3}$-symmetric loop $l_{\beta}$ encircling K. Third row: schematic configuration of symmetry protected nodal points within the HBZ $S_{\beta}$ bounded by $l_{\beta}$ (nodal points in red have $W_I=+1$, and in blue $W_I=-1$). Crystalline winding numbers are (a) $W_{I}=\gamma_{\beta}/\pi = +1$, (b) $W_{I}=-2$, (c) $W_{I}=+4$, and (d) $W_{I}=-5$.
}
\end{figure*}

In order to have a finer definition of the Berry phase which takes advantage also of the other crystalline symmetries, let us now consider the HBZ patch $S_{\beta}$ bounded by the oriented loop $l_{\beta}$ (green dashed line in Fig.~\ref{fig_BZ}). Contrary to $l_{\alpha}$, $l_{\beta}$ exhibits the $C_{3z}$-symmetry of the K point which allows us to split it into three symmetric sections. A third of the loop can therefore be taken as the path beginning at M$_2$ crossing $\Gamma$ and ending at M$_1$, i.e.~$l'_{\beta} = l_{\text{M}_1\Gamma} \circ l_{\Gamma\text{M}_2}$. Then using a smooth reference gauge for the eigenstates, we find the non-cyclic Berry phase over the loop segment $l'_{\beta}$ (see derivation in Appendix \ref{noncyclic_BP})
\begin{equation}
	e^{i \gamma'_{\beta}}   = \dfrac{2 - (1+i \sqrt{3}) \xi^{\Gamma}_{2} \xi^{\text{M}}_{2} }{2 - (1-i \sqrt{3}) \xi^{\Gamma}_{2} \xi^{\text{M}}_{2} } \;,
\end{equation}
with the symmetry eigenvalues of the occupied eigenstate determined by 
\begin{equation}
\label{IRREP_ordering}
\begin{aligned}
	\xi^{\Gamma}_{2}  & = -\mathrm{sign}\{ \mathrm{Re} [h_{AB}(\Gamma)] \} \;, \\
	\xi^{\text{M}}_{2}  & =  \mathrm{sign}\{ \mathrm{Re} [h_{AB}(\text{M})]\} \;. 
\end{aligned}
\end{equation}
Here $h_{AB}(\boldsymbol{k})$ is the off-diagonal element of the tight-binding model written in the Bloch basis $\sum_{\boldsymbol{R}_n} e^{i \boldsymbol{k} (\boldsymbol{R}_n+\boldsymbol{r}_{A(B)}) } \vert w_{A(B)} , \boldsymbol{R}_n \rangle $ where $\boldsymbol{R}_n$ labels the unit cells, $\boldsymbol{r}_{A(B)}$ is the sub-lattice position within one unit cell, and $\langle \boldsymbol{r}\vert w_{A(B)} , \boldsymbol{R}_n \rangle=w(\boldsymbol{r}-\boldsymbol{R}_n-\boldsymbol{r}_{A(B)})$ gives a complete basis set of localized Wannier functions, see Appendix \ref{spinless_ap} for more details. 
Therefore, since by rotational symmetry $e^{i \gamma_{\beta}} = (e^{i \gamma'_{\beta}})^3$, we find for the phase winding over the entire loop $l_{\beta}$
\begin{equation}
\label{berry_phase_lg}
	\gamma_{\beta} = \left\{ 
	\begin{array}{rcr} \pi  ~(\mathrm{mod} ~6\pi) &\quad& \mathrm{if~~}\xi^{\Gamma}_{2} \xi^{\text{M}}_{2}= -1 \\
								- 2\pi ~(\mathrm{mod} ~6\pi)  &\quad& \mathrm{if~~} \xi^{\Gamma}_{2} \xi^{\text{M}}_{2} = +1
	\end{array}\right. \;,
\end{equation}
from which we define the winding number $W_{I}=\gamma_{\beta}/\pi$. 

$W_{I}$ is the topological invariant that classifies all symmetry protected topological semimetallic phases of the spinless Graphene model with a single orbital per site. We illustrated this with the numerical examples of Fig.~\ref{fig_nodal_points} that realize $W_{I}=+1, -2,+4$, and $-5$. For each case, 
we show the band structure, the non-cyclic Berry phase over the loop segment $l'_{\beta}$, and the schematic configuration of NPs within the HBZ patch $S_{\beta}$. Starting from a band structure with a single NP at K for which $W_I=+1$ shown in Fig.~\ref{fig_nodal_points}(a), the other topological sectors are reached through successive band inversions at $\Gamma$ and M triggered by tight-binding parameters of increasing range.\footnote{We also find a topological transition with no band inversion at $\Gamma$ or M happening through the closing of a whole nodal line encircling $\Gamma$ (or similarly three nodal lines encircling the M points). However, such a nodal line is realized only at fine tuned values of the microscopic parameters.} For instance, restricting the tight-binding model to the nearest-neighbor parameters, only the single NP at K can be formed, while we included up to the 7th layer of neighbors in order to generate the higher winding of Fig.~\ref{fig_nodal_points}(d). Each band inversion leads to a jump of the winding number by $\pm$3, as predicted by Eq.~(\ref{berry_phase_lg}). 
We conclude that the two-band model of the honeycomb lattice at half-filling is classified by
\begin{equation}
\label{spinless_homotopy} 
	\pi^{(l_{\beta})}_1(\mathcal{H}^{1+1}_{\mathrm{AI+L80}(2b)}) \cong \boldsymbol{1}+3\mathbb{Z} \quad\ni \quad W_{I} = \dfrac{\gamma_{\beta}}{\pi}\;,
\end{equation}
where $\pi_1^{(l_{\beta})}$ means the first homotopy group restricted to the $C_{3}$-symmetric base loops $l_{\beta}$ connecting $\Gamma$ and the three inequivalent M points, $\mathcal{H}^{1+1}_{\mathrm{AI+L80}(2b)}$ is the classifying space (Grassmannian) of the two-band Hamiltonian at half-filling (1 occupied state + 1 unoccupied state) for the symmetry class AI+L80($2b$). The offset $\boldsymbol{1}$ marks the obstruction due to the essential NP at K that forbids the trivial phase. 

It is important to note that the topological content of Eq.~(\ref{berry_phase_lg}) and Eq.~(\ref{spinless_homotopy}) is \textit{relative} and not \textit{absolute}.\cite{Gauge_dependence} This means that they hold under the following relative assumptions. (i) A reference trivialization of the total Bloch bundle has been fixed. Practically this is done through the choice of a complete basis set of Bloch functions, here obtained from a complete basis set of localized Wannier functions.\cite{reference_trivialization} (ii) The eigenstates are defined within the same smooth reference gauge. This is done in the appendix through the choice of the global analytical ansatz for the wave function (see also the discussion on gauge transformation in Appendix \ref{gauge_dependence}). (iii) Large gauge transformations are excluded.\footnote{While large gauge transformations are allowed within equivalence classes defined up to bundle isomorphisms, they carry their own non-trivial windings and hence permit to jump between different homotopy equivalence classes, see Ref.~\onlinecite{Gauge_dependence}} (iv) The HSP K$_1$ has been chosen as a representative (choosing K$_2$ instead, while keeping fixed the origin of the Bravais lattice, reverses the winding numbers with the offset `$-\boldsymbol{1}$'). We also note that for a given winding number $W_I$ of the occupied subspace we find the reversed winding $-W_I$ for the unoccupied subspace. It is also worth noting that our approach is global in the sense that the results Eq.~(\ref{berry_phase_lg}) and Eq.~(\ref{spinless_homotopy}) are derived from a wave function that is smooth over the whole BZ (discarding the singularities at the NPs),\footnote{A global smooth gauge is allowed by the vanishing of the first Chern class in the gapless phase.\cite{global_top_semi}} see the details in Appendix \ref{ansatz}, hence allowing us to go significantly beyond the local $\boldsymbol{k}\cdot\boldsymbol{p}$-approach.\cite{BernevigHughes}

Finally we note that the $\mathbb{Z}$-type structure of Eq.~(\ref{spinless_homotopy}) can be understood as inherited from the few-band result $\pi_1(\mathcal{H}^{1+1}_{\mathrm{AI}+\mathcal{I}}) = \mathbb{Z}$.\cite{BzduSigristRobust} In our case, the crystalline symmetries act as an obstruction within the classifying superspace $\mathcal{H}^{1+1}_{\mathrm{AI}+\mathcal{I}}$ such that only a subset of the topological sectors are allowed ($ \boldsymbol{1}+3\mathbb{Z} \subset \mathbb{Z}$) excluding the trivial phase. In general, for a two-dimensional spinless system (or a two-dimensional momentum subspace) with no inversion symmetry, we can instead use the symmetry class $\mathrm{AI}+C_{2\perp}\mathcal{T}$ (with the $C_{2\perp}$-axis perpendicular to the system, here $C_{2z}$), since $C_{2\perp}\mathcal{T}$ similarly imposes a reality condition on the classifying space within the $\sigma_h$-mirror invariant plane ($\sigma_h=C_{2\perp}\mathcal{I}$) leading again to $\pi_1(\mathcal{H}^{1+1}_{\mathrm{AI}+C_{2\perp}\mathcal{T}}) = \mathbb{Z}$.\cite{FuC2T,BzduConversion}

\section{Spinful case with inversion symmetry}\label{spinful_inversion}

Having shown that the non-cyclic Berry phase is able to characterize the extra structure from the impact of crystalline symmetries in the case of spinless semi-metallic phases,  we now move to the spinful AZ symmetry class AII that is relevant when SOC is turned on.\cite{Schnyder08} Due to the spin degree of freedom the Hamiltonian is now four-by-four and we concentrate on the fully gapped phase at half filling.

In this section we focus on the case with inversion symmetry, i.e.~we characterize the topology of the symmetry class AII+L80($2b$). Basal mirror symmetry ($\sigma_h = C_{2z}\mathcal{I}$) forbids spin-flip terms in the four-band Hamiltonian such that it can be separated into spin-up and spin-down sectors leading to the decomposition $\mathcal{H}^{2+2}_{\mathrm{AII+L80}(2b)} = \mathcal{H}^{1+1}_{\uparrow} \oplus \mathcal{H}^{1+1}_{\downarrow}$ which makes the expectation value of the $z$-spin component $s_z$ a good quantum number over the whole BZ. Practically, the generic tight-binding Hamiltonian of this symmetry class corresponds to a generalized Kane-Mele model, i.e.~including arbitrary many neighbors, with zero Rashba SOC and no staggered potential (as to preserve inversion and $C_{2z}$ symmetry).

\begin{figure}[t]
\centering
\begin{tabular}{cc}
	\includegraphics[width=0.48\linewidth]{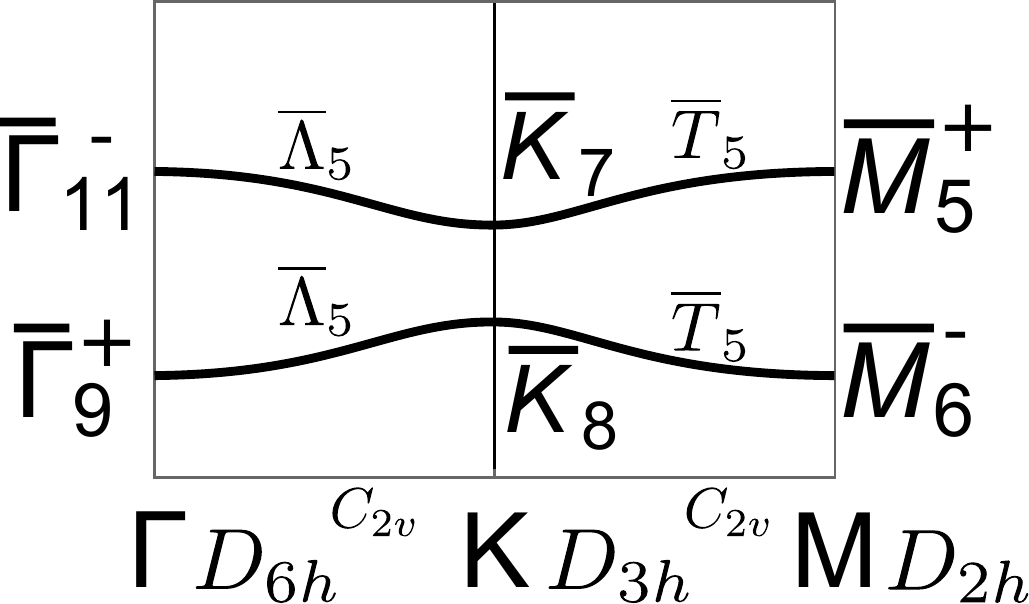} &
	\includegraphics[width=0.52\linewidth]{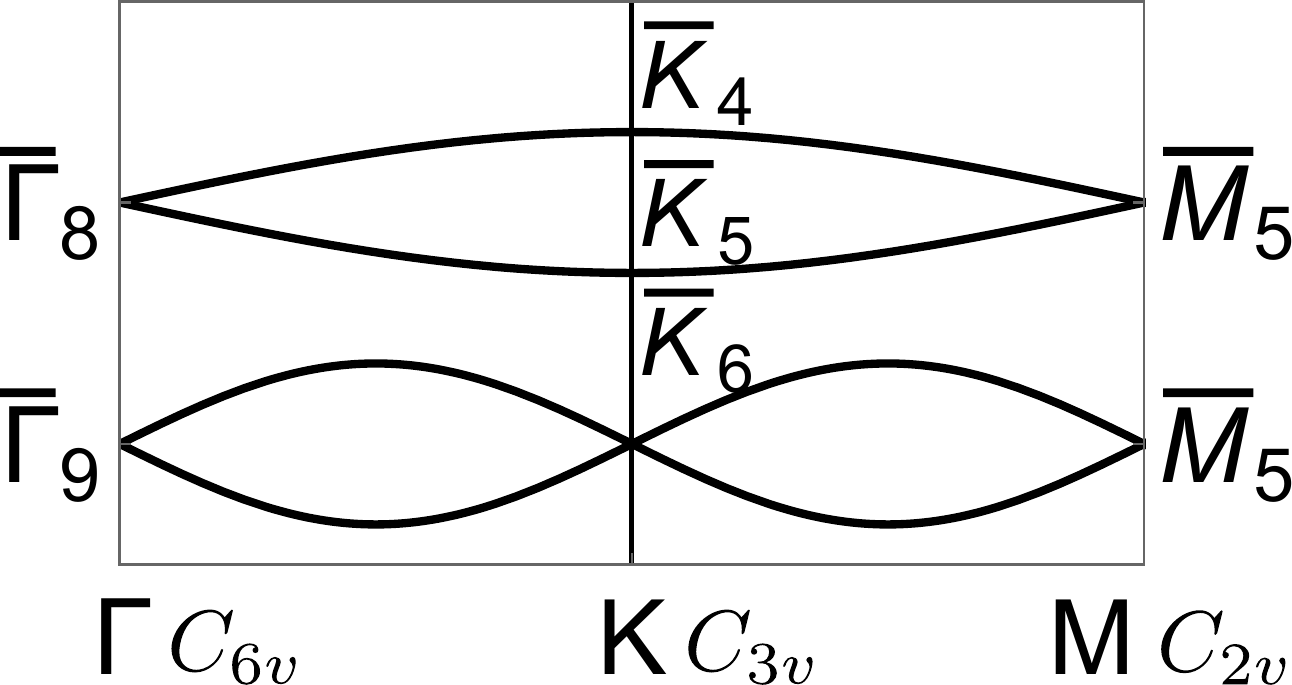} \\
	(a) & (b) 
\end{tabular}
\caption{\label{fig_BS_IRREPs_spinful} EBRs for (a) spinful symmetry class AII+L80$(2b)$, and (b) spinful symmetry class AII+L77$(2b)$, when restricted to the Wyckoff's position $2b$. Notations from the Bilbao Crystallographic Server\cite{Bilbao} for the IRREPs of the space groups.
}
\end{figure}

A direct consequence of the spin separation is that the topological classification is reduced to that one of the spin-polarized subspaces. Let us address the spinless symmetries of each spin-polarized subspace, i.e.~when we forget about the spin. Contrary to the spinless case (AI+L80) the spin polarized subspaces $\mathcal{H}^{1+1}_{\sigma}$ have no spinless TRS, i.e.~the AZ symmetry class is lowered as AI$\rightarrow$A, nor spinless vertical mirror symmetries, i.e.~the lattice symmetry class is lowered as L80($D_{6h}$)$\rightarrow$ L75($C_{6h}$) which is effectively similar to the effect of an external magnetic field perpendicular to the basal plane. As a consequence, the spin polarized subspaces have no essential degeneracy at K (see Appendix \ref{spinful_inv_ap} for details) and the band structure at half-filling is fully gapped, see Fig.~\ref{fig_BS_IRREPs}(b) showing the EBR of one spin subspace and Fig.~\ref{fig_BS_IRREPs_spinful}(a) showing the EBR of the parent spinful class. 
It follows that the topology of each gapped spin polarized subspace can be characterized by a Chern number, or equivalently by the flow of Berry phase over the BZ.\cite{Kiritsis} We write this as $2\pi C_1[\mathrm{BZ}] = \gamma[\partial \mathrm{BZ}] -  \gamma[\partial \boldsymbol{0}] = \gamma[\partial \mathrm{BZ}]$ where $\partial \mathrm{BZ}$ is an oriented boundary of the BZ. Since the BZ is a closed manifold, $\partial \mathrm{BZ}$ is equivalent to a point and in the previous expression we assume that we have kept track of the Berry phase as we deform the base point $\partial \boldsymbol{0}$ into a loop that we then sweep over the whole BZ into $\partial \mathrm{BZ}$, that is (topologically) a point again. It then follows that $\gamma[\partial \mathrm{BZ}] = n 2\pi$ (as a phase is defined modulo $2\pi$) and the Chern number is given by the integer $n \in \mathbb{Z}$. 

We now take advantage of the $C_{6}$-symmetry of the system and compute the Berry phase winding over the patch $S_{\rho}$ that is one sixth of the whole BZ (a similar construction has already been used in Ref.~\onlinecite{HolAlex_Bloch_Oscillations} giving the constraint due to $C_{6}$-symmetry on the Chern number of a single isolated band). $S_{\rho}$ is bounded by the oriented loop $l_{\rho} = l^{-1}_{\Gamma\text{K}} \circ l_{\Gamma\text{M}}$, see Fig.~\ref{fig_BZ}, and by folding $ l_{\Gamma\text{M}}$ with $C_{2z}$ and $l_{\Gamma\text{K}}$ with $C_{3z}$ we find (see the algebraic derivation using Wilson loop techniques in Appendix \ref{spinful_Berry_ap})
\begin{equation}
\label{gamma_rho_IRREP}
	e^{i \gamma_{\rho}} =\xi^{\Gamma}_{2} \xi^{\text{K}}_{3}  (\xi^{\text{M}}_{2}  \xi^{\Gamma}_{3} )^{-1} = \pm e^{ \pm i \pi/3}\;,
\end{equation}
or $\gamma_{\rho} = \pm \pi/3~ (\mathrm{mod}~\pi)$, where all possible permutations of the IRREPs composing the EBR (Fig.~\ref{fig_BS_IRREPs}(b) and see also Table \ref{T_comp}) have been taken into account. This result is consistent with Ref.~\onlinecite{HolAlex_Bloch_Oscillations} with the difference that we also included the phases with a trivial Fu-Kane-Mele index (see below). Phase transitions between distinct topological sectors can be engineered through band inversions at $\Gamma$, K and M triggered by tight-binding parameters of increasing range. Since all $\sigma_h$-symmetric SOC terms vanish at $\Gamma$ and M the energy ordering of the IRREPs at these points is still determined by Eq.~(\ref{IRREP_ordering}), while the IRREP of the occupied eigenstate at K is now dictated by $\mathrm{sign} \{h_{AA,\sigma}(\text{K}) - h_{BB,\sigma}(\text{K})\}$, see Appendix \ref{spinful_inv_ap}.  

\begin{figure}[t]
\centering
\begin{tabular}{cc}
	\includegraphics[width=0.4\linewidth]{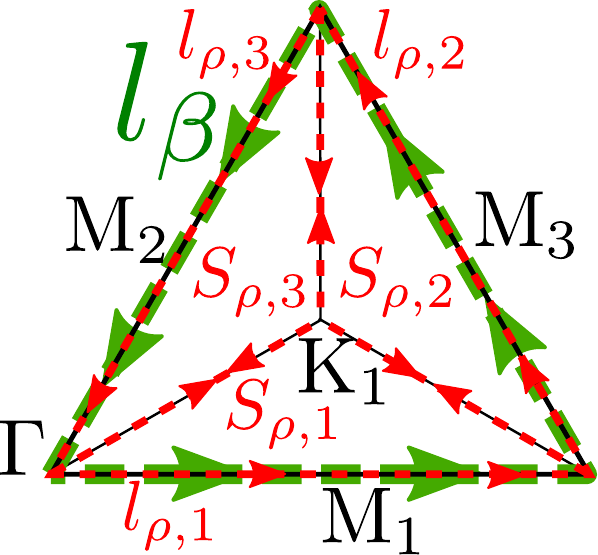} &
	\includegraphics[width=0.4\linewidth]{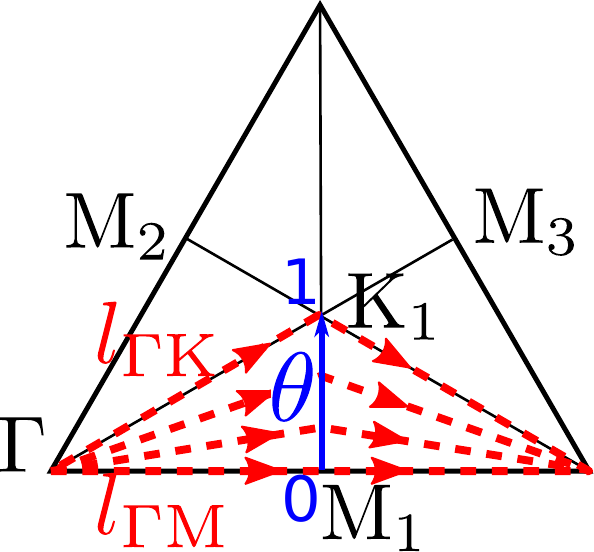} \\
	(a) & (b)
\end{tabular}
\caption{\label{fig_patch} (a) Symmetry decomposition of the HBZ patch $S_{\beta} = S_{\rho,1}+S_{\rho,2}+S_{\rho,3}$ bounded by $l_{\beta} \sim l_{\rho,3} \circ l_{\rho,2} \circ l_{\rho,1} $. (b) Loop parametrization of the patch $S_{\rho} = \bigcup_{\theta\in[0,1]}l_{\theta}$ with $l_0 = l_{\Gamma\text{M}}$ and $l_1 = l_{\Gamma\text{K}}$. 
}
\end{figure}

Combining the images of $S_{\rho}$ under successive rotations by $C_{3z}$ we recover the HBZ patch $S_{\beta}$ introduced in the previous section, i.e.~$S_{\beta} = S_{\rho,1}+S_{\rho,2}+S_{\rho,3}$ as shown in Fig.~\ref{fig_patch}(a) with the boundary $l_{\beta} \sim  l_{\rho,3} \circ l_{\rho,2} \circ l_{\rho,1} $ where $ l_{\rho,2} \sim C_{3z} l_{\rho,1}$ and $ l_{\rho,3} \sim C^2_{3z} l_{\rho,1}$ ($\sim$ here means equal up to a translation by a reciprocal lattice vector). The symmetry of the Berry curvature under $C_{3z}$ rotation gives $e^{i \gamma_{\beta}} = (e^{i \gamma_{\rho}})^{3}$ and we conclude
\begin{equation}
\label{spinful_inversion_homotopy}
	\pi_2(\mathcal{H}_{\mathrm{AII+L80}(2b)}^{2+2})  \cong \pm \mathbf{1}+3\mathbb{Z}  \ni W^{\mathcal{I}}_{II} = \dfrac{\gamma_{\beta}}{\pi} \;, 
\end{equation}
here the second homotopy group $\pi_2$ refers to the continuous maps from the whole BZ to the classifying space $\mathcal{H}_{\mathrm{AII+L80}(2b)}^{2+2}$ with the approximation $\mathrm{BZ}\cong \mathbb{T}^2 \rightarrow \mathbb{S}^2$. Therefore the winding number $W^{\mathcal{I}}_{II}$ classifies all symmetry protected topological insulating phases of the honeycomb lattice with SOC and inversion symmetry for a single orbital per site at Wyckoff's position $2b$. For instance, the Fu-Kane-Mele $\mathbb{Z}_2$ invariant is readily obtained as the parity $\nu_{\mathrm{FKM}} = W^{\mathcal{I}}_{II} ~\mathrm{mod}~2$. Concretely, the value of the winding number depends on the (spinless) IRREPs of the spin-polarized occupied eigenstate according to Eq.~(\ref{gamma_rho_IRREP}) and the winding increases with an increasing range of the tight-binding parameters. In particular, we can switch the sign of the winding number by switching the sign of $\{h_{AA,\sigma}(\text{K})-h_{BB,\sigma}(\text{K})\}$ that is nonzero only with SOC. Reintroducing the spin degrees of freedom, we have $\xi^{\bar{\boldsymbol{k}}}_{2,\downarrow}  = (\xi^{\bar{\boldsymbol{k}}}_{2,\uparrow})^*$ and $\xi^{\bar{\boldsymbol{k}}}_{3,\downarrow}  = (\xi^{\bar{\boldsymbol{k}}}_{3,\uparrow})^*$ from which follows $\gamma_{\rho,\downarrow} = - \gamma_{\rho,\uparrow}$ and $W^{\mathcal{I}}_{II,\downarrow} = - W^{\mathcal{I}}_{II,\uparrow}$. Similarly to Eq.~(\ref{spinless_homotopy}) the classification of Eq.~(\ref{spinful_inversion_homotopy}) has been derived under the assumption that large gauge transformations are discarded, however the explicit use of a smooth reference gauge is not needed in the algebraic computation of Appendix \ref{spinful_Berry_ap}. Nevertheless, Eq.~(\ref{gamma_rho_IRREP}) should still be interpreted in a relative sense, i.e.~it gives the actual quantity obtained from the surface integral of the Berry connection derived from a smooth reference gauge over the patch $S_{\rho}$. Using the analytical ansatz of the spin-polarized eigenstate given in Appendix \ref{spinful_inv_ap} we have verified numerically that both computations give the same results.

Importantly, we conclude from Eq.~(\ref{spinful_inversion_homotopy}) that the classification of the spinless system Eq.~(\ref{spinless_homotopy}) is lifted into the classification of inversion symmetric spinful gapped phases with a doubling that is due to the spin grading, i.e.~$ W_I \in \mathbf{1}+3\mathbb{Z} \rightarrow W^{\mathcal{I}}_{II} \in \{ \mathbf{1}+3\mathbb{Z}, -\mathbf{1}+3\mathbb{Z}\} $. We stress that the band structure is now gapped and that the winding number $W^{\mathcal{I}}_{II}$ captures a flow of Berry phase over the BZ, which then gives the Chern number of the two-dimensional gapped phase. Therefore, even though the form of Eq.~(\ref{spinful_inversion_homotopy}) closely resembles Eq.~(\ref{spinless_homotopy}) of the gapless phase, the interpretation of the winding number is different. It is also important to note that $W^{\mathcal{I}}_{II} = +1 + 3n $ and $W^{\mathcal{I}}_{II} = -1 - 3n $ do refer to two distinct topological classes in our classification. For instance, a reversal of $W^{\mathcal{I}}_{II}$ happens when the bands $K_4$ and $K_6$ are inverted in Fig.~\ref{fig_BS_IRREPs}(b) that requires a closing of the band gap and thus constitutes a topological phase transition.  

We note that the $\mathbb{Z}$-type classification is robust in the many-band case. This is the case when the system is symmetric under $C_2$ spin rotation along the $z$ axis guaranteeing the separation of the spin sectors and leading to a classification of the topology in terms of the spin Chern number.\cite{spin_chern,Shiozaki14} A more physical symmetry however is the basal mirror which also leads to a $\mathbb{Z}$ classification in the many-band case.\cite{TeoFuKane_mirror,Shiozaki14} We here readily obtain the (spinful) mirror Chern number as $C^{+}_1 = 2W^{\mathcal{I}}_{II}\pi /(2\pi)= W^{\mathcal{I}}_{II} = \pm 1 ~\mathrm{mod}~3$ for the even mirror subspace, and $C^-_1 = -C^+_1$ for the odd mirror subspace. It is the special case of the four-by-four Hamiltonian with AII+L80 symmetry that the spin Chern number here matches with the mirror Chern number, i.e. there are not enough bands to break the spin rotation symmetry without breaking the symmetries of L80.

\section{Wilson loop winding}\label{WLW}

Breaking inversion (equivalently basal mirror) symmetry allows spin-flip Rashba SOC terms that forbid the spin decomposition of Section \ref{spinful_inversion}. Nevertheless, the topological classification must be preserved when inversion symmetry is broken adiabatically, i.e.~without closing the band gap.\cite{Prodan_spin_chern} However the question remains of a direct verification that goes beyond the previous derivation of the Berry phase based on $s_z$ being a good quantum number over the whole BZ. Because of the spin grading, we must switch from a $U(1)$ Berry phase to a $U(2)$ Wilson loop\cite{WindingKMZ2, InvTIBernevig, PointGroupsTI, Wi1, Wi3, hourglass, Alex_BerryPhase} description of the two-band occupied subspace.\footnote{\unexpanded{The $U(2)$ Berry-Wilczek-Zee connection matrix is defined as $\mathcal{A}^{mn}_{\mu} = \langle u_{m}, \boldsymbol{k} \vert \partial_{k^{\mu}}u_{n}  ,\boldsymbol{k} \rangle$ with $\vert u_{n} ,\boldsymbol{k} \rangle$ the $n$-th occupied cell-periodic Bloch eigenstate and the Wilson loop over a base loop $l$ is given as $\mathcal{W}[l] = \exp\left\{- \oint_{l} d\boldsymbol{k} \cdot \boldsymbol{\mathcal{A}}(\boldsymbol{k}) \right\}$.}}

In this section we first review known properties of the Wilson loop that follow from TRS and $C_{2z}\mathcal{T}$-symmetry,\cite{Wi3} from which we motivate the existence of nontrivial windings of the Wilson loop over the BZ. Wilson loop winding has been used in a gauge-invariant computation of the Fu-Kane-Mele $\mathbb{Z}_2$ invariant.\cite{WindingKMZ2} It has also been shown to capture a crystalline invariant of systems with inversion symmetry,\cite{Wi1} and Ref.~\onlinecite{EBR_2} has recently shown one example of a spinless Wilson loop winding protected by $C_2$-symmetry. We derive here the expression of the spinful Wilson loop winding over one patch of the BZ that captures the effect of crystalline symmetries, i.e.~here the point group $D_{6h}$ or $C_{6h}$. The Wilson loop characterization of the topological sectors will then allow us to treat the cases when $s_z$ is not a good quantum number leading to a complete classification of the symmetry class AII+L77($2b$), i.e.~including the inversion symmetry breaking topological phases that are not adiabatically connected to the phases classified by Eq.~(\ref{spinful_inversion_homotopy}).

Let us first introduce some notations. We write the Wilson loop computed over a momentum base loop $l$ as $\mathcal{W}_l \equiv \mathcal{W}[l] $. The Wilson loop spectrum is gauge invariant due to the invariance of the eigenvalues under unitary transformations. Writing it as $\mathrm{eig}\{\mathcal{W}_l\} = [e^{i\varphi_{l,1}},e^{i\varphi_{l,2}}] $ this defines the Wilsonian phases $\varphi_{l,1}$ and $\varphi_{l,2}$. 

Whenever a loop satisfies the symmetry $\mathcal{I}l=l^{-1}$ we have $\mathcal{W}_{\mathcal{I}l} = \mathcal{W}^{-1}_{l}$ and spinful TRS ($\mathcal{T}^2=-1$) gives $A \mathcal{W}^*_{l} A^{-1} = \mathcal{W}_{\mathcal{I}l} = \mathcal{W}^{-1}_{l}$ where $A\mathcal{K}$ is the anti-unitary representation of TRS in the occupied band basis ($A$ is unitary and $\mathcal{K}$ is the complex conjugation) and $(A\mathcal{K})^2=-\mathbb{I}$. Such base loops must connect time reversal invariant momenta (TRIMPs) leading to a Wilson loop spectrum that is composed of Kramers pairs, i.e.~pairs of mutually orthogonal eigenstates with the same Wilsonian eigenvalue.\cite{Wi3} This is true for the loops $l_{\Gamma\text{M}}$ and $l_{\text{M}\text{M}}$ (Fig.~\ref{fig_BZ}). A formulation of the Fu-Kane-Mele $\mathbb{Z}_2$ invariant\cite{KaneMeleZ2,FuKaneZ2} then follows from the spectral flow of Wilson loop over the HBZ patch $S_{\alpha}$ bounded by $l^{-1}_{\text{M}\text{M}} \circ l_{\Gamma\text{M}}$.\cite{WindingKMZ2} We note that the edges $l_{\Gamma M_2}$ and $l^{-1}_{\Gamma-\boldsymbol{b}_1, M_2-\boldsymbol{b}_1}$ of $l_{\alpha}$ can be neglected when large gauge transformations are excluded. Practically this is done by imposing the periodic gauge (see Appendix \ref{spinless_ap}). Also the HBZ patch must be chosen such that the composition of the oriented boundary (e.g.~$l_{\alpha}$ for $S_{\alpha}$) with its image under inversion ($\mathcal{I}l_{\alpha}$) gives an oriented boundary of the whole BZ, i.e.~$l_{\alpha} \circ \mathcal{I} l_{\alpha} \cong \partial BZ$.

Considering now the combined symmetry $ C_{2z}\mathcal{T} = \sigma_h \mathcal{K}$, the Wilson loop satisfies $U_{\sigma_h}\mathcal{W}^*_l U_{\sigma_h}^{\dagger} = \mathcal{W}_{l}  $ where $U_{\sigma_h}\mathcal{K}$ is the anti-unitary representation of $\sigma_h \mathcal{K}$ in the occupied band basis.\cite{Wi3} It follows that the eigenstates of the Wilson loop are composed of mutually orthogonal pairs with their eigenvalues being partners under complex conjugation, i.e.~$\mathrm{eig}\{\mathcal{W}_l\} = [e^{i \varphi_l},e^{-i \varphi_l}]$.\footnote{If $V_l$ is an eigenvector of $\mathcal{W}_l$ with the Wilsonian eigenvalue $e^{i\varphi_l}$, then $U_{\sigma_h}^T V^*_l$ is also an eigenvector with the Wilsonian eigenvalue $e^{-i\varphi_l}$.} This feature has an important consequence for the topological classification of the flows of Wilson loop over the BZ. In order to see this let us parameterize the BZ through loop sections as $\bigcup_{\theta \in \mathbb{S}^1} l_{\theta} \cong \mathbb{S}^1 \times \mathbb{S}^1 \cong \mathbb{T}^2$ as follows naturally from the parametrization of the 2-torus $\mathbb{T}^2$ by two circles $\mathbb{S}^1$, where each $\theta$ labels one loop $l_{\theta} \cong \mathbb{S}^1$ of the BZ. By choosing one \textit{representative} Wilsonian phase $\varphi(\theta)$, we can then classify the spectral flow of Wilson loop from the parameter base space $\mathbb{S}^1 \ni \theta$ into the classifying space $U(1) \ni e^{i \varphi(\theta)}$ by a winding number since $\pi_1(U(1)) = \mathbb{Z}$ (indeed, $\mathrm{eig}\{\mathcal{W}_l\}\in SO(2) \cong U(1)$). Note here that the choice of $\varphi(\theta)$ instead of $-\varphi(\theta)$ is arbitrary.\footnote{This arbitrariness is related to the fact that the real IRREP of a positive rotation within $SO(2)$ is equivalent to the irreducible representation of a negative rotation. Similarly $U_{\sigma_h} \mathcal{W}^*_l U^{\dagger}_{\sigma_h} = \mathcal{W}_{l} $ is a relation of equivalence between two Wilson loops with opposite phases.}

An important consequence of $C_{2z}\mathcal{T}$-symmetry is the double degeneracy of $0$ and $\pm \pi$ Wilsonian phases. An other important feature that comes from the combined constraints of TRS and $C_{2z}\mathcal{T}$-symmetry is that for any loop connecting two TRIMPs and satisfying $l^{-1} = \mathcal{I} l$, the Wilson loop spectrum must be both doubly degenerated and symmetric under complex conjugation. Therefore, the two-band Wilson loop spectrum over $l_{\Gamma\text{M}}$ and $l_{\text{M}\text{M}}$ must either be $[+1,+1]$ or $[-1,-1]$. It is worth noting that while we expect the $\mathbb{Z}$-type Wilson loop winding to be present in two-band subspaces with TRS and $C_{2z}\mathcal{T}$-symmetry we do not expect that it is robust in subspaces with more bands. Indeed, the homotopy classification of flows of Wilson loop over the BZ, i.e. $\mathrm{eig}\{\mathcal{W}_{l_{\theta}}\} \in SO(2n)$ with $\theta\in\mathbb{S}^1$, for $n>1$ gives $\pi_1(SO(2n)) = \mathbb{Z}_2$.\cite{BzduSigristRobust} This is confirmed in Section \ref{six_band} and Section \ref{discussion} where we analyze six-band models and find a fragile topology of spit EBRs.

Let us now consider the effect of the other crystalline symmetries. This is again revealed by considering the patch $S_{\rho}$ bounded by $l_{\rho} = l^{-1}_{\Gamma\text{K}} \circ l_{\Gamma\text{M}}$ (Fig.~\ref{fig_BZ}). Similarly to the above parameterization of the whole BZ, we parameterize the patch $S_{\rho}$ through loop sections as $l_{\theta} : [0,1] \rightarrow S_{\rho} : \theta\mapsto l_{\theta} $ with $l_{0} = l_{\Gamma\text{M}}$ and $l_{1} =  l_{\Gamma\text{K}}$, see Fig.~\ref{fig_patch}(b). Therefore, for each loop $l_{\theta}$ we can define a representative Wilsonian phase $\varphi(\theta)= \varphi[l_{\theta}]$ such that the winding of the Wilson loop is captured by the winding of $\varphi(\theta)$ over the patch $S_{\rho}$, i.e.~
\begin{align}
\label{winding_rho}
	 W[S_{\rho}] &= \int_{0}^{1}  \dfrac{d\theta}{\pi} \dfrac{d \varphi(\theta)}{d\theta}  \in \dfrac{1}{\pi} \left[\overline{\varphi}(1)-\overline{\varphi}(0)\right](\mathrm{mod}~2) \;,\\
	\overline{\varphi}(0) &=\varphi(0)\in [-\pi,\pi) \;,\nonumber\\
	 \overline{\varphi}(1) &= (\varphi(1)~\mathrm{modulo}~ 2\pi ) \in [-\pi,\pi)\;, \nonumber
\end{align} 
with $\varphi(0) =\varphi[l_{\Gamma\text{M}}]$ and $\varphi(1) = \varphi[l_{\Gamma\text{K}}]$, and where we have defined reference Wilsonian phases $\overline{\varphi}(0)$ and $\overline{\varphi}(1)$ with respect to which we compute the winding of the Wilson loop and where ``modulo'' here refers to the operation of taking the remainder as opposed to ``mod'' used to define a set.\footnote{While the representative phase is smooth and takes value in a unbounded domain, i.e.~$\varphi(\theta)/\pi \in \mathbb{R}$, the reference phases $\overline{\varphi}(0)$ and $\overline{\varphi}(1)$ are fixed constants within $[-\pi,\pi)$.} Similarly to the computation of the Fu-Kane-Mele $\mathbb{Z}_2$ invariant from the spectral flow of the Wilson loop,\cite{WindingKMZ2} TRS makes it sufficient to consider the Wilsonian flow over one HBZ. From the decomposition of the HBZ $S_{\beta}$ into $S_{\rho}$ patches as shown in Fig.~\ref{fig_patch}(a), we find that the Wilson loop winding over $S_{\beta}$ is given as $W[S_{\beta}] = 3 W[S_{\rho}] $. Now by conservation of the total flow over the BZ, the flow over the HBZ patch $S_{\alpha}$ must match with the flow over the HBZ patch $S_{\beta}$ (note also that, as for the HBZ patch $S_{\alpha}$, $\mathcal{I}l_{\beta} \circ l_{\beta} \cong \partial BZ$) and we define the Wilson loop winding number
\begin{equation}
\label{windings}
	W_{II} \equiv 3 W[S_{\rho}] = W[S_{\beta}] = W[S_{\alpha}] \;. 
\end{equation}
The Fu-Kane-Mele $\mathbb{Z}_2$ invariant is then obtained by taking the parity of this winding number, i.e.~$\nu_{\mathrm{FKM}} = W_{II} \mod  2$. 

We importantly note that the Wilson loop winding is supported by the homotopy of the flow of Wilson loop protected by $C_2\mathcal{T}$ symmetry ($\pi_1(SO(2)) = \mathbb{Z}$). It follows that while we use special cuts of the BZ to derive the constraints from crystalline symmetries on the Wilson loop, \textit{the total Wilson loop winding (protected by $C_2\mathcal{T}$-symmetry) is independent of the parameterization of the flow over the BZ}, as shown explicitly in Fig.~\ref{adiab_SOC}.

\section{Spinful case with adiabatic breaking of inversion symmetry}\label{spinful_ad_inv}
We are now ready to address in detail the spinful case where inversion symmetry is only broken adiabatically.
Inversion symmetry together with $C_{2z}$ gives the basal mirror symmetry $\sigma_h$ that makes $s_z$ a good quantum number over the whole BZ, as we have seen in Section \ref{spinful_inversion}. It then follows that with $\sigma_h$ symmetry the Wilson loop spectrum takes the form $\mathrm{eig}\{\mathcal{W}_l\} = [e^{i\gamma_{l,\uparrow}},e^{i\gamma_{l,\downarrow}}] = [e^{i\gamma_{l,\uparrow}},e^{-i\gamma_{l,\uparrow}}] $ in the spin basis and we readily find the Wilson loop winding $W_{II} = W^{\mathcal{I}}_{II}$ with the winding number defined in Eq.~(\ref{spinful_inversion_homotopy}). 

When Rashba SOC is switched on adiabatically the symmetry protected windings of the spin polarized Berry phase Eq.~(\ref{spinful_inversion_homotopy}) is lifted into the spectral flow of the Wilson loop through an adiabatic mapping, $ \gamma_{l,\sigma} \mapsto \varphi_l $, leading again to $W_{II} = W^{\mathcal{I}}_{II}$.\cite{Prodan_spin_chern}  
\begin{figure*}[t]
\centering
\begin{tabular}{c|c|c|c} 
	\includegraphics[width=0.25\linewidth]{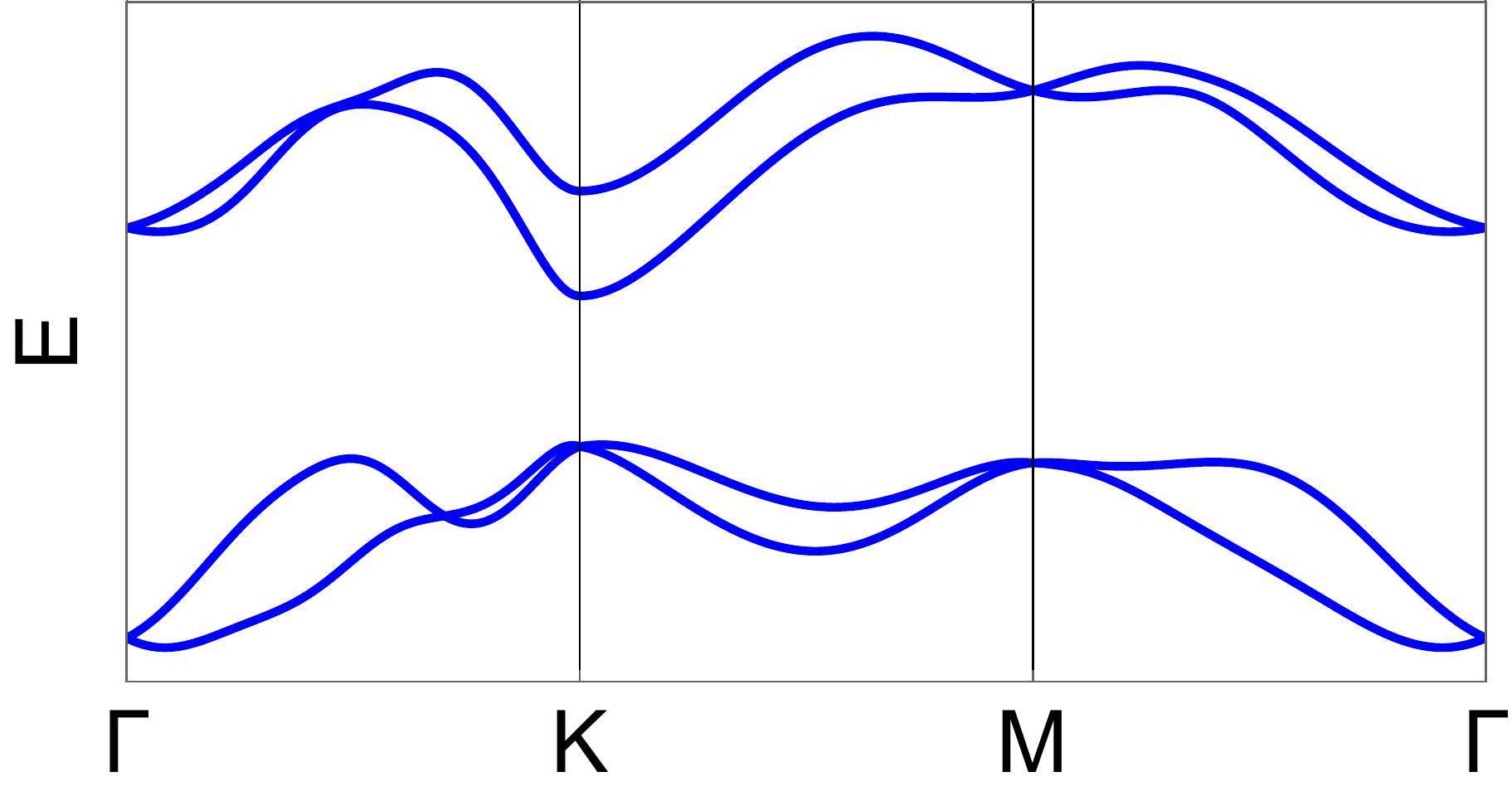} &
	\includegraphics[width=0.25\linewidth]{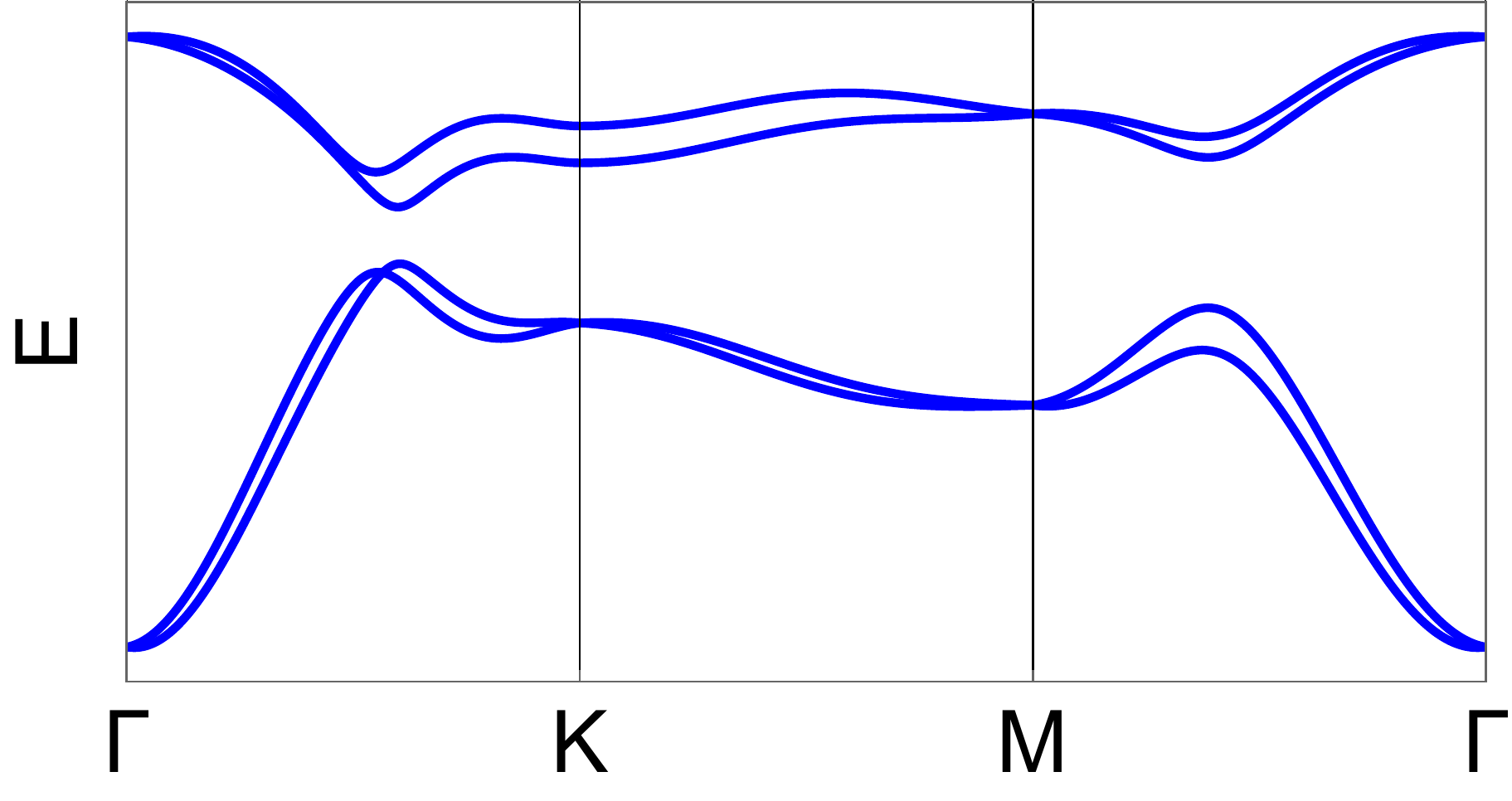} &
	\includegraphics[width=0.25\linewidth]{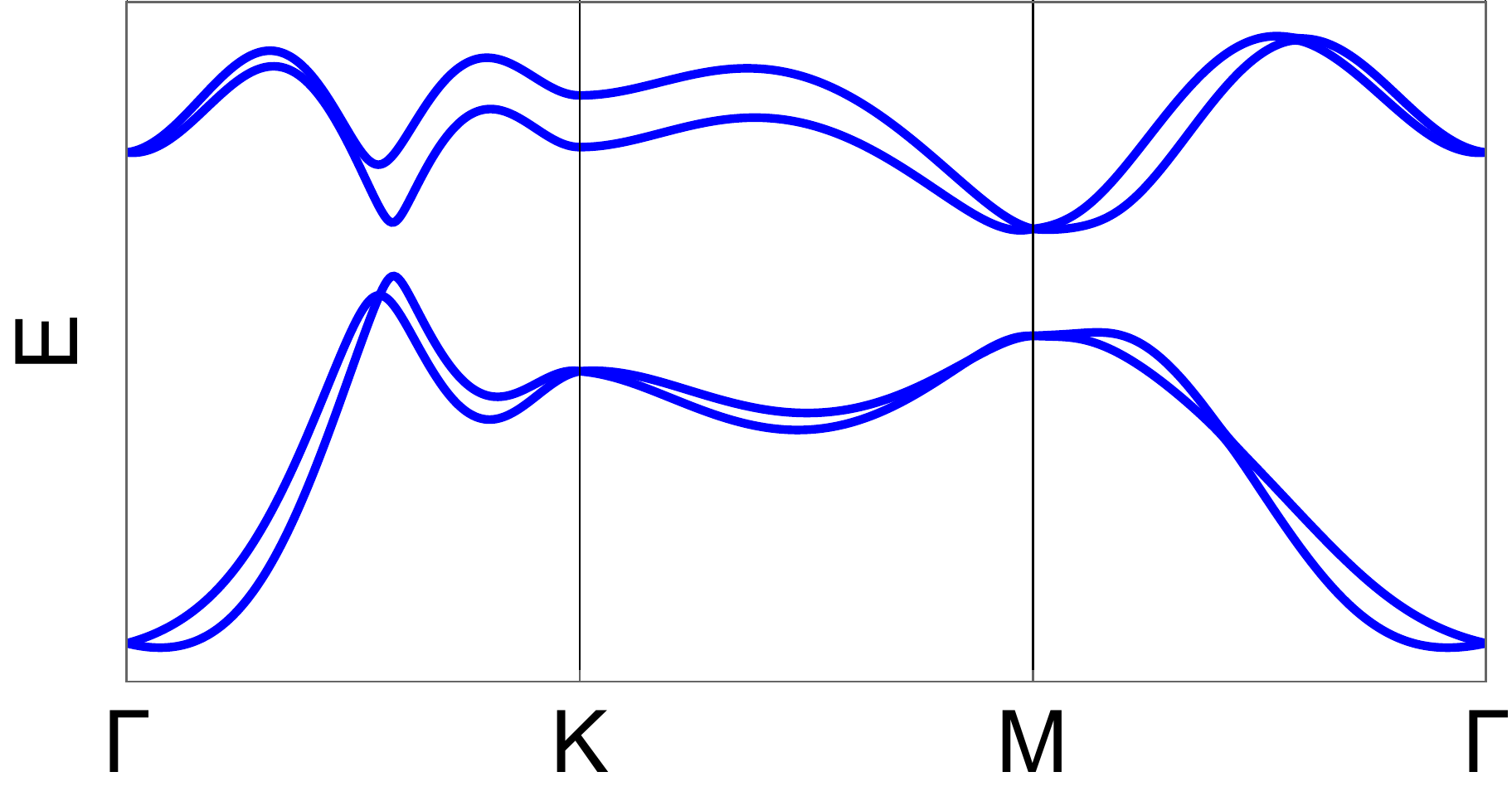} &
	\includegraphics[width=0.25\linewidth]{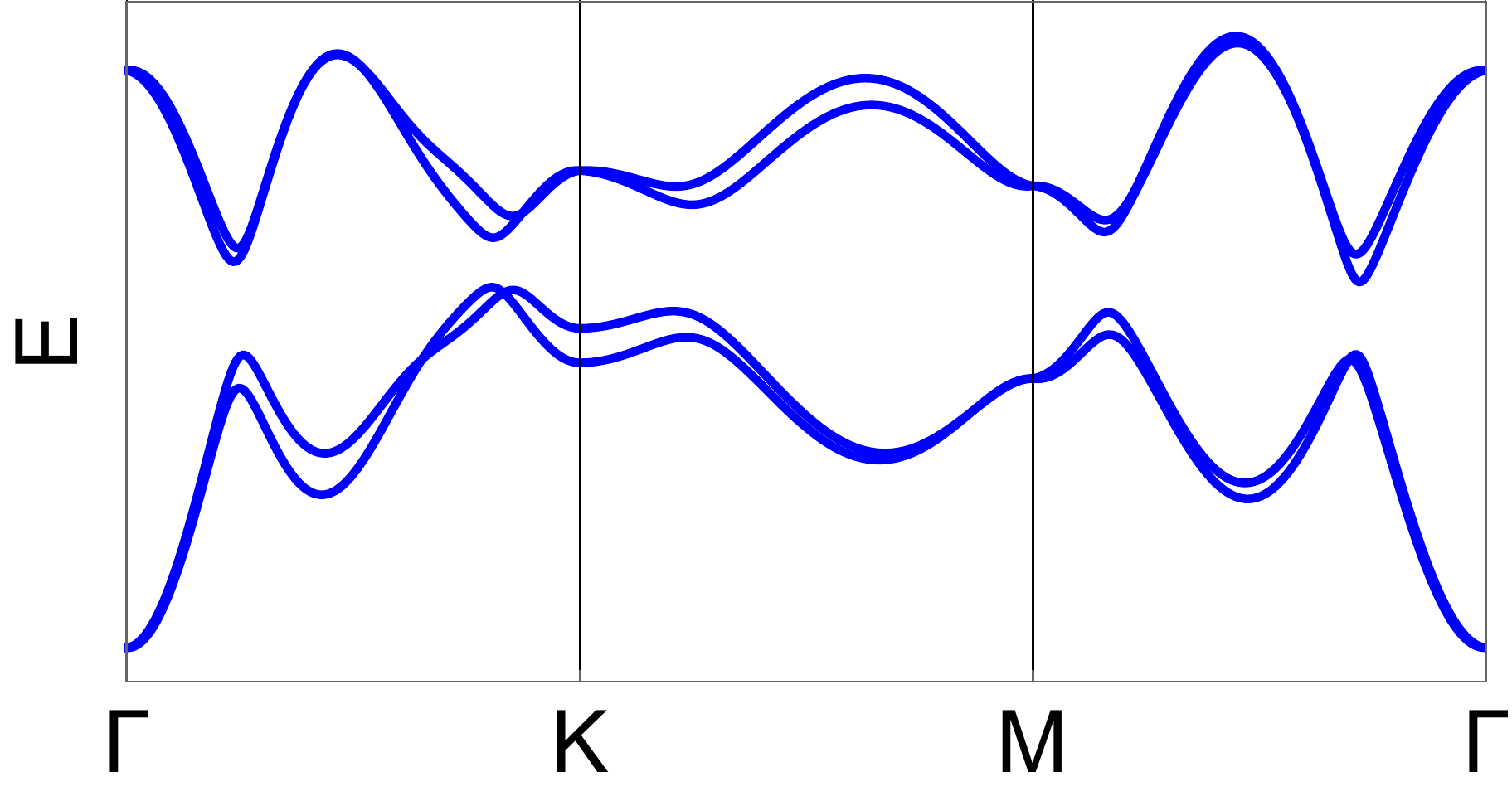} \\
	\includegraphics[width=0.25\linewidth]{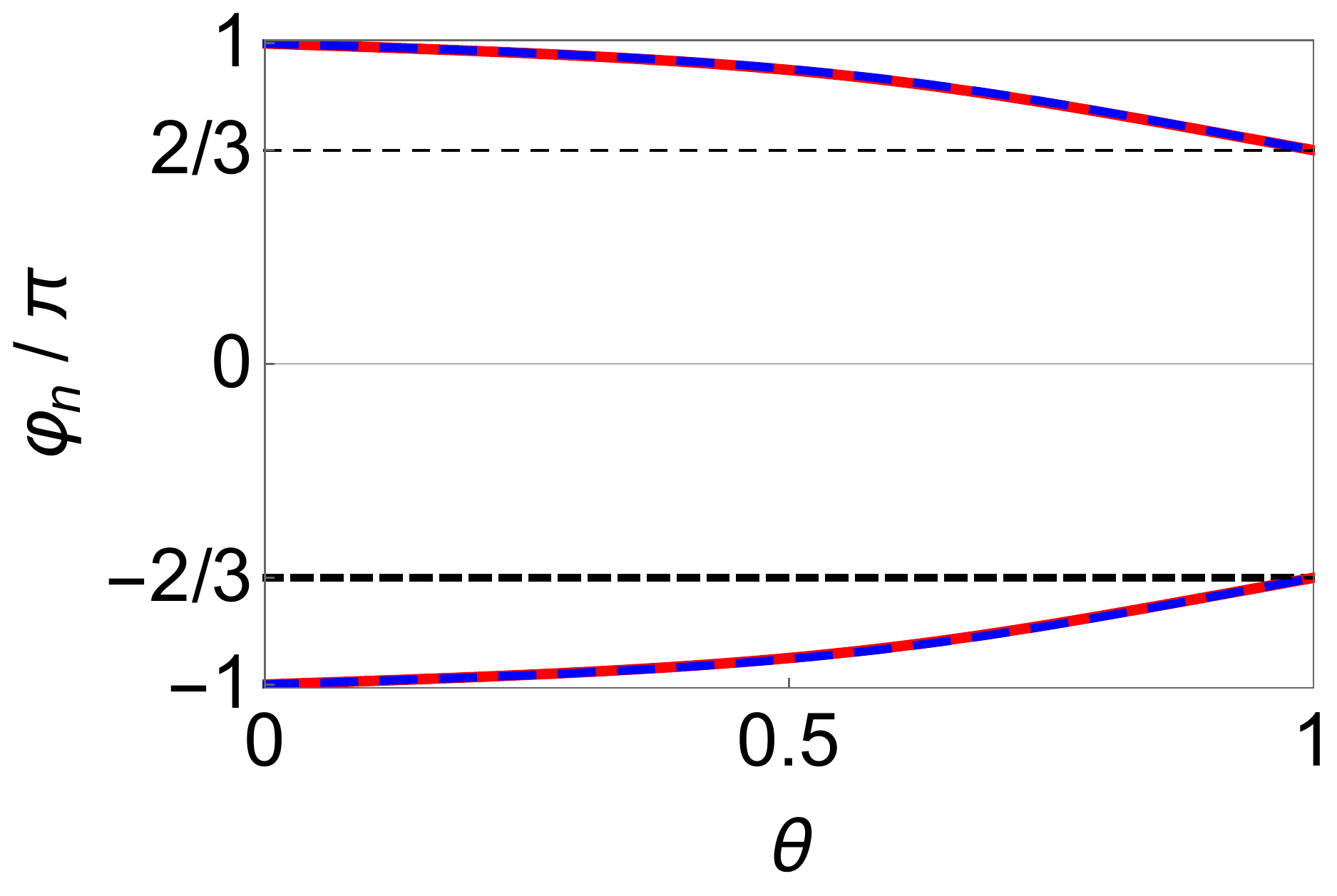} &
	\includegraphics[width=0.25\linewidth]{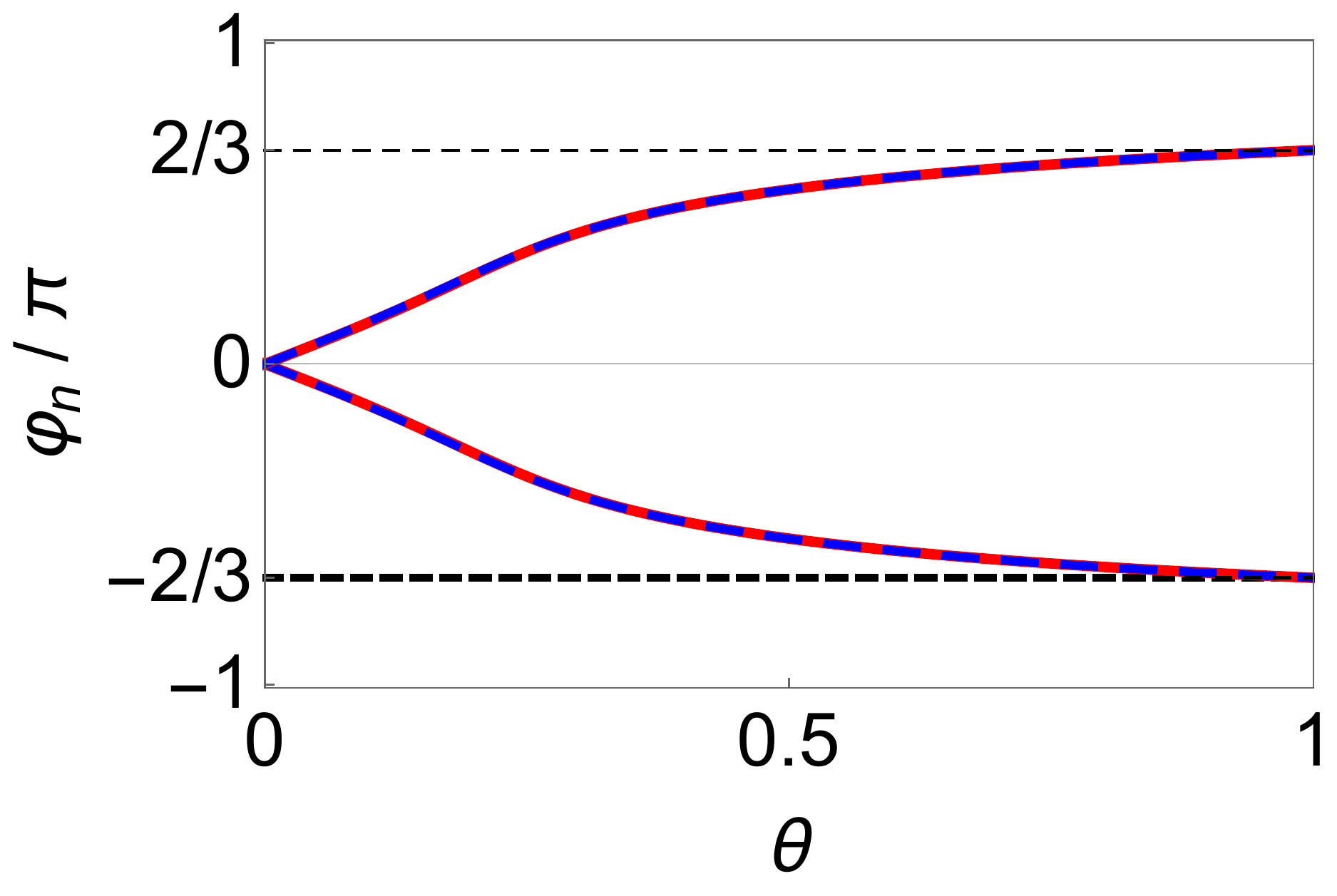} &
	 \includegraphics[width=0.25\linewidth]{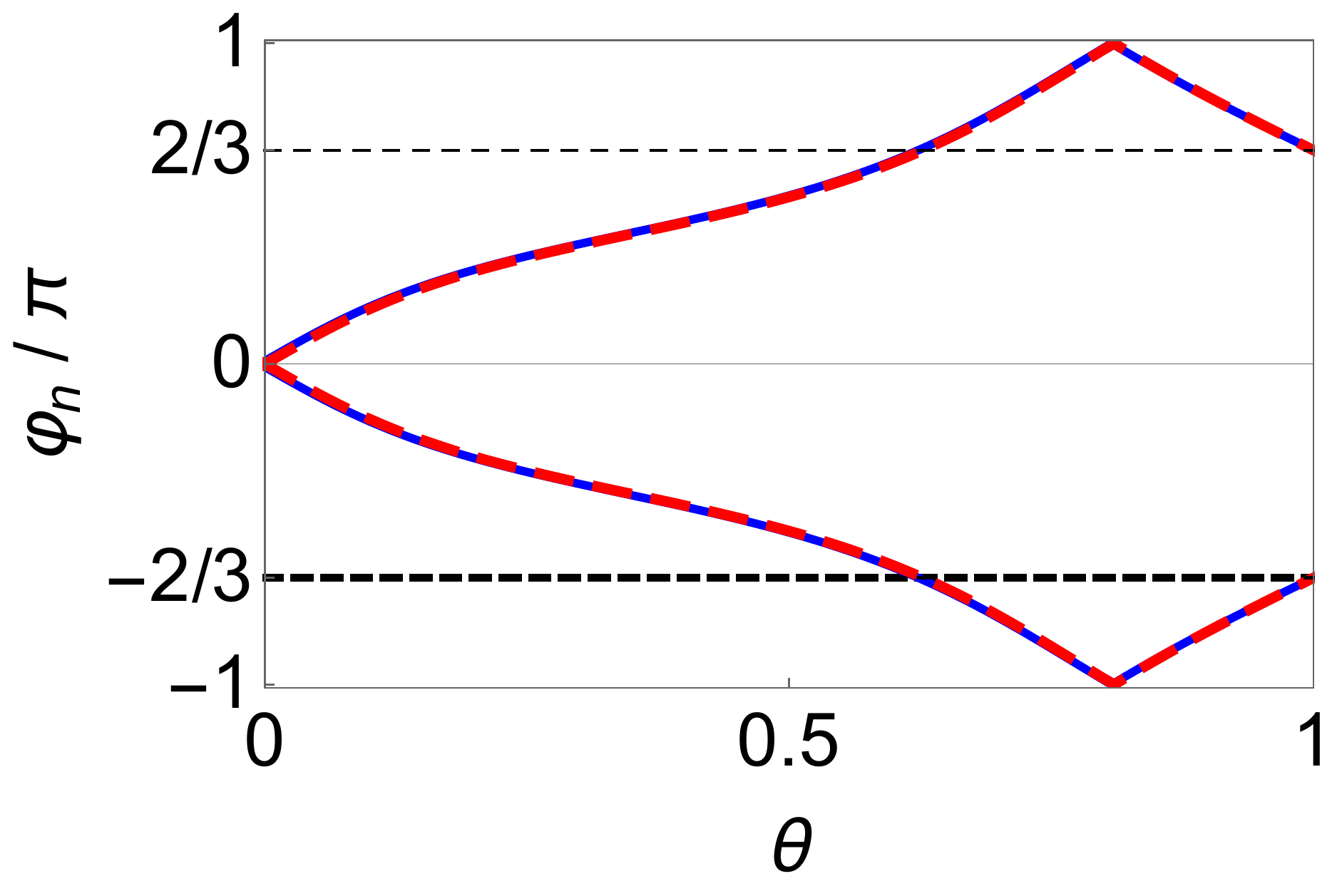} & 
	 \includegraphics[width=0.25\linewidth]{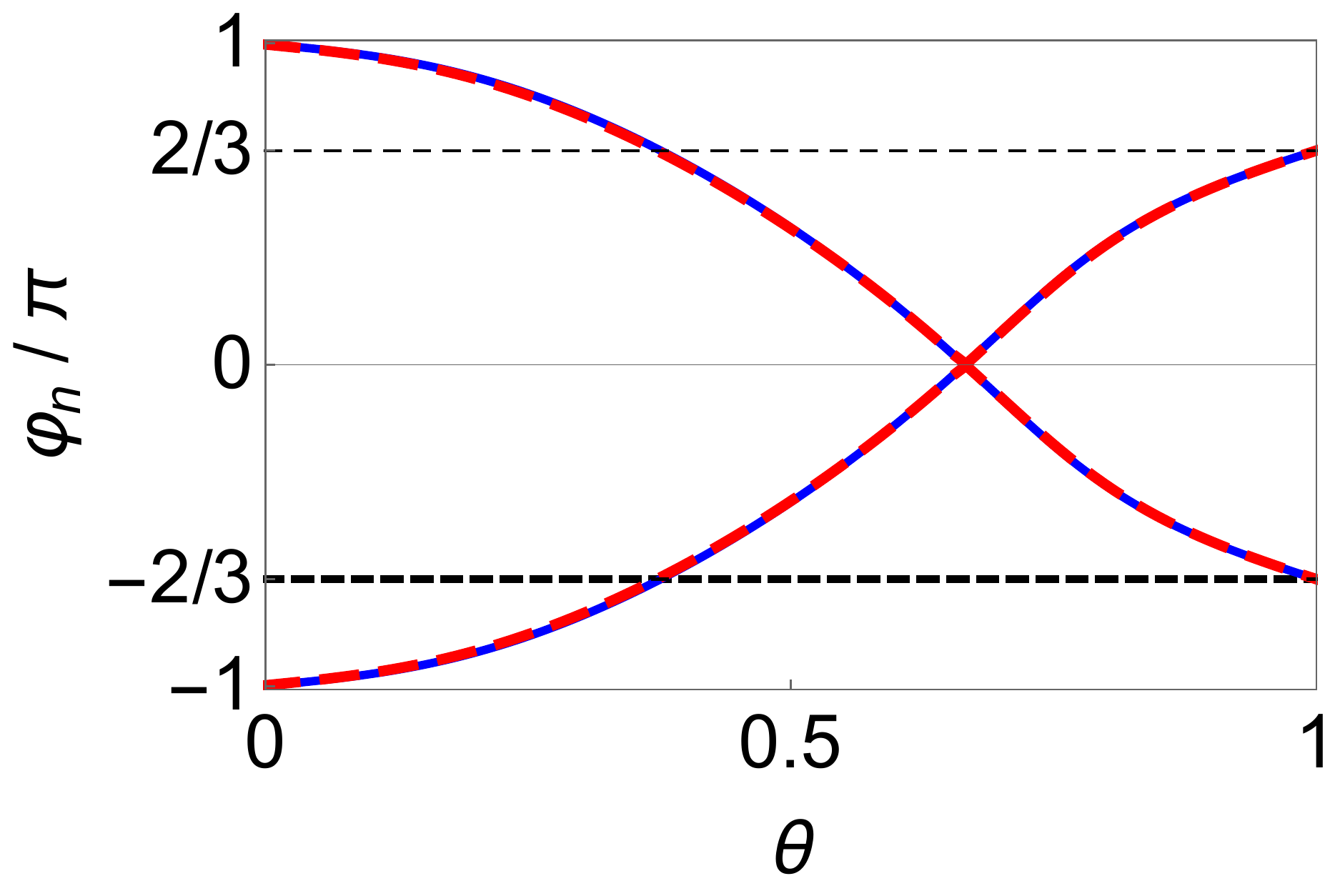} \\
	 $W[S_{\rho}] = +1/3$ & $W[S_{\rho}] = -2/3$ & $W[S_{\rho}] = +4/3$ & $W[S_{\rho}] = -5/3$ \\
	\includegraphics[width=0.25\linewidth]{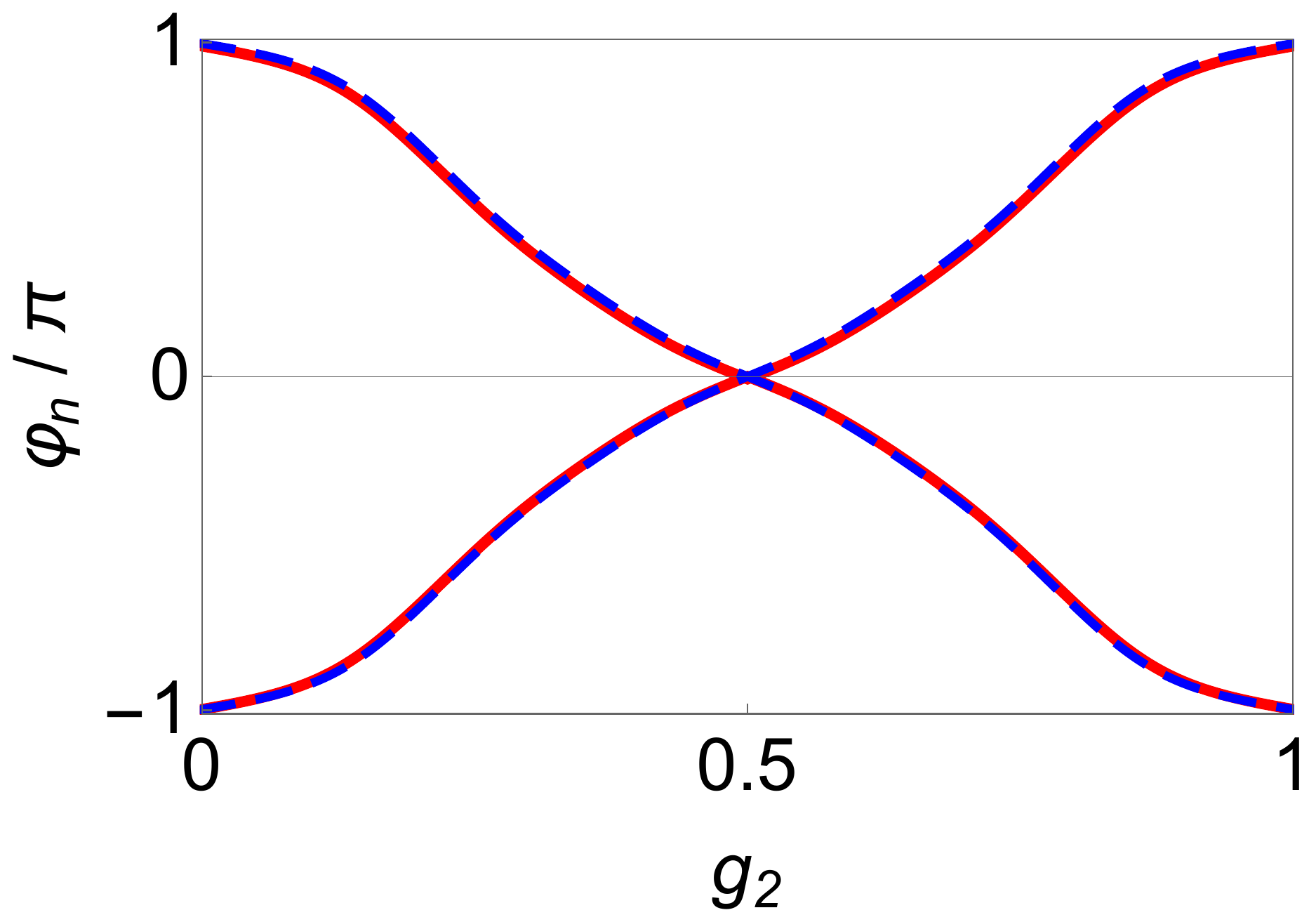} &
	\includegraphics[width=0.25\linewidth]{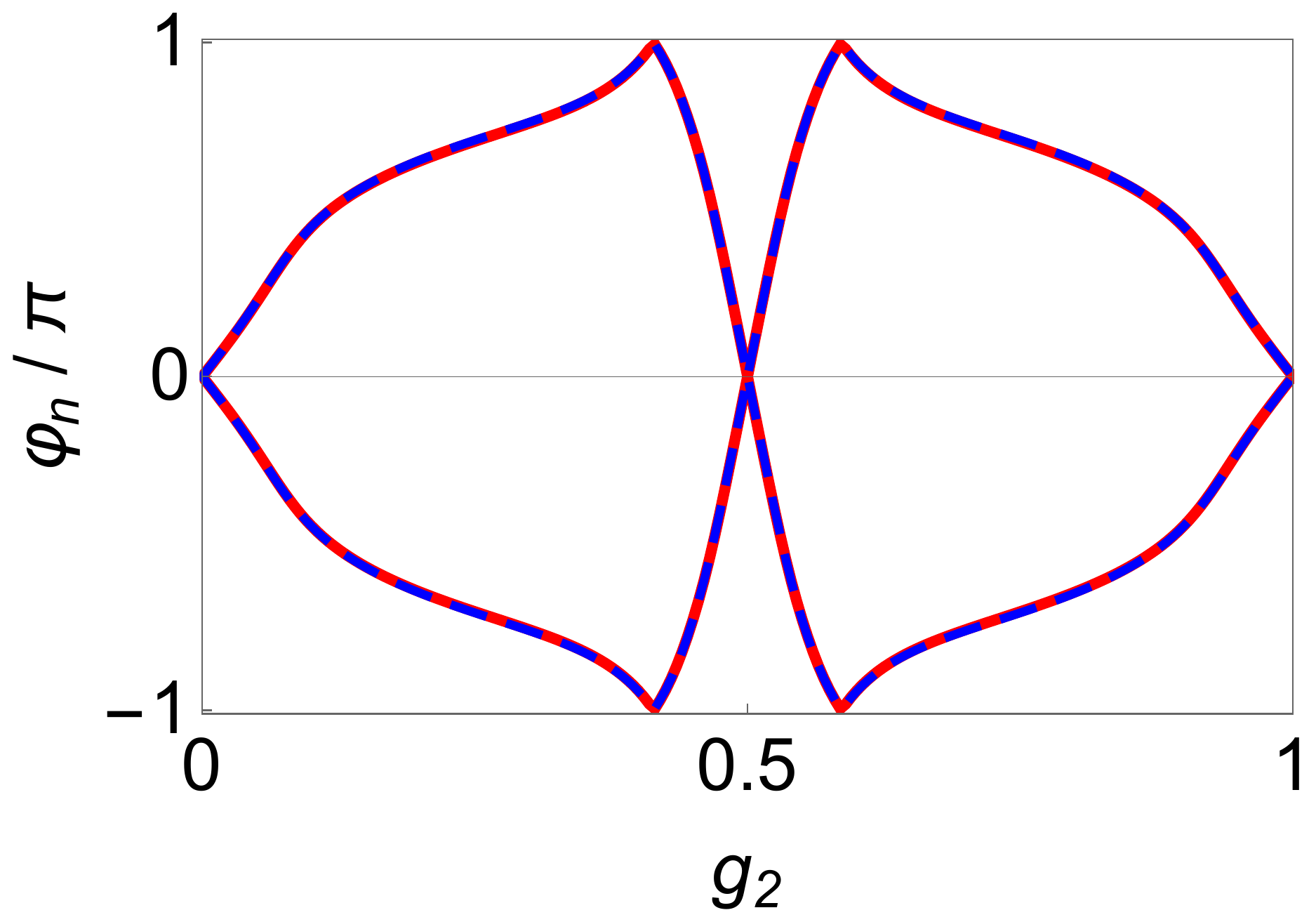} &  
	\includegraphics[width=0.25\linewidth]{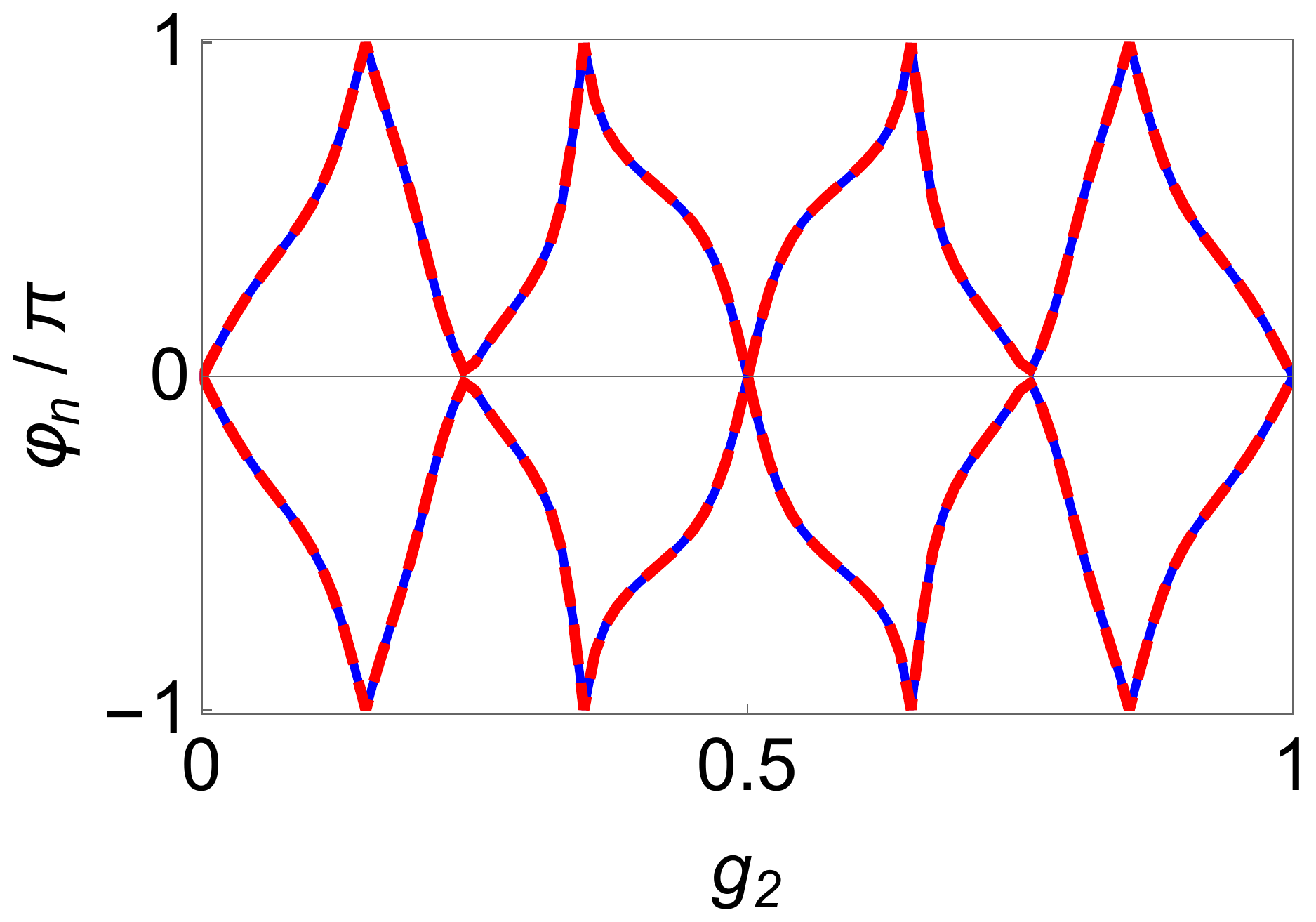} &
	\includegraphics[width=0.25\linewidth]{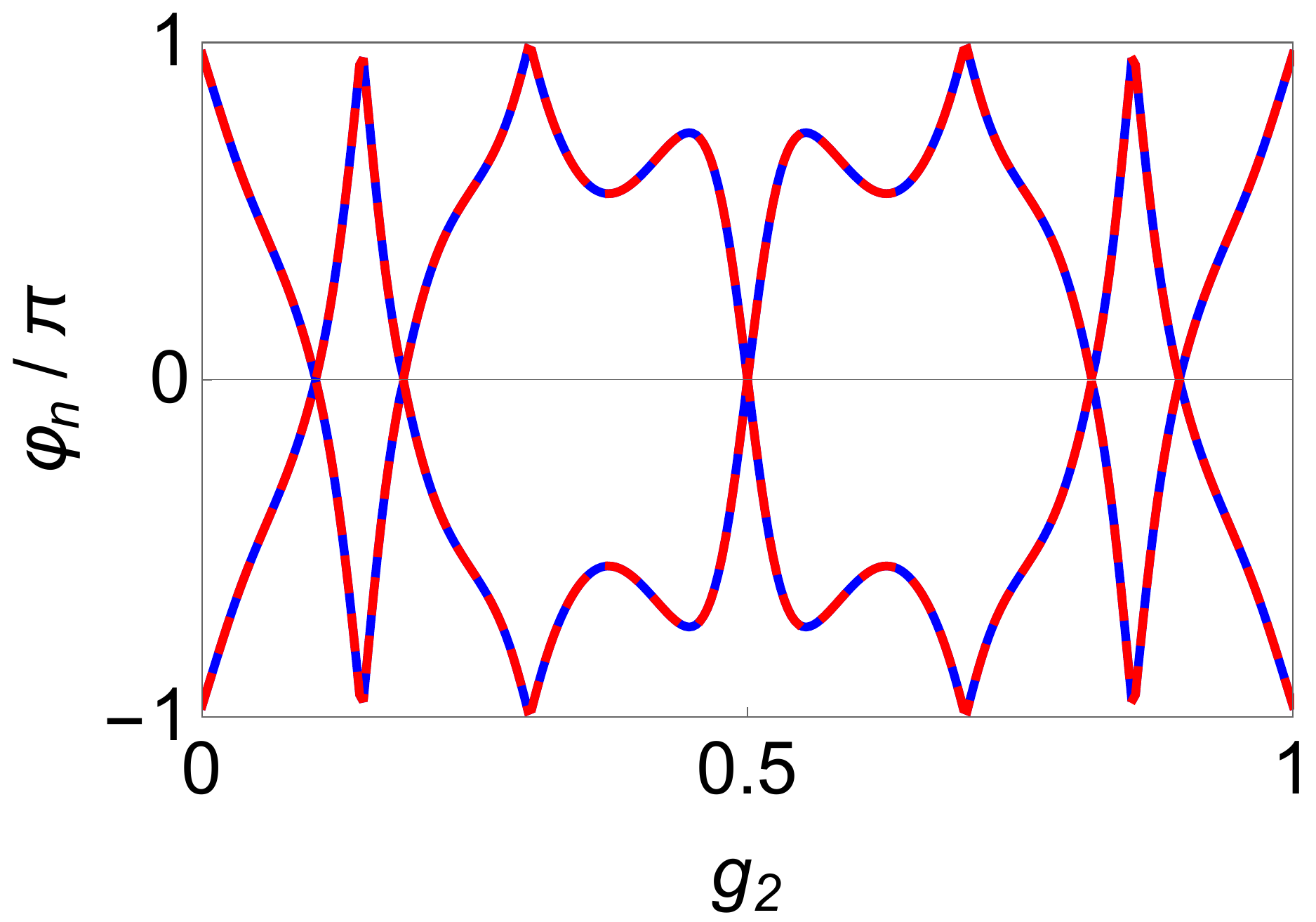} \\
	(a) $W_{II} = +1$, $\nu_{\mathrm{FKM}} = 1$ &  (b)  $W_{II}  = -2$, $\nu_{\mathrm{FKM}} = 0$ & 
	 (c)  $W_{II}  = +4$, $\nu_{\mathrm{FKM}} = 0$ & (d)   $W_{II}  = -5$,$\nu_{\mathrm{FKM}} = 1$
\end{tabular}
\caption{\label{adiab_SOC} Spinful topological phases obtained for SOC that breaks inversion symmetry only adiabatically. Columns (a-d) correspond to those of Fig.~\ref{fig_nodal_points}. First row: band structures along the high symmetry lines of the BZ. Second row: spectral flow of the Wilson loop over one patch $S_{\rho}$ parameterized by loop sections $l_{\theta}$ from $ l_{\Gamma\text{M}}$ ($\theta=0$) to $l_{\Gamma\text{K}}$ ($\theta=1$). Third row: spectral flow of the Wilson loop over the whole BZ parameterized by loop sections $\{l_{g_2}\}_{g_2\in[0,1]}$ with $l_{g_2=0,1} = l_{\Gamma\text{M}}$ and $l_{g_2=0.5} = l_{\text{M}\text{M}}$. Blue (red) dashed lines give the Wilson loop of the two-band occupied (unoccupied) subspace.  For each case the Wilson loop winding over a sixth BZ patch $W[S_{\rho}]$ (computed with respect to reference Wilsonian phase marked by a thick black line), the Wilson loop winding $W_{II}$, and the Fu-Kane-Mele $\mathbb{Z}_2$ invariant is given. Importantly, the agreement between the second and the third rows confirms that the total winding of the Wilson loop over the BZ (protected by $C_2\mathcal{T}$ symmetry) does not depend on the parameterization of the flow. 
}
\end{figure*}
We verify this with several numerical examples in Fig.~\ref{adiab_SOC} with Wilson loop windings up to $-5$ (in Fig.~\ref{seven_winding} in Appendix \ref{winding_seven} we show one example of topological phase with a Wilson loop winding of $+7$). Starting from the spinless cases of Fig.~\ref{fig_nodal_points}, Fig.~\ref{adiab_SOC} shows the effect of first switching on $s_z$-preserving SOC (which opens a band gap over the whole BZ) and then switching on adiabatically Rashba SOC. The first row shows the band structures. The second row shows the flow of Wilsonian phases over the patch $S_{\rho}$ as we smoothly sweep the base loop from $l_{\Gamma\text{M}}$ ($\theta=0$) to $l_{\Gamma\text{K}}$ ($\theta=1$) as illustrated in Fig.~\ref{fig_patch}(b). The third row shows the flow of Wilsonian phases over the whole BZ as we sweep a base loop $l_{g_2}$ parallel to the primitive reciprocal vector $\boldsymbol{b}_2$ over the whole BZ, i.e.~we parametrize the BZ as $\mathrm{BZ} \cong \cup_{g_2\in[0,1]} l_{g_2}$ with $l_{g_2} \cong \{g_1 \boldsymbol{b}_1 + g_2 \boldsymbol{b}_2 \vert g_1 \in [0,1]\}$. The flow then connects the Wilson loop over $ l_{\Gamma\text{M}}$ ($g_2=0$), $ l_{\text{M}\text{M}}$ ($g_2=0.5$), and $ l_{\Gamma\text{M}}+\boldsymbol{b}_1 \sim l_{\Gamma\text{M}}$ ($g_2=1$), see Fig.~\ref{fig_BZ}. We have used the periodic gauge in the numerical computation of the Wilson loop which prevents large gauge transformations, see Appendix \ref{spinless_ap}. The only two allowed values $\overline{\varphi}(0)=0$ and $\overline{\varphi}(0)=-\pi$ follow from the combined constraints of TRS and $C_{2z}\mathcal{T}$-symmetry that enforce the Wilson loop spectrum on $l_{\Gamma\text{M}}$ (and $l_{\text{M}\text{M}}$) to be $[+1,+1]$ or $[-1,-1]$. The other lattice symmetries enforce the reference Wilsonian phase on $l_{\Gamma\text{K}}$ to be $\overline{\varphi}(1) = \pm 2\pi/3$ (see derivation in the next section). In each case we have written in the bottom of Fig.~\ref{adiab_SOC} the Wilson loop winding over the patch $S_{\rho}$ ($W[S_{\rho}]$) for the choice $\overline{\varphi}(1) = - 2\pi/3$ as the reference Wilsonian phase, the winding of the Wilson loop over one HBZ ($W_{II}$), and the Fu-Kane-Mele $\mathbb{Z}_2$ invariant. We see a perfect confirmation of $ W_{II} = 3 W[S_{\rho}]  = W^{\mathcal{I}}_{II} = W_{I}$ and $\nu_{\mathrm{FKM}} = W[S_{\alpha}] \mod 2 = W_{II} \mod 2 $. We note that the robustness of higher Wilson loop windings comes from the $C_{2z}\mathcal{T}$-symmetry protection of the crossings of Wilson loop branches (of Wilsonian phases) at $0$ and $\pm \pi$. Furthermore, the agreement in Fig.~\ref{adiab_SOC} between the winding obtained from a symmetric flow over the patch $S_{\rho}$ and from a non-symmetric flow over the whole Brillouin zone confirms our claim that the Wilson loop winding protected by $C_2\mathcal{T}$ symmetry is independent of the parameterization of the flow.

Importantly, the spinful insulating phases always split an EBR, see the examples of Fig.~\ref{fig_BS_IRREPs_spinful}(a) and Fig.~\ref{fig_BS_IRREPs_spinful}(b) characterized by their IRREPs at the HSPs $\Gamma$, K, and M. We give in Table \ref{table_IRREP_77} the relevant IRREPs for the generators of the little co-groups at the HSPs.\footnote{\unexpanded{In this work we have chosen the convention $C^+_{n}\vert \uparrow\rangle  = e^{- i 2\pi/n} \vert \uparrow\rangle$ and $C^+_{n}\vert \downarrow\rangle  = e^{+i 2\pi/n} \vert \downarrow\rangle$.}} In the case with inversion symmetry we have the split EBR $\mathrm{B}_1$+$\mathrm{B}_2$ where the allowed qBRs (quasiband representation\cite{EBR_1}) $\mathrm{B}_{1,2}$ can be formed from all the permutations among the IRREPs of Fig.~\ref{fig_BS_IRREPs_spinful}(a). In the case without inversion symmetry (AII+L77), we have the split EBR $\mathrm{B}_1$+$\mathrm{B}_2$ where now the qBRs are constrained by the compatibility relations among the IRREPs as $\mathrm{B}_1=(\overline{\Gamma}_{9(8)},\overline{K}_6,\overline{M}_5)$ and $\mathrm{B}_2=(\overline{\Gamma}_{8(9)},\overline{K}_4,\overline{K}_5,\overline{M}_5)$, i.e.~only $\overline{\Gamma}_{8}$ and $\overline{\Gamma}_{9}$ can now be freely permuted. Furthermore, the Wilsonian phases can be interpreted as the expectation values of the position operator within the basis of Wannier functions generated only by the occupied Bloch eigenstates.\cite{Zak1,Zak2,Van1,Van2,WindingKMZ2,Vanderbilt_WCC,Wi1,Wi3, HolAlex_Bloch_Oscillations} Therefore, whenever the Wilson loop has a symmetry protected nontrivial winding there is an obstruction to the definition of a basis set of localized and \textit{symmetric} Wannier functions, since then the Wannier functions cannot have well defined Wyckoff's positions. The symmetric condition means that the set of Wannier must be closed under all the symmetries of the system, otherwise some symmetry operations map some of these Wannier functions to the Wannier functions obtained from the Bloch functions of the unoccupied subspace. The offset of Eq.~(\ref{spinful_inversion_homotopy}) marks that all the insulating phases of these symmetry classes must be topologically non-trivial. Hence by considering a band structure that is composed only of the bands of a single EBR (here the four bands of the spinful honeycomb lattice), we quantitatively corroborate with a concrete example the claim of ``Topological Quantum Chemistry" \cite{Clas5,EBR_1} that whenever an EBR is split the system must be topologically nontrivial. In Section \ref{six_band} we analyze the effect of adding (in the vector bundle sense) two EBRs and reveal the fragile topology of split EBRs.

{\def\arraystretch{1.3}  
\begin{table}[thb]
\caption{\label{table_IRREP_77} Spinful IRREPs for the generators of the little co-group\cite{BradCrack} at the high-symmetry points $\Gamma$, K, and M, for L77 and L80 with TRS. Retrieved from the Bilbao Crystallographic Server.\cite{Double_IRREPs} We have marked as a sup-script the parity eigenvalues ($\pm$) of the IRREPs of L80.}
\begin{tabular*}{\linewidth}{ @{\extracolsep{\fill}} l |  c c | cc}
\hline 
\hline 
	  $\Gamma$ & $C_{6v}$ & $D_{6h}$ & $C^+_{6z}$ &  $m_x$ \\
	\hline 
	& $\overline{\Gamma}_9$ & $\overline{\Gamma}_9^{(+)}$ & $\left(\begin{array}{cc} e^{-i \pi/6} & 0\\0 & e^{i \pi/6} \end{array}\right)$  & 
		$\left(\begin{array}{cc}  i & 0\\0 & -i  \end{array}\right)$  \\
	& $\overline{\Gamma}_8$ & $\overline{\Gamma}_{11}^{(-)}$ & $\left(\begin{array}{cc} e^{i 5\pi/6} & 0\\0 & e^{-i 5\pi/6} \end{array}\right)$  & 
		$\left(\begin{array}{cc} i & 0\\0 & -i \end{array}\right)$  \\
	\hline
	$\text{K}$ & $C_{3v}$ & $D_{3h}$ & $C^+_{3z}$ & $m_x$  \\
	\hline 
	& $ \begin{array}{c} \overline{\text{K}}_4 \\ \overline{\text{K}}_5 \end{array}$ & $\overline{\text{K}}_7$  & $ \begin{array}{c} -1 \\ -1 \end{array}$  & 
		$ \begin{array}{c} -i \\ i \end{array}$  \\
	& $\overline{\text{K}}_6$ &  $\overline{\text{K}}_8$ &$\left(\begin{array}{cc} e^{ i \pi/3} & 0\\0 & e^{- i \pi/3} \end{array}\right)$  & 
		$\left(\begin{array}{cc} i & 0\\0  & -i \end{array}\right)$  \\
	\hline
	$\text{M}$ & $C_{2v}$  &  $D_{2h}$  & $C^+_{2z}$ & $m_x$   \\
	\hline 
	& $\overline{\text{M}}_5$ & $\overline{\text{M}}_5^{(+)}$  &$\left(\begin{array}{cc} i & 0\\0 & -i \end{array}\right)$  & 
		$\left(\begin{array}{cc} i & 0\\0 & -i \end{array}\right)$  \\
	& $\overline{\text{M}}_5$ & $\overline{\text{M}}_6^{(-)}$  &$\left(\begin{array}{cc} i & 0\\0 & -i \end{array}\right)$  & 
		$\left(\begin{array}{cc} i & 0\\0 & -i \end{array}\right)$  \\
	\hline
	\hline
\end{tabular*}
\end{table} 
}

\section{Spinful case with non-adiabatic breaking of inversion symmetry}\label{non_adiabatic}

We now address the case beyond the adiabatic breaking of inversion symmetry. Phases with adiabatically broken inversion symmetry require Rashba SOC to be weak as compared to $s_z$-preserving SOC. When Rashba SOC is instead strong enough, additional topological phase transitions can occur. Putting aside the topological semimetallic phases at half filling, the insulating phases of AII+L77($2b$) always splits the EBR into $\mathrm{B}_1$ (with the IRREP $\overline{K}_6$ at K) and $\mathrm{B}_2$ (with the IRREPs $\overline{K}_4$ and $\overline{K}_5$ at K). We thus continue the characterization of TQC\cite{Clas5,EBR_1,EBR_2} relating the split of an EBR and nontrivial topology. 

Towards that aim, we use an alternative approach to the symmetry protected quantization of Wilson loop spectra over $l_{\Gamma\text{K}}$ and $l_{\Gamma\text{M}}$ based on the (pseudo-)spin locking at $\Gamma$, K, and M. This not only provides a more direct proof that $W_{II} = W^{\mathcal{I}}_{II}$ when inversion symmetry is broken adiabatically, but further nicely leads to the interpretation of the topological phases beyond the adiabatic breaking of inversion symmetry. The $\hat{z}$-axis of rotational symmetries at the HSPs $\Gamma$, K, and M naturally defines a quantization axis for the pseudo-spin degrees of freedom combining the orbital and the bare spin structures. Fewer band models have the additional feature that spin-flip terms vanish at some HSPs leading to a bare spin composition of the eigenstates that diagonalize the matrix representations of the rotation symmetries. For instance, all the spin-dependent terms vanish at $\Gamma$ such that each Kramers doublet can be decomposed into pure spin components. By inspection of the Hamiltonian at $\Gamma$ we find that the Bloch eigenstates form the doublets $(\vert \phi_{\Gamma_j},\uparrow , \Gamma \rangle ,\vert \phi_{\Gamma_j},\downarrow , \Gamma \rangle)^T$, with $j=1$ for the IRREP $\overline{\Gamma}_9$ ($\phi_{\Gamma_1} \propto \varphi_A+\varphi_B$) and $j=4$ for the IRREP $\overline{\Gamma}_8$ ($\phi_{\Gamma_4} \propto \varphi_A-\varphi_B$), written in the basis that makes the matrix representation of the rotation symmetries at $\Gamma$ diagonal (this is the symmetry-Bloch basis defined in Appendix \ref{spinless_ap}). We find $\mathrm{diag}(e^{- i \pi/3},e^{ i \pi/3})$ as the representation of $C^+_{3z}$-symmetry for both doublets. By inspection of the Hamiltonian at K (say at $K_1$), we find that the doublet for the IRREP $\overline{K}_6$ can again be decomposed into pure spin components as $\vert \boldsymbol{\psi}_{\overline{K}_6} , K_1 \rangle = (\vert \varphi_A ,\uparrow, K_1 \rangle , \vert \varphi_B,\downarrow , K_1 \rangle )^T$ which gives $\mathrm{diag}(e^{ i \pi/3},e^{- i \pi/3})$ as the representation of $C^+_{3z}$-symmetry. The partner doublet at the inverted momentum, $-K_1$, can be taken as $\vert \boldsymbol{\psi}_{\overline{K}_6} , -K_1 \rangle = (\vert \varphi_B ,\uparrow, -K_1 \rangle , \vert \varphi_A,\downarrow , -K_1 \rangle )^T$ such that it has the same matrix representation of rotations. We also verify that these two doublets are partners under TRS through 
\begin{equation}
\label{TRS_relation}
		 \langle \boldsymbol{\psi}_{\overline{K}_6} , -K_1 \vert \mathcal{T} 
		\vert \boldsymbol{\psi}_{\overline{K}_6} , K_1 \rangle = \left( \begin{array}{cc} 0& -1 \\
		1 &0 \end{array}\right) \mathcal{K}\;.
\end{equation} 
The most general rotations of the doublets that preserves this form of the TRS corresponds to a $SU(2)$ transformation.\cite{RoyZ2} If we further want to preserve the diagonal form of the matrix representation of rotations then only transformations of the form $\mathrm{diag}(e^{i \theta},e^{- i \theta})$ are allowed. The Bloch eigenstates for the IRREPs $\overline{K}_4$ and $\overline{K}_5$ have a mixed spin structure due to the presence of spin-flip terms at K, i.e.~we find 
\begin{equation}
\begin{aligned}
	\vert \psi_{\overline{K}_4}, K_1 \rangle &= (\omega \vert \varphi_A , \downarrow, K_1 \rangle + \vert \varphi_B , \uparrow, K_1 \rangle )/\sqrt{2} , \\
	\vert \psi_{\overline{K}_5}, K_1 \rangle &=  (-\omega \vert \varphi_A , \downarrow, K_1 \rangle + \vert \varphi_B , \uparrow, K_1 \rangle )/\sqrt{2} ,
\end{aligned}
\end{equation}
with $\omega = e^{i 2\pi/3}$, and at the inverted momentum,
\begin{equation}
\begin{aligned}
	\vert \psi_{\overline{K}_4}, -K_1 \rangle &= (-\omega^* \vert \varphi_A , \uparrow, -K_1 \rangle + \vert \varphi_B , \downarrow, -K_1 \rangle )/\sqrt{2} , \\
	\vert \psi_{\overline{K}_5}, -K_1 \rangle &=  (\omega^* \vert \varphi_A , \uparrow, -K_1 \rangle + \vert \varphi_B , \downarrow, -K_1 \rangle )/\sqrt{2} .
\end{aligned}
\end{equation}
The $C_{3z}^+$-symmetry eigenvalue of the eigenstates $\overline{K}_4$ and $\overline{K}_5$ is $-1$. Allowing a rotation among the eigenstates, the basis $\vert \varphi_{B} , \uparrow, K_1 \rangle \propto \vert \psi_{\overline{K}_4}, K_1 \rangle +\vert \psi_{\overline{K}_5}, K_1 \rangle $ and $\vert \varphi_{A} , \downarrow, K_1 \rangle \propto \vert \psi_{\overline{K}_4}, K_1 \rangle - \vert \psi_{\overline{K}_5}, K_1 \rangle $ has a pure spin composition and conserves the $C_{3z}^+$-symmetry eigenvalue. Contrary to $\Gamma$ and K, the eigenstates at M do not have a simple form and there is not a direct relation between the pseudo-spin basis that diagonalizes the representation of $C_{2z}$ symmetry and the bare spin components.

In the spirit of Ref.~[\onlinecite{FuKaneZ2}] we characterize the topology of the two-band occupied subspace through the construction of a basis set of smooth (cell-periodic) Bloch functions $\vert \boldsymbol{v}, \boldsymbol{k} \rangle =(\vert v_1, \boldsymbol{k} \rangle , \vert  v_2, \boldsymbol{k}\rangle)^T$ spanning the same Hilbert space. We call it a smooth frame for the occupied subspace. Such a frame always exists over one-dimensional loops in momentum space (actually it always exists over the BZ by the vanishing of the first Chern class due to TRS).\cite{Panati_chern} Furthermore, we can make it periodic over a loop that threads the BZ, i.e.~where the final point of the loop is given by a shift of the base point by reciprocal lattice vector as for $l_{\Gamma\text{K}}$ and $l_{\Gamma\text{M}}$. Therefore, Wilson loops are diagonal and thus gauge invariant within this frame. More precisely, the Wilson loop operator takes the form $\widetilde{W}_{l} = \mathcal{P}_{1} +  \mathcal{P}_{2}$ with the product of projection operators written in the smooth and periodic frame $\mathcal{P}_{i} = \prod\limits_{\boldsymbol{k}}^{l} \vert v_i, \boldsymbol{k} \rangle \langle v_i, \boldsymbol{k} \vert $ for $i=1,2$. The Wilson loop over $l_{\Gamma\text{K}}$, connecting $\Gamma$, $\text{K}$, and $\Gamma' = \Gamma +\boldsymbol{b}_1+\boldsymbol{b}_2$, is then given by $\widetilde{\mathcal{W}}_{l_{\Gamma\text{K}}} = \langle \boldsymbol{v}, \Gamma' \vert \widetilde{W}_{l_{\Gamma\text{K}}} \vert \boldsymbol{v}, \Gamma \rangle  = \mathrm{diag}(e^{i \varphi_1},e^{i \varphi_2})$ where $\varphi_{1,2}$ are the Berry phases of each component of the frame. Similarly for $l_{\Gamma\text{M}}$ that connects $\Gamma$, M, and $\Gamma' = \Gamma +\boldsymbol{b}_1+\boldsymbol{b}_2$.  

In order to derive the symmetry protected quantization of the Wilson loop spectrum over $l_{\Gamma\text{K}}$ we further require that the matrix representation of rotations at $\Gamma$, K, and $\Gamma'$ be diagonal within the smooth and periodic frame. We call this last condition the rotation-symmetric gauge. 
Following the parallel transport method of Soluyanov and Vanderbilt in Ref.~\onlinecite{Vanderbilt_smooth_gauge} we have verified numerically that a smooth and periodic frame that also satisfies the rotation-symmetric gauge always exists in two-band subspaces.\footnote{Alternatively to Ref.~\onlinecite{Vanderbilt_smooth_gauge} that uses a singular value decomposition, we mention the early construction of Ref.~\onlinecite{Erik_polar} based on a polar decomposition instead.} While this points to a general proof of existence an analytical solution of the four-band Hamiltonian is missing and we leave it as future work. In the next section we argue that such a frame in general does not exist for a four-band subspace of a six-band Hamiltonian. 

Within the smooth, periodic, and rotation-symmetric frame for the two-band occupied subspace, $\boldsymbol{v} = (v_1,v_2)$, the Wilson loop over $l_{\Gamma\text{K}} = l_{b}\circ l_a$, with the segments $l_a = l_{K_1\leftarrow\Gamma}$ and $l_b = l_{\Gamma'\leftarrow K_1}$, is given by 
\begin{align*}
	\widetilde{\mathcal{W}}_{l_{\Gamma\text{K}}}	&= \langle \boldsymbol{v}, \Gamma'\vert \widetilde{W}_{b} \vert  \boldsymbol{v}, K_1 \rangle \langle \boldsymbol{v}, K_1\vert \widetilde{W}_{a} \vert  \boldsymbol{v}, \Gamma \rangle \;,\\
	&= \widetilde{\mathcal{W}}_{b}  \widetilde{\mathcal{W}}_{a} \;,\\
	&= \tilde{R}^{\Gamma}_{3} 	\widetilde{\mathcal{W}}^{-1}_{a} \left(\tilde{R}^{\text{K}}_{3}\right)^{-1}  \widetilde{\mathcal{W}}_{a} \;,\\
	&= \tilde{R}^{\Gamma}_{3} \left(\begin{array}{cc} e^{-i\varphi_a} & 0 \\ 0 & e^{i\varphi_a} \end{array}\right) \tilde{R}^{\text{K}}_{\bar{3}} \left(\begin{array}{cc} e^{i\varphi_a} & 0 \\ 0 & e^{-i\varphi_a} \end{array}\right)\;,
\end{align*}
where $\tilde{R}^{\bar{\boldsymbol{k}}}_{3(\bar{3})}$ is the diagonal matrix representation of the symmetry $C_{3z}^+$ ($C_{3z}^-$) at the HSP $\bar{\boldsymbol{k}}$. Therefore, 
\begin{equation}
	\widetilde{\mathcal{W}}_{l_{\Gamma\text{K}}} = \left(\begin{array}{cc} \xi^{\Gamma(1)}_{3} \xi^{K(1)}_{\bar{3}} & 0 \\ 0 & \xi^{\Gamma(2)}_{3} \xi^{K(2)}_{\bar{3}} \end{array}\right)\;,
\end{equation}
where $\xi^{\bar{\boldsymbol{k}}(n)}_{3(\bar{3})}$ are the $C_{3z}^+$-($C_{3z}^-$-)symmetry eigenvalues of the $n$-th component of the doublets at $\bar{\boldsymbol{k}} = \Gamma,\text{K}$, i.e.~$\tilde{R}^{\bar{\boldsymbol{k}}}_{3(\bar{3})} = \mathrm{diag}\left(\xi^{\bar{\boldsymbol{k}}(1)}_{3(\bar{3})},\xi^{\bar{\boldsymbol{k}}(2)}_{3(\bar{3})}\right)$. Proceeding similarly for $l_{\Gamma\text{M}}$, we find
\begin{equation}
	\widetilde{\mathcal{W}}_{l_{\Gamma\text{M}}} = \left(\begin{array}{cc} \xi^{\Gamma(1)}_{2} \xi^{M(1)}_{\bar{2}} & 0 \\ 0 & \xi^{\Gamma(2)}_{2} \xi^{M(2)}_{\bar{2}} \end{array}\right)\;,
\end{equation}
where $\xi_{2(\bar{2})}$ are the $C_{2z}^+$-($C_{2z}^-$-)symmetry eigenvalues. 

It is now straightforward to obtain the symmetry protected Wilson loop spectrum for all possible configurations of the two-band occupied subspace of the insulating phases of AII+L77($2b$). For the qBR $\mathrm{B}_1$, i.e.~connecting $\overline{\Gamma}_9$ or $\overline{\Gamma}_8$ to $\overline{K}_6$, and assuming that within the smooth frame the spin components of the doublets are aligned at $\Gamma$ and K (which we write $\mathrm{B}_{1,\uparrow\uparrow}$), we find  
\begin{equation}
\label{spin_aligned}
	\widetilde{\mathcal{W}}^{\mathrm{B}_{1,\uparrow\uparrow}}_{l_{\Gamma\text{K}}} = \left(\begin{array}{cc} e^{- i \frac{2\pi}{3}} & 0 \\ 0 & e^{ i \frac{2\pi}{3}} \end{array}\right)\;,
\end{equation}
i.e.~the smooth frame connects a doublet $( \uparrow  , \downarrow )$ at $\Gamma$ to a doublet $( \uparrow  , \downarrow )$ at K. For the qBR $\mathrm{B}_2$, i.e.~connecting $\overline{\Gamma}_9$ or $\overline{\Gamma}_8$ to $(\overline{K}_4,\overline{K}_5)$, either spin alignments gives 
\begin{equation}
\label{BR2}
	\widetilde{\mathcal{W}}^{\mathrm{B}_2}_{l_{\Gamma\text{K}}} =  \left(\begin{array}{cc} -e^{-i \frac{\pi}{3}} & 0 \\ 0 & - e^{ i \frac{\pi}{3}} \end{array}\right)=  \left(\begin{array}{cc} e^{i \frac{2\pi}{3}} & 0 \\ 0 &  e^{- i \frac{2\pi}{3}} \end{array}\right) \;,
\end{equation}
i.e.~the smooth frame connects a doublet $( \uparrow  , \downarrow )$ at $\Gamma$ to a doublet $( \uparrow , \downarrow )$ or $(  \downarrow ,\uparrow)$ at K. When we compute the smooth and rotation-symmetric frame for $\mathrm{B}_2$ we find that it is composed of pure spin components at K, i.e.~$(v_1,v_2)\propto(\varphi_{B,\uparrow}(K_1),\varphi_{A,\downarrow}(K_1))$, which have the same rotation eigenvalue of $-1$ as the Bloch eigenstates under a $C_{3z}$ rotation. We note however that the smooth frame is not really needed for $\mathrm{B}_2$ since with $\tilde{R}^{\text{K}}_{\bar{3}} = -\mathbb{I}$ we have\cite{HolAlex_Bloch_Oscillations} $\mathcal{W}^{-1}_a \tilde{R}^{\text{K}}_{\bar{3}} \mathcal{W}_a = -\mathbb{I}$ which readily leads to the quantization of the Wilson loop $\mathcal{W}^{\mathrm{B}_2}_{l_{\Gamma\text{K}}}=-\tilde{R}^{\Gamma}_{3}$ independently of the smoothness of the chosen Bloch frame (although it must satisfy the rotation-symmetric gauge in order for the rotation matrices to be diagonal). As for $l_{\Gamma\text{M}}$, since there is a single IRREP at M ($\overline{M}_5$) and the IRREPs at $\Gamma$ ($\overline{\Gamma}_8,\overline{\Gamma}_9$) have identical $C_{2z}$-eigenvalues, we find  
\begin{equation}
	\widetilde{\mathcal{W}}_{l_{\Gamma\text{M}}} =  \left(\begin{array}{cc} -i  (\pm i) & 0 \\ 0 & i  (\mp i) \end{array}\right)=  \left(\begin{array}{cc} \pm1 & 0 \\ 0 &  \pm1 \end{array}\right) \;,
\end{equation}
both allowed for $\mathrm{B}_1$ and $\mathrm{B}_2$. This result based on the construction of a smooth (periodic) and rotation-symmetric Bloch frame recovers the algebraic derivation of the quantization in Section \ref{WLW} based on the combined constraints of TRS and $C_{2z}T$ on the Wilson loop spectrum. 

We conclude that $\mathrm{B}_{1,\uparrow\uparrow}$ and $\mathrm{B}_2$ give the same symmetry protected reference Wilsonian phases $\overline{\varphi}(1) = \pm 2\pi/3$, see definition in Eq.~(\ref{winding_rho}). Also $\widetilde{\mathcal{W}}_{l_{\Gamma\text{M}}} = [\pm1 , \pm1]$ leads to the symmetry protected reference Wilsonian phases $\overline{\varphi}(0)=0,\pm\pi$ in Eq.~(\ref{winding_rho}). We hence readily find a classification of the Wilson loop winding $W_{II}$ (Eq.~(\ref{windings})) equivalent to Eq.~(\ref{spinful_inversion_homotopy}) but here without assuming inversion symmetry. Having argued earlier that the classification of Eq.~(\ref{spinful_inversion_homotopy}) must be conserved when inversion symmetry is broken only adiabatically (without closing the band gap), we then conclude that the condition of aligned spins at $\Gamma$ and K (within the same smooth Bloch branch) in $\mathrm{B}_1$ is equivalent to this adiabatic condition (i.e. adiabatic breaking of inversion symmetry) and $W_{II} = W^{\mathcal{I}}_{II}$.

Instead, taking the smooth frame and assuming that the spin components of the doublets of $\mathrm{B}_1$ are flipped between $\Gamma$ and K, which we write $\mathrm{B}_{1,\uparrow\downarrow}$, we get
\begin{equation}
\label{spin_flip}
	\tilde{\mathcal{W}}^{\mathrm{B}_{1,\uparrow\downarrow}}_{l_{\Gamma\text{K}}} = \left(\begin{array}{cc} +1 & 0 \\ 0 & +1 \end{array}\right) \;,
\end{equation} 
i.e.~the symmetry protected Wilsonian phase is now $\overline{\varphi}(1) = 0$. Therefore, taking this together with the unchanged $\overline{\varphi}(0) = 0,\pm\pi$ within Eq.~(\ref{winding_rho}), gives $W^{\mathrm{B}_{1,\uparrow\downarrow}}[S_{\rho}] \in [\overline{\varphi}(1) -\overline{\varphi}(0)]/\pi~ (\mathrm{mod}~2)  = \{0,\pm1\} (\mathrm{mod}~2) = 0 ~(\mathrm{mod}~1) \cong \mathbb{Z}$. Hence the set of symmetry allowed Wilson loop windings Eq.~(\ref{windings}) would now be $ 3\mathbb{Z} \ni W^{\mathrm{B}_{1,\uparrow\downarrow}}_{II}$. While this could suggest that many more topological insulating sectors can be realized we actually find only one additional topological insulating phase that is not contained in the classification of $\mathrm{B}_{1,\uparrow\uparrow}$. Indeed, after exploring the phase diagram of the tight-binding model for the symmetry class AII+L77($2b$) including up to the 10th layer of neighbors we find that all the insulating phases beyond those classified by Eq.~(\ref{spinful_inversion_homotopy}) have a zero winding of the Wilson loop for $\mathrm{B}_1$, i.e.~$W^{\mathrm{B}_{1,\uparrow\downarrow}}_{II}=0$. We show below that this phase exactly corresponds to the ``fragile'' topological phase that has been very recently reported in Ref.~\onlinecite{Ft1}. A formal proof that $\Gamma$-K spin-flip phases can only have a zero Wilson loop winding of $\mathrm{B}_{1}$ is still missing and is kept for the future.

Let us analyze in more detail the two cases allowed by $\mathrm{B}_1$. The pure spin structure of the doublets at $\Gamma$ and K (for $\mathrm{B}_1$) is a result of the vanishing of spin-flip terms of the four-band tight-binding Hamiltonian at $\Gamma$ and K. On the one hand, the $\Gamma$-K spin-aligned configuration ($\mathrm{B}_{1,\uparrow\uparrow}$) is naturally adiabatically connected to the inversion symmetric case for which $s_z$ is a good quantum number over the whole BZ. On the other hand, the $\Gamma$-K spin-flip configuration ($\mathrm{B}_{1,\uparrow\downarrow}$) is allowed by Rashba spin-flip terms that unlock the quantization axis of the spins between $\Gamma$ and K. We naturally expect that this is realized only when Rashba SOC is strong enough. Actually, since the rotation-symmetric gauge can always be satisfied by the smooth frame, any adiabatic transformation of the Hamiltonian cannot induce a transition between the spin-aligned and the spin-flipped configurations since it would require the continuous rotation of the spin quantization axis for the doublet at K which is not supported by the four-band Hamiltonian of AII+L77($2b$). Therefore we conclude that the phase with zero Wilson loop winding of $\mathrm{B}_1$ can only be realized when Rashba SOC is strong enough to close and reopen the band gap. The classification of Wilson loop windings obtained from Eq.~(\ref{spin_aligned}) and Eq.~(\ref{BR2}) is in one-to-one correspondence with Eq.~(\ref{spinful_inversion_homotopy}), which we have argued covers all the topological sectors that are adiabatically connected to the inversion symmetric phases, i.e.~with zero Rashba SOC. Then, since Eq.~(\ref{spin_flip}) implies a classification of the Wilson loop windings that is not contained in Eq.~(\ref{spinful_inversion_homotopy}), it must relate to different topological sectors than Eq.~(\ref{spin_aligned}). Therefore, the phase with a trivial $\mathrm{B}_1$ subspace and captured by Eq.~(\ref{spin_flip}) cannot be adiabatically mapped to any of the inversion symmetric phases. 

\begin{figure}[t!]
\centering
\begin{tabular}{c|c}
	\includegraphics[width=0.5\linewidth]{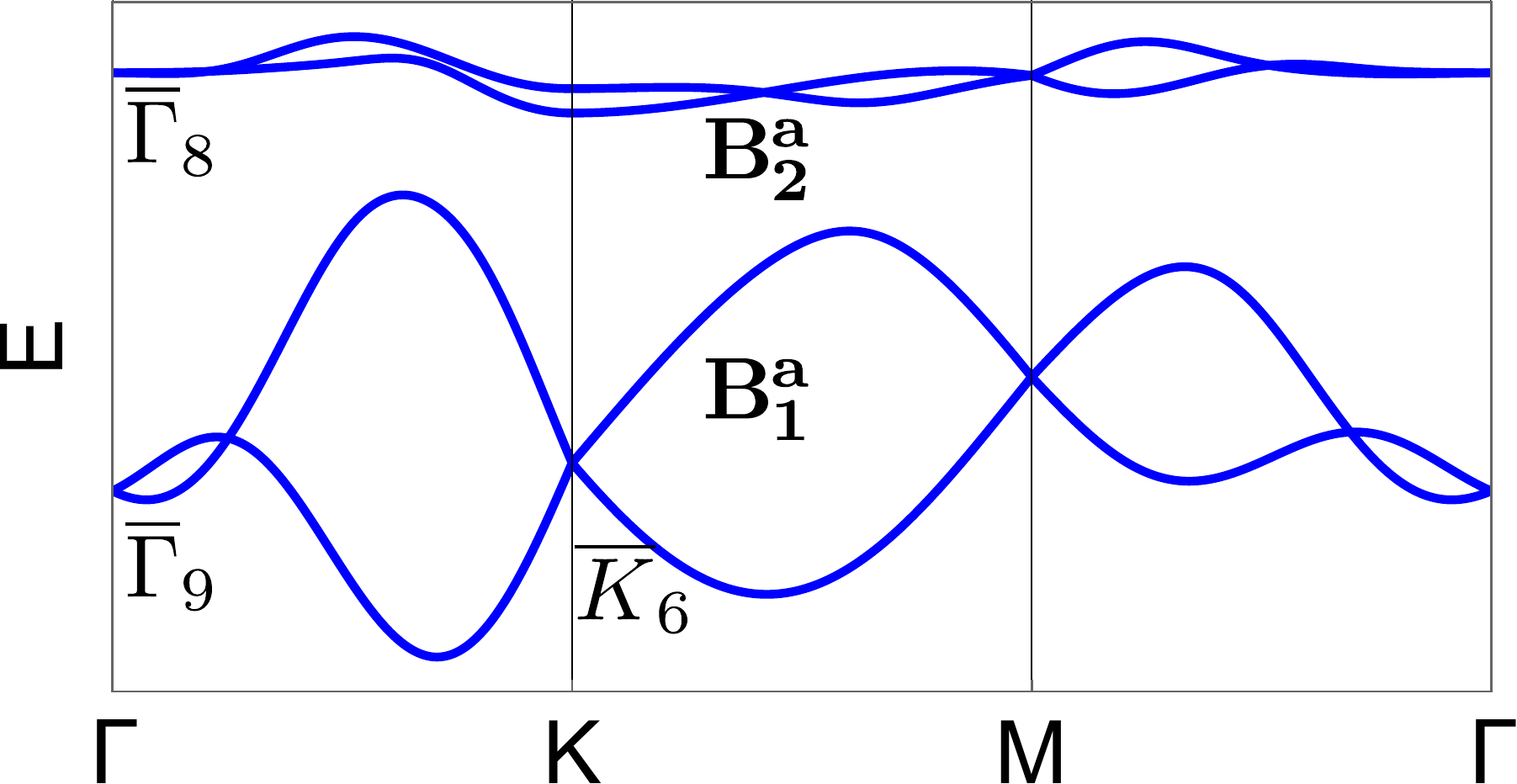} &  
	\includegraphics[width=0.5\linewidth]{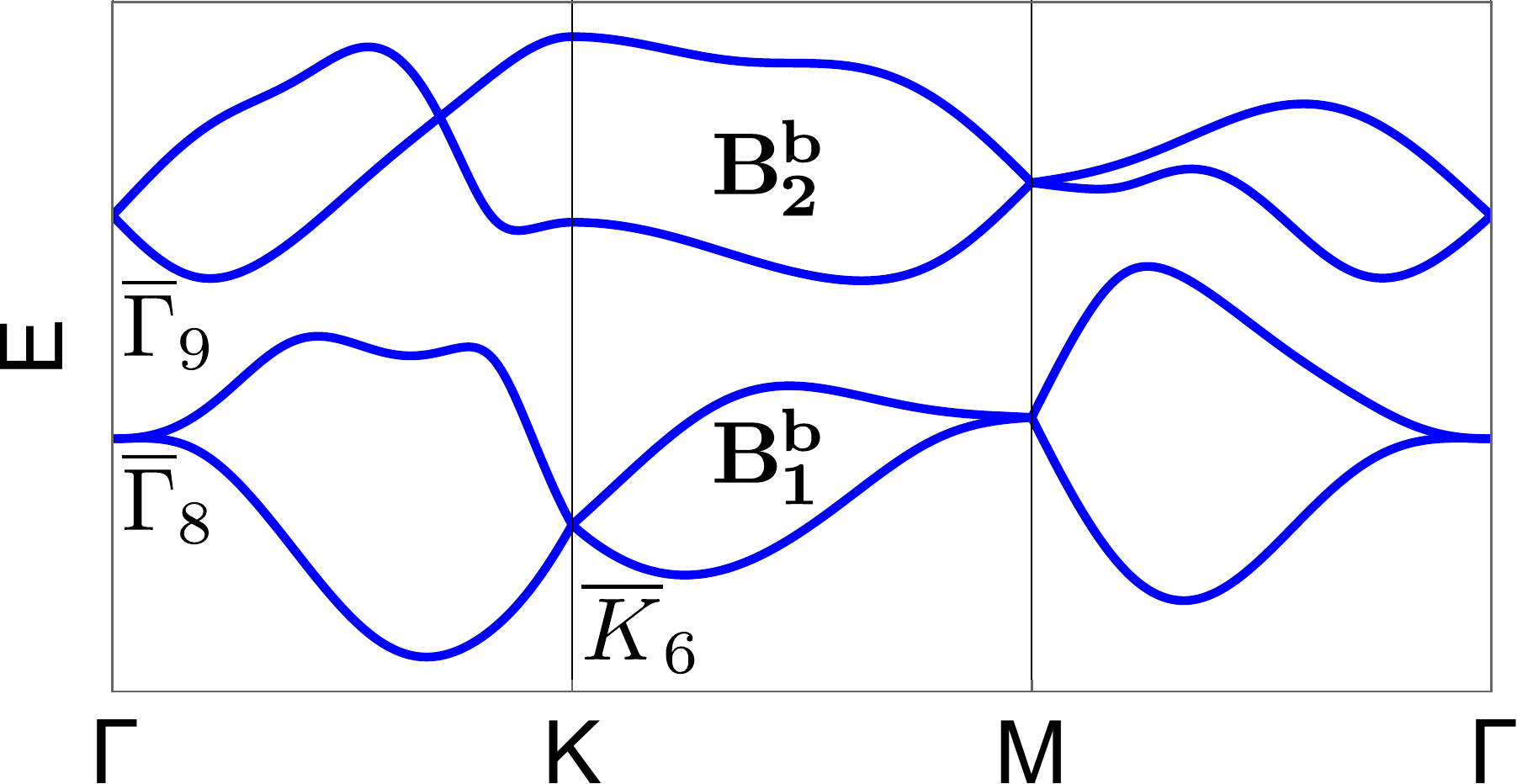} \\
	\includegraphics[width=0.5\linewidth]{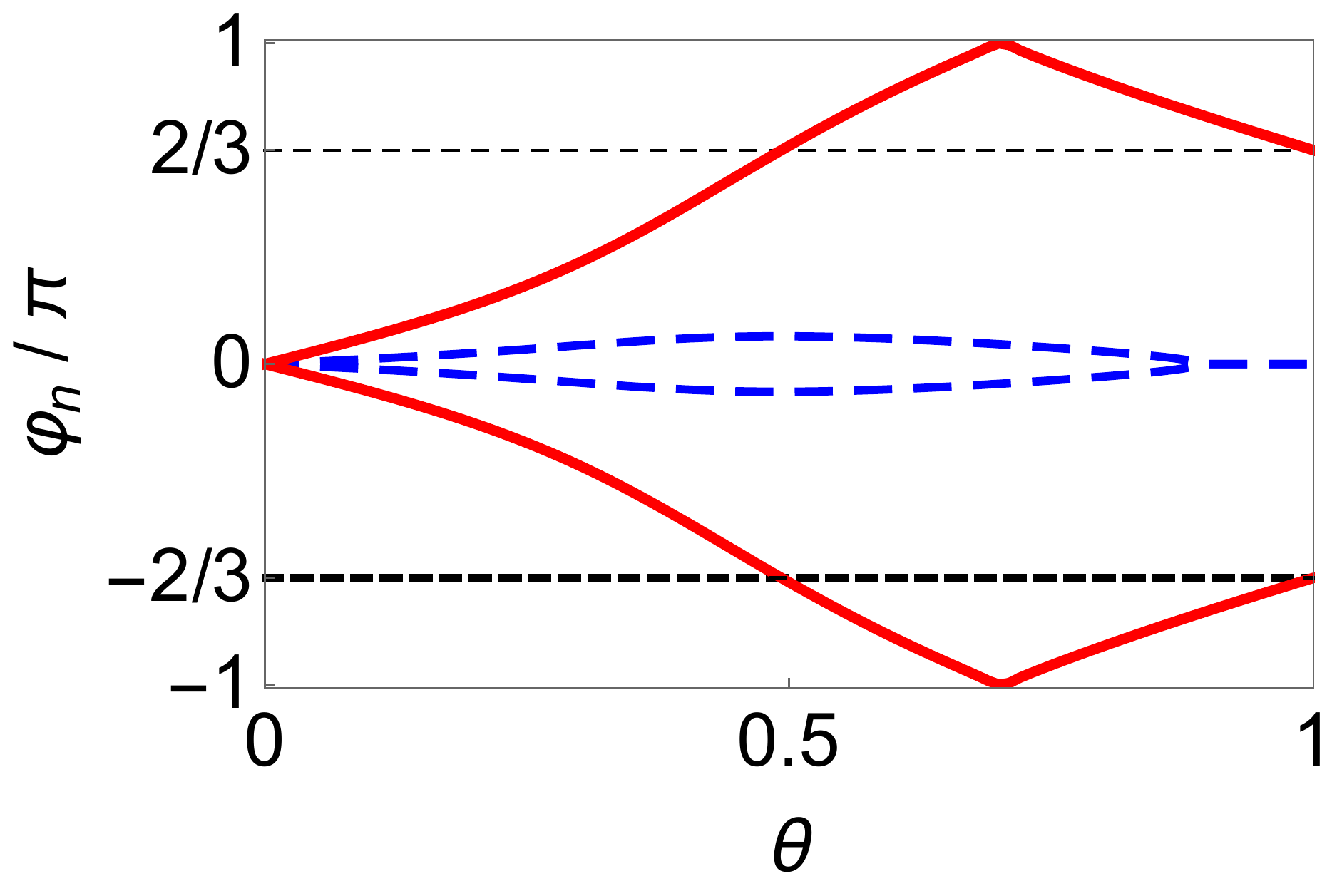} & 
	\includegraphics[width=0.5\linewidth]{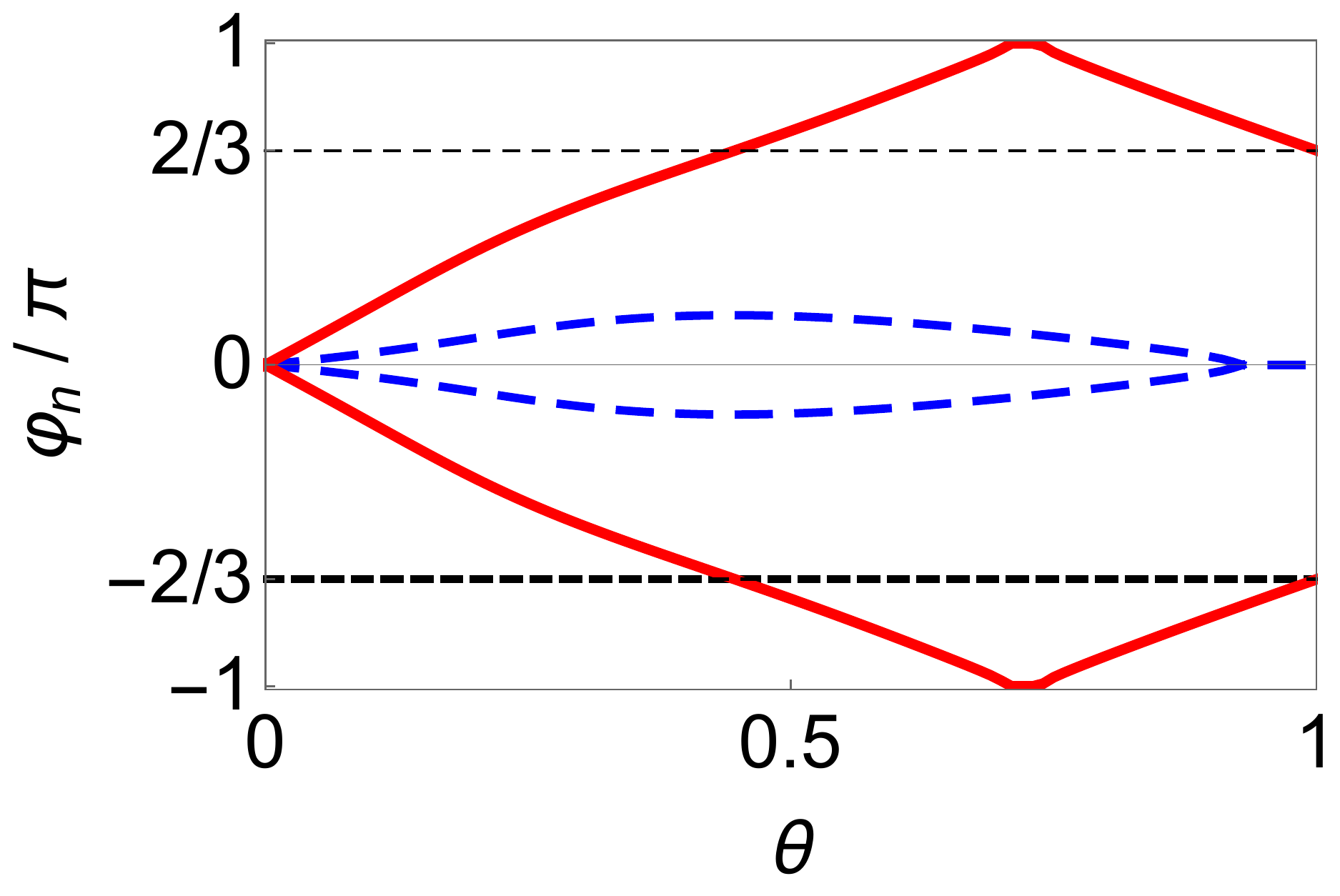} \\
	\includegraphics[width=0.5\linewidth]{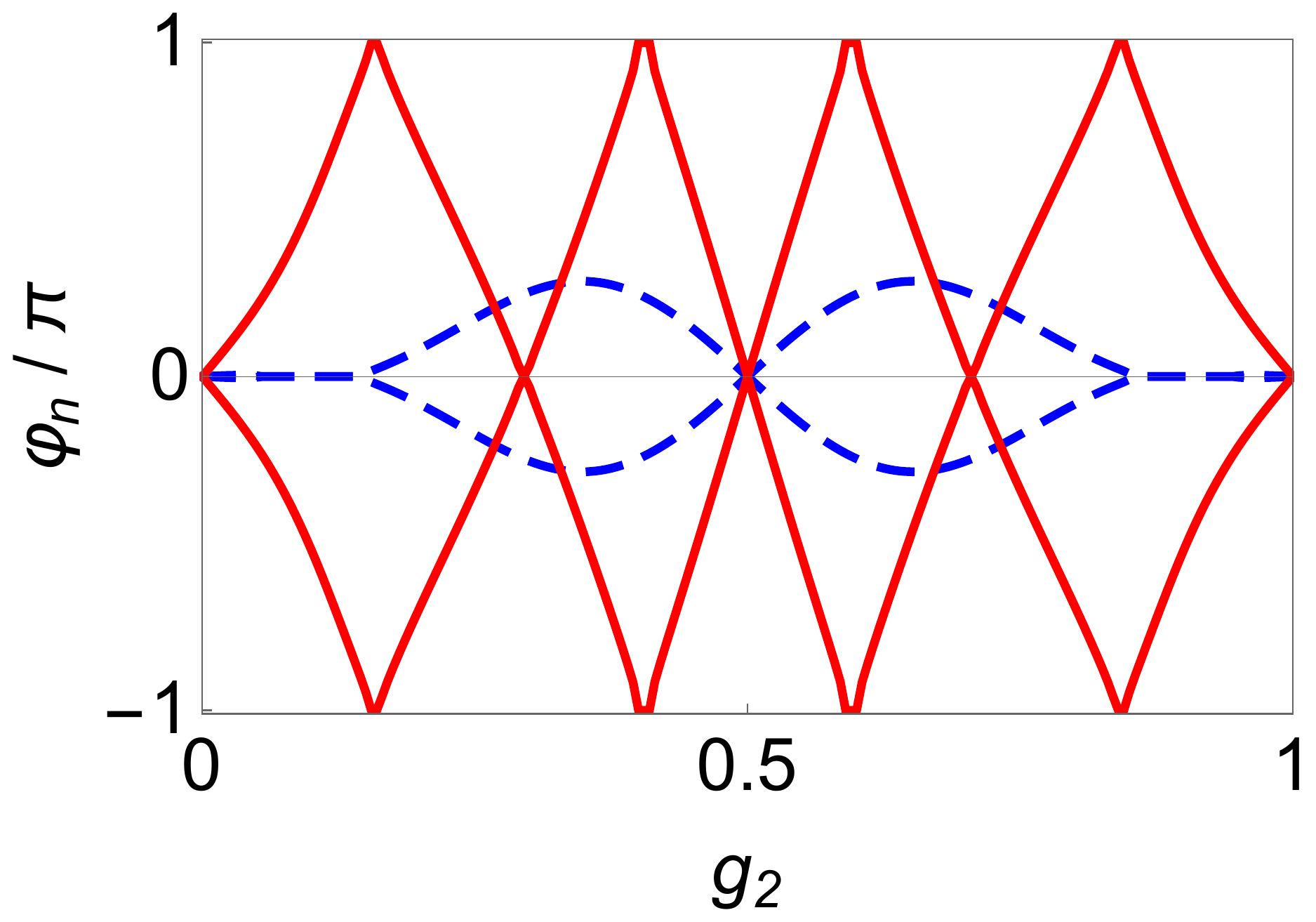} & 
	\includegraphics[width=0.5\linewidth]{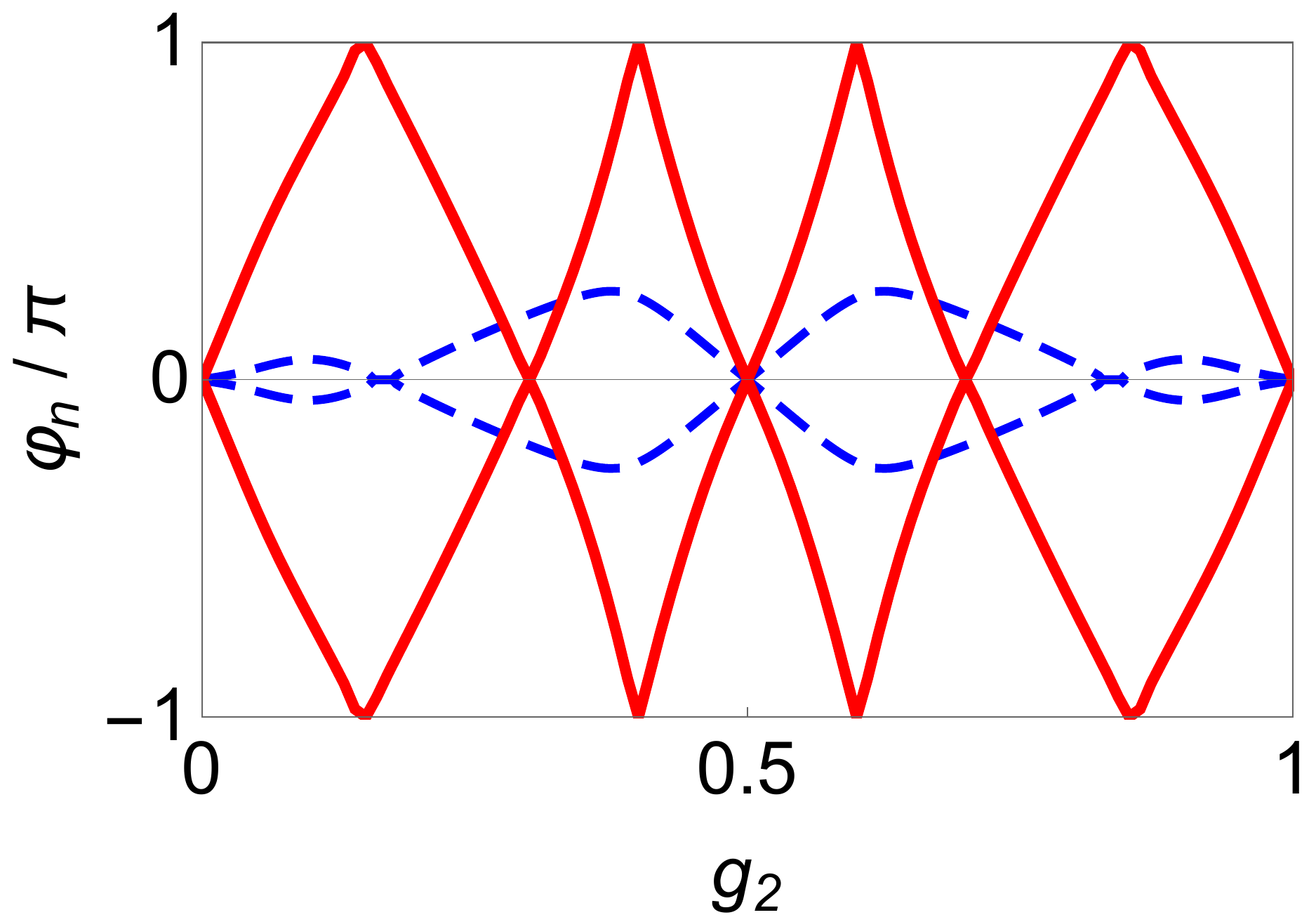} \\
	(a) $W^{\mathrm{B}^a_1}[S_{\rho}] = W^{\mathrm{B}^a_1}_{II} = 0$, & (b) $W^{\mathrm{B}^b_1}[S_{\rho}] = W^{\mathrm{B}^b_1}_{II} = 0$, \\
	 $W^{\mathrm{B}^a_2}_{II} = 4$, $\nu_{\mathrm{FKM}}=0$ & $ W^{\mathrm{B}^b_2}_{II} = 4$, $\nu_{\mathrm{FKM}}=0$ 
\end{tabular}
\caption{\label{fig_PWV} Spinful topological phases with Rashba SOC breaking inversion symmetry non-adiabatically. First, second and third rows are similar to Fig.~\ref{adiab_SOC}. (a) EBR split as $\mathrm{B}_1^a+\mathrm{B}_2^a$ with $\mathrm{B}^a_1=(\overline{\Gamma}_9,\overline{K}_6,\overline{M}_5)$ and $\mathrm{B}^a_2=(\overline{\Gamma}_8,\overline{K}_4,\overline{K}_5,\overline{M}_5)$, with the parameters chosen as in Ref.~\onlinecite{Ft1} for comparison. (b) EBR split as $\mathrm{B}_1^b+\mathrm{B}_2^b$ with $\mathrm{B}_i^b = \mathrm{B}_i^a[\overline{\Gamma}_9 \leftrightarrow \overline{\Gamma}_8]$. }
\end{figure}

Fig.~\ref{fig_PWV}(a) shows one numerical example where the EBR is split into an occupied subspace with $\mathrm{B}^a_1=(\overline{\Gamma}_9,\overline{K}_6,\overline{M}_5)$ and an unoccupied subspace with $\mathrm{B}^a_2=(\overline{\Gamma}_8,\overline{K}_4,\overline{K}_5,\overline{M}_5)$. This example corresponds to the fragile topological phase of Ref.~\onlinecite{Ft1} (we have used the same parameters for comparison). Fig.~\ref{fig_PWV}(b) shows one example where the EBR is split into $\mathrm{B}^b_1=(\overline{\Gamma}_8,\overline{K}_6,\overline{M}_5)$ and $\mathrm{B}^b_2=(\overline{\Gamma}_9,\overline{K}_4,\overline{K}_5,\overline{M}_5)$. In both cases, the Wilson loop spectrum of the $\mathrm{B}_1$ occupied subspace has a zero winding (blue dashed lines), while the unoccupied subspace with $\mathrm{B}_2$ conserves a nontrivial winding of $W_{II} = \pm 4$ (red lines). Ref.~\onlinecite{Ft1} has shown the triviality of $\mathrm{B}^a_1$ in the example of Fig.~\ref{fig_PWV}(a) by computing, from the two occupied Bloch eigenstates, a set of two localized Wannier functions that are both centered at the lattice site $C$, and are mapped on each-other under all the symmetries of the system (i.e.~it is a closed basis set under symmetry). We arrive here at the same conclusion by straightforwardly computing the flow of Wilson loop revealing a zero Wilson loop winding for $\mathrm{B}_1$. Furthermore, it is worth noting that these localized Wannier functions\cite{Ft1} exhibit a strong spin mixture. This nicely supports our spin locking argument which predicts that the trivialized $\mathrm{B}_1$ is characterized by a spin flip between $\Gamma$ and K within the Bloch functions composing the smooth frame.

\section{Six-band case}\label{six_band}

\begin{figure*}[t!]
\centering
\begin{tabular}{c|c|c|c} 
	\includegraphics[width=0.25\linewidth]{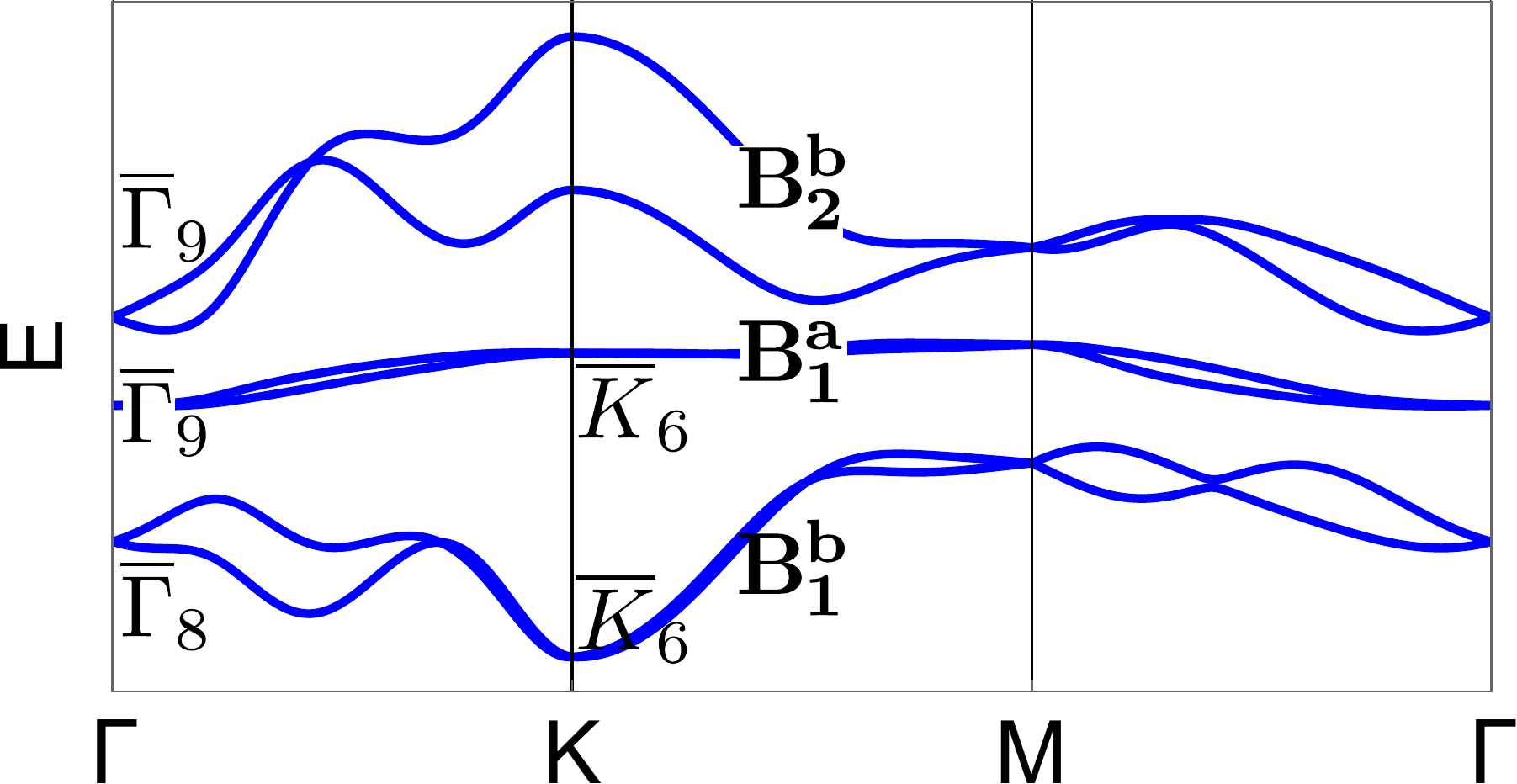} &
	\includegraphics[width=0.25\linewidth]{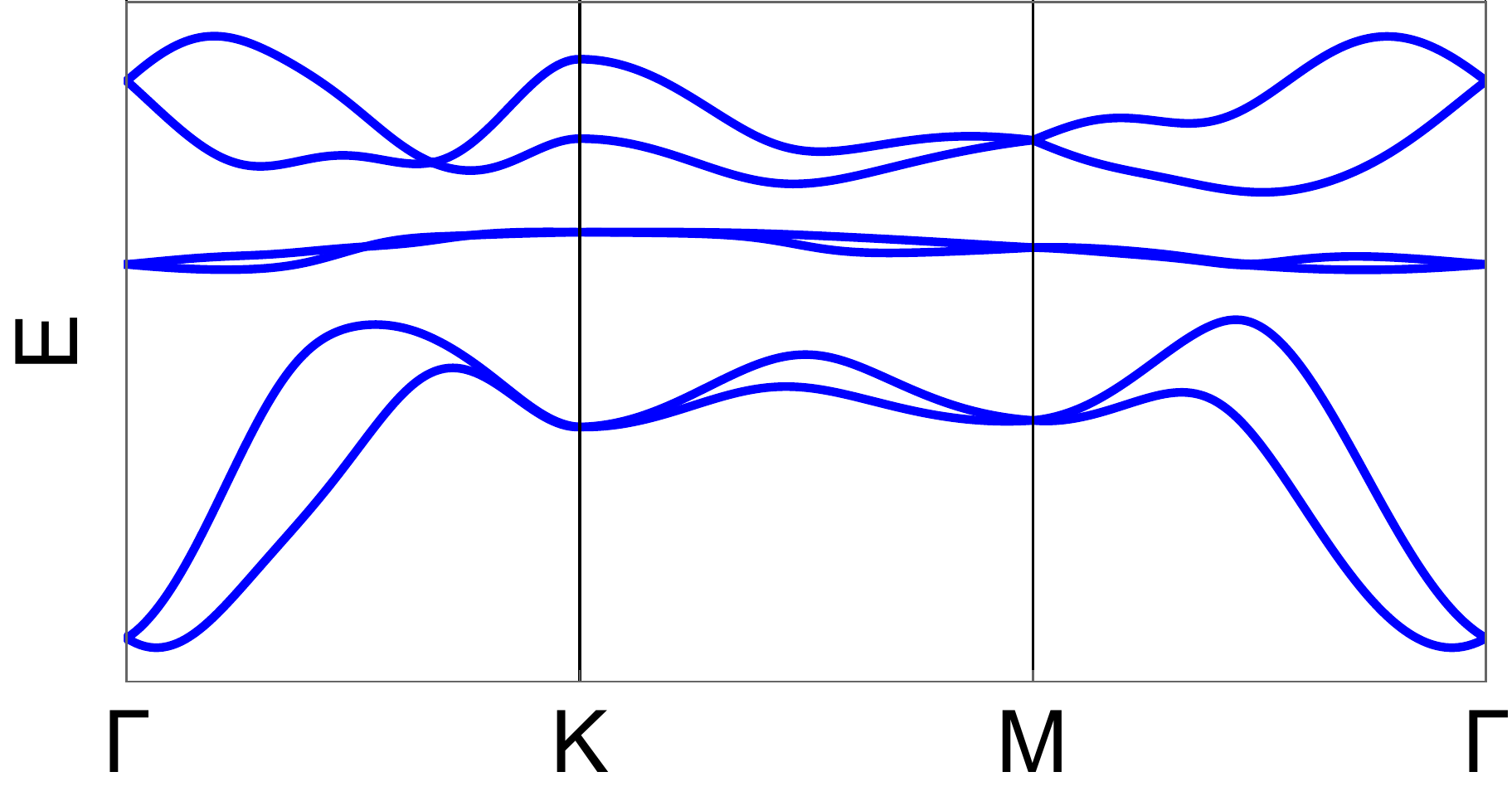} &
	\includegraphics[width=0.25\linewidth]{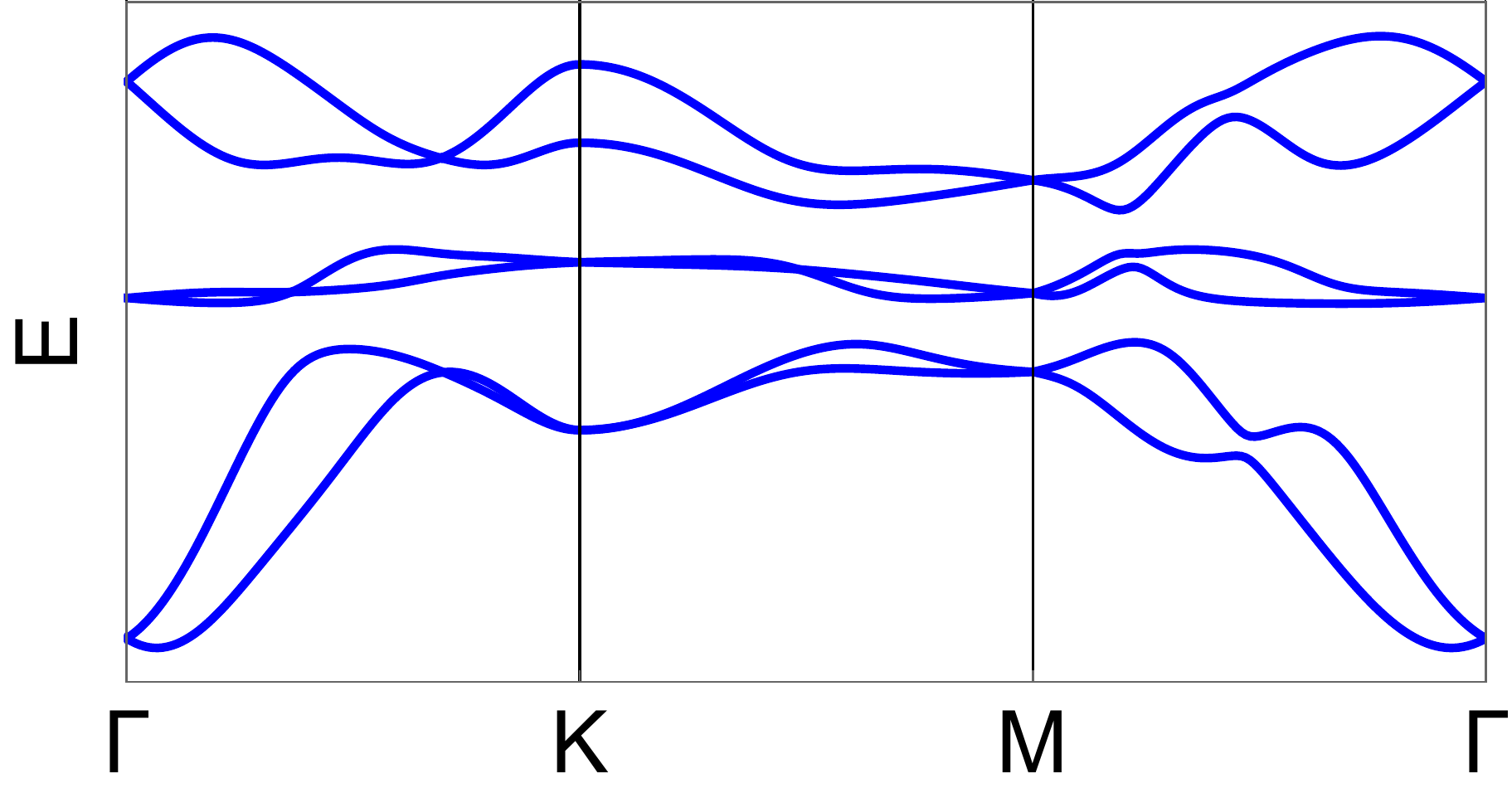} &
	\includegraphics[width=0.25\linewidth]{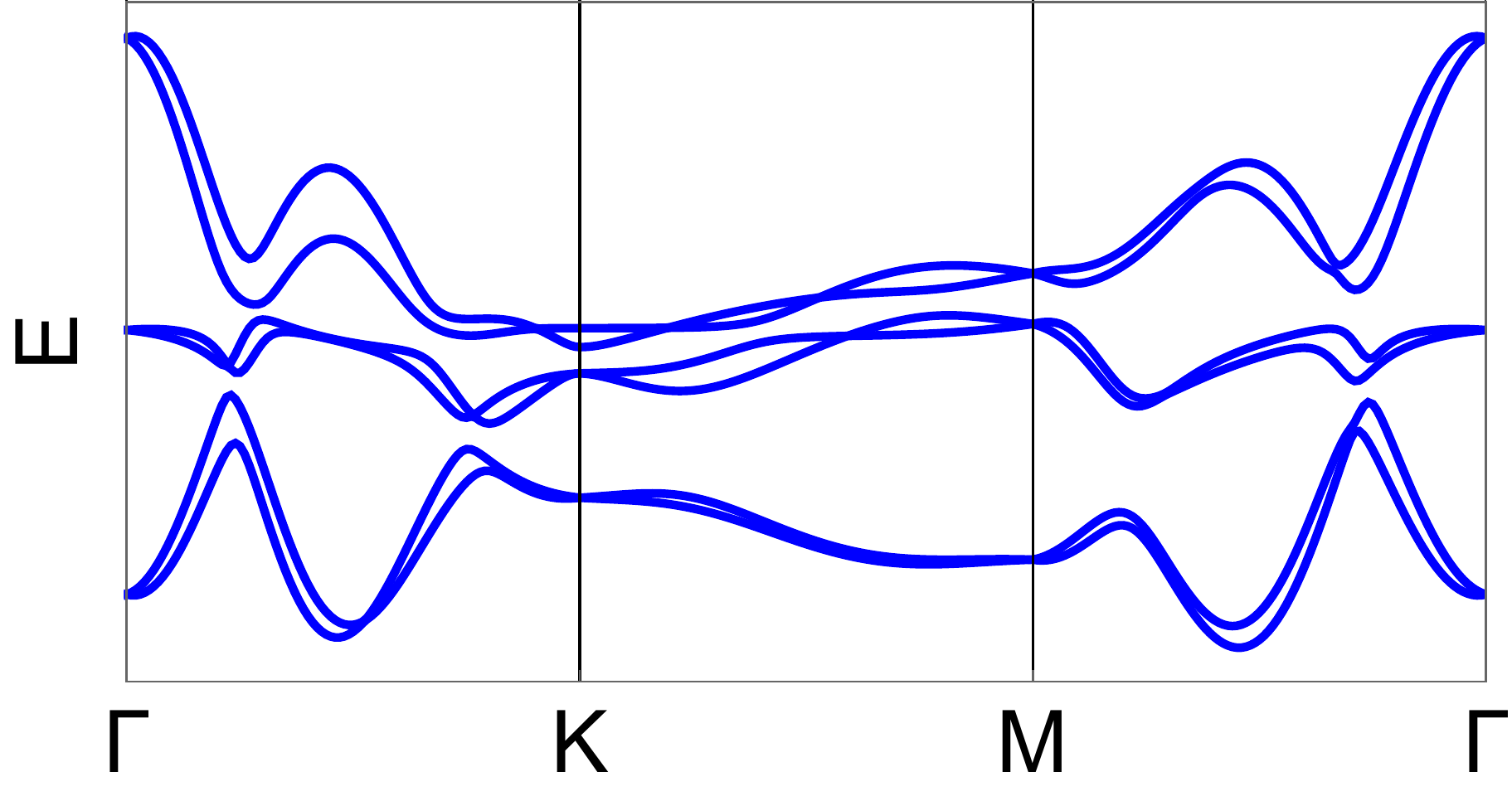} \\
	\includegraphics[width=0.25\linewidth]{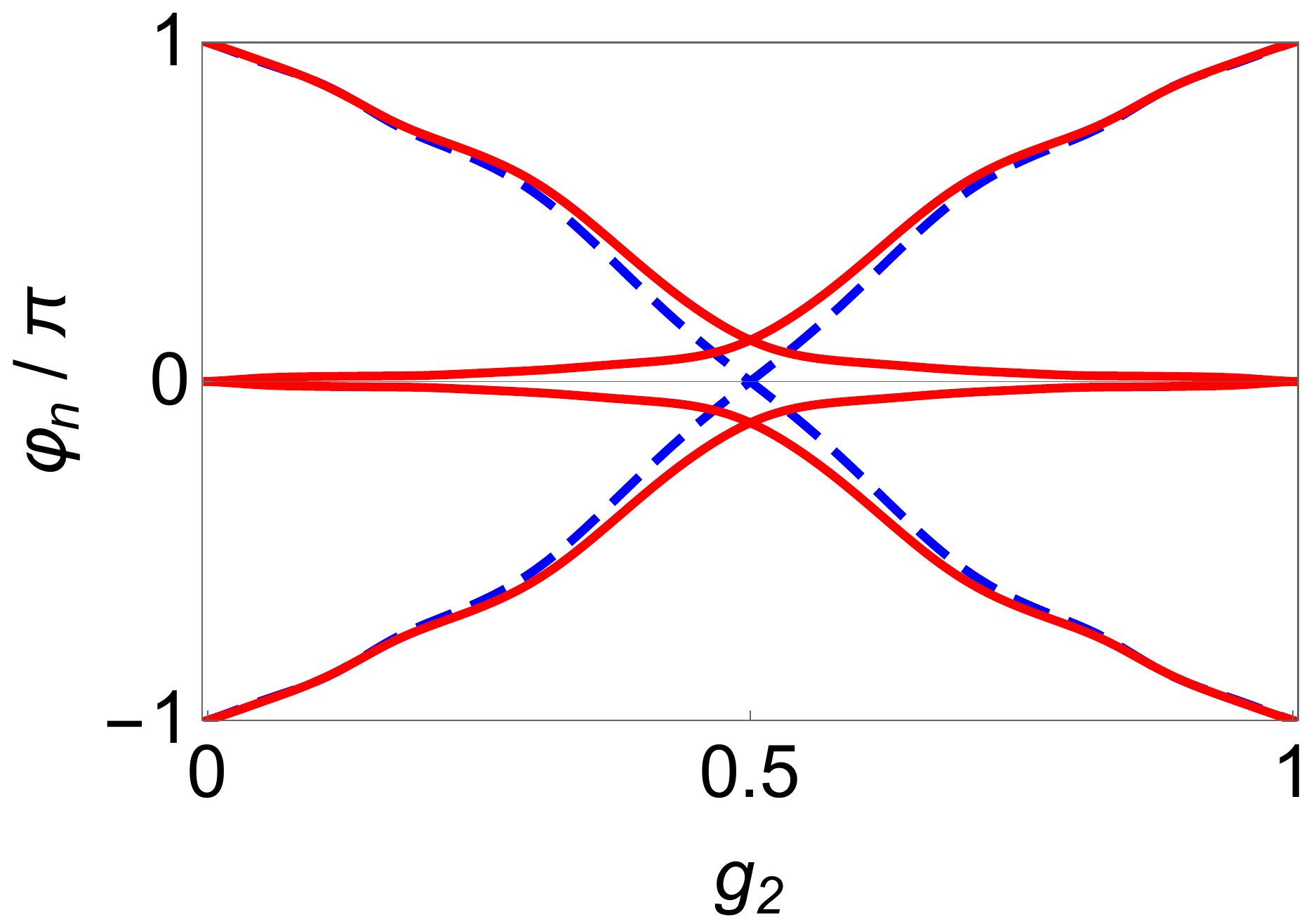} &
	\includegraphics[width=0.25\linewidth]{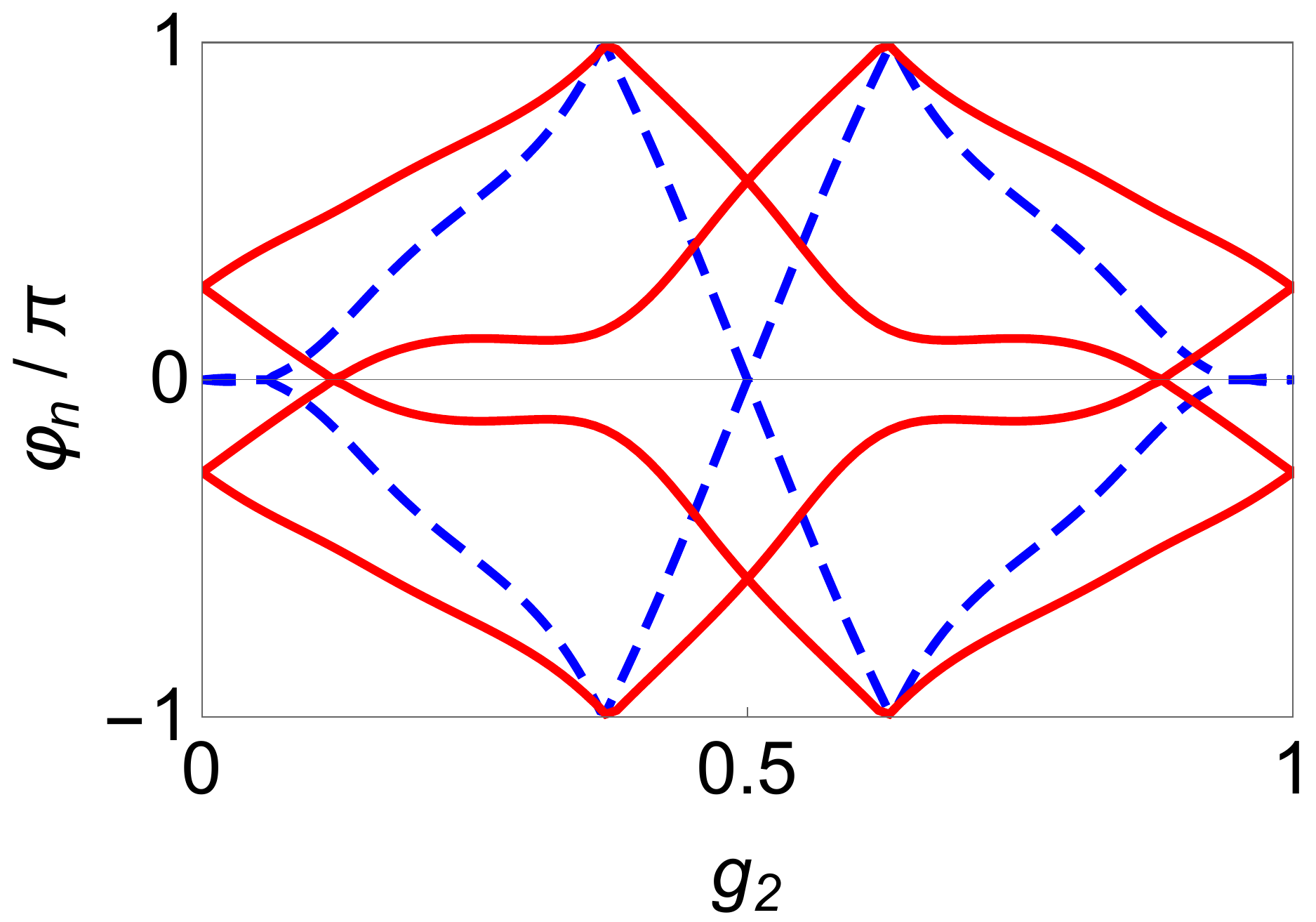} &  
	\includegraphics[width=0.25\linewidth]{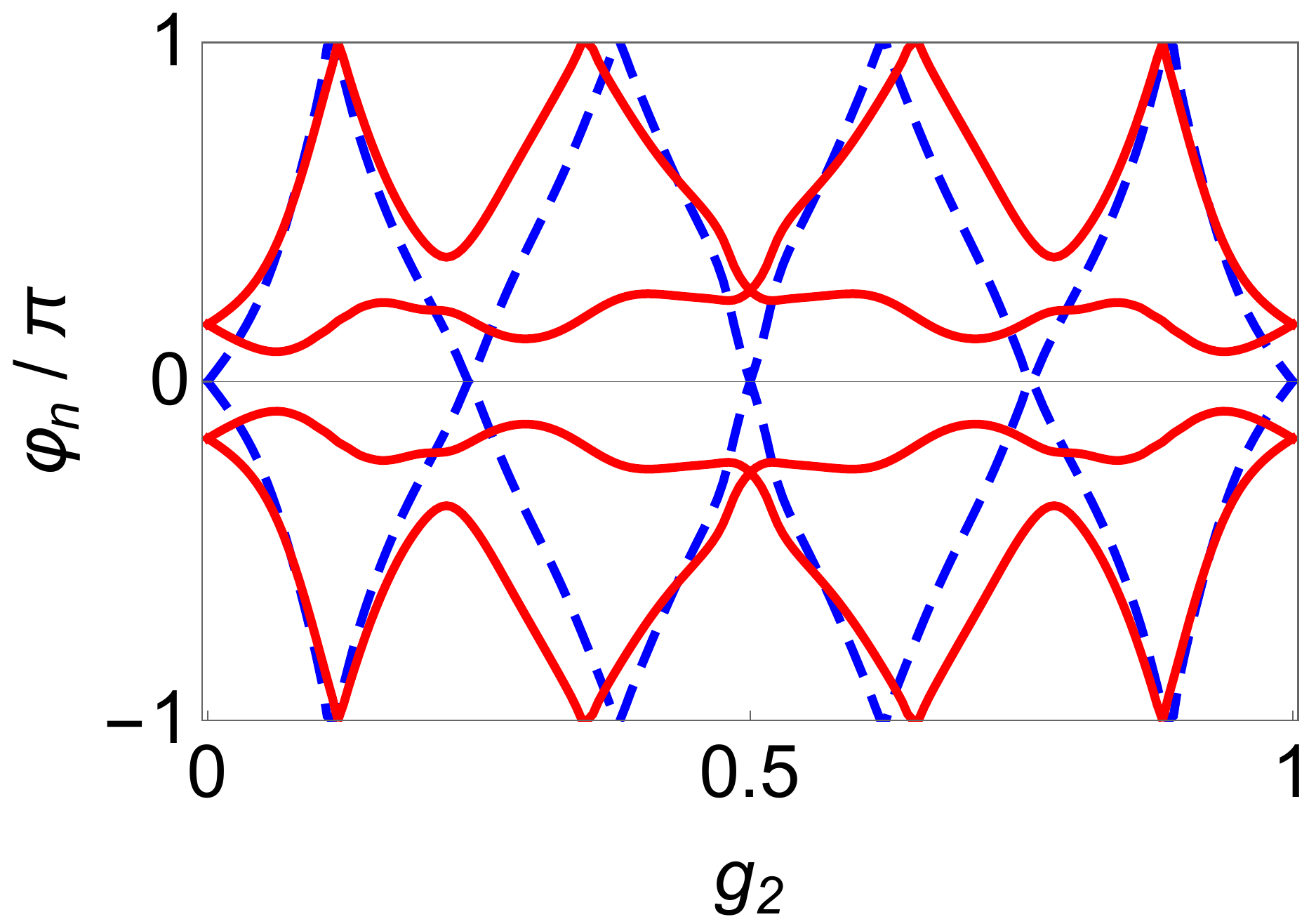} & 
	\includegraphics[width=0.25\linewidth]{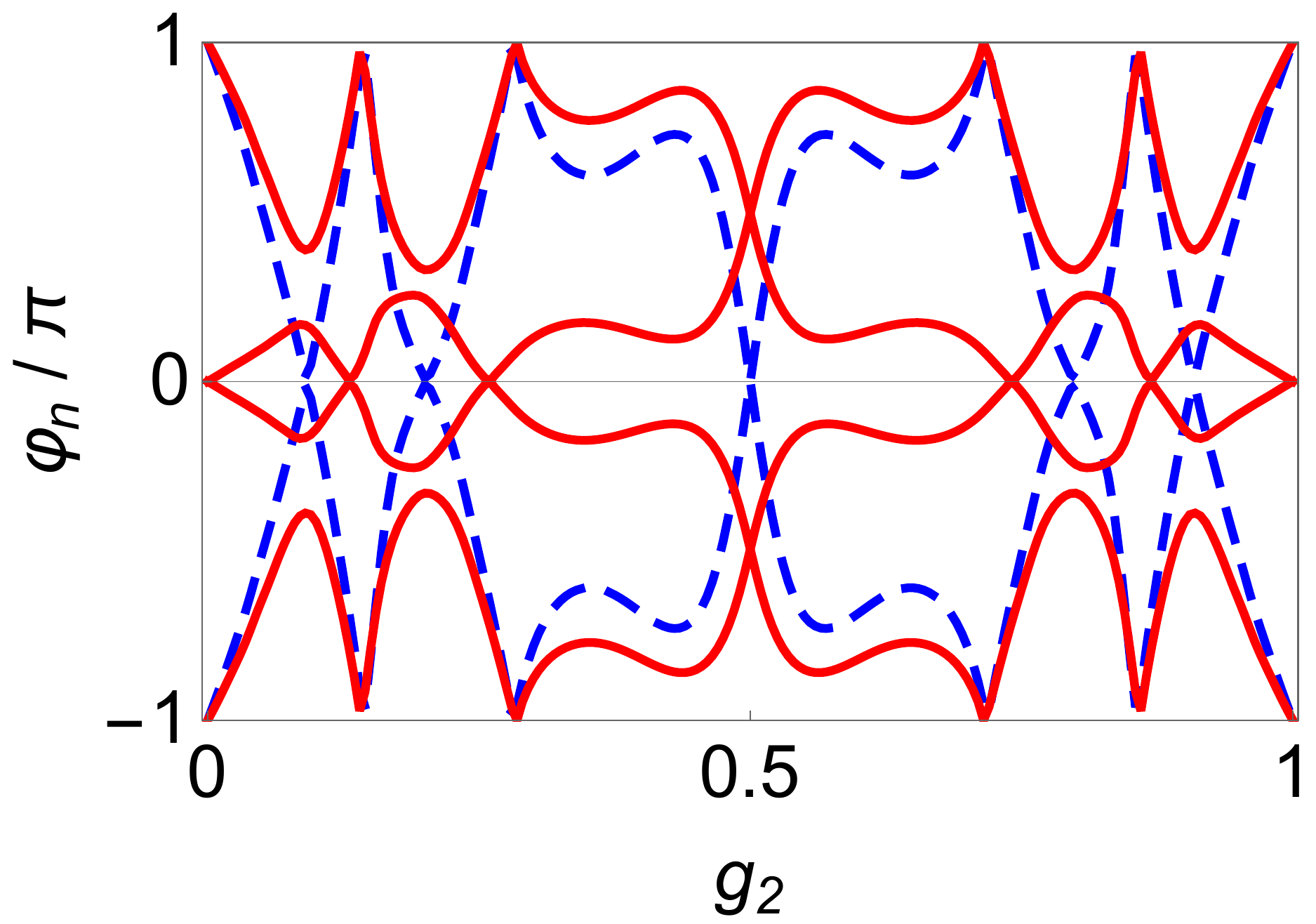} \\
	(a) $W^{\mathrm{B}_1^b}_{II} =1$, $\nu_{\mathrm{FKM}} = 1$, &  
	(b) $W^{\mathrm{B}_1^b}_{II} =2$, $\nu_{\mathrm{FKM}} = 0$, & 
	 (c) $W^{\mathrm{B}_1^b}_{II} = 4$, $\nu_{\mathrm{FKM}} = 0$, & 
	 (d) $W^{\mathrm{B}_1^b}_{II} = 5$,  $\nu_{\mathrm{FKM}} = 1$, \\
	 $\nu_{21'} = 1$ & 
	 $\nu_{21'} = 2$ &
	 $\nu_{21'} = 0$ & 
	$\nu_{21'} = 1$ 
\end{tabular} 
\caption{\label{KM_1_899} Six-band topological phases from the coupling of the split EBR $\mathrm{B}_1+\mathrm{B}_2=\mathrm{B}^b_1(2b)+\mathrm{B}^b_2(2b)$ (Fig.~\ref{adiab_SOC}) with $\mathrm{B}_1'=\mathrm{B}_1^a(1a)$. Each six-band phase is generated from the corresponding four-band phase of Fig.~\ref{adiab_SOC}. First row: band structures where all columns have the same energy ordering of IRREPs as in column (a). Second row: Wilson loop flows over the BZ for the occupied subspace $\mathrm{B}_1^b$ (blue dashed line) and for the unoccupied subspace $\mathrm{B}_1^a+\mathrm{B}_2^b$ (red line). For each case we give the Wilson loop winding ($W_{II}$) and the Fu-Kane-Mele invariant ($\nu_{\mathrm{FKM}}$) of the two-band occupied subspace, as well as the four-band index $\nu_{21'} \in \mathbb{Z}_3$ protected by TRS and $C_{2z}\mathcal{T}$. 
}
\end{figure*}
Having extensively treated the spinless two-band and spinful four-band case, we now study the effect of including one extra sub-lattice site ($C$) located at the center of the unit cell, resulting in a total of six bands. The lattice sites $C$ form a triangular lattice that corresponds to the Wyckoff's position $1a$ of L77.\footnote{We have derived our model using a generalization of Dresselhaus method for expanding global tight-binding models from the crystallographic space group.\cite{Dressel_GT} Details of the method and the model will be given elsewhere. When restricted to the seventh layer of neighbors, our tight-binding model matches with the model given in Ref.~\onlinecite{Ft1}. } Assuming a single orbital per site (e.g.~an $s$-orbital) the two spins on each site $C$ gives rise to the two-band EBR $\mathrm{B}^a_1(1a) = (\overline{\Gamma}_9,\overline{K}_6,\overline{M}_5)$ (we could also form $\mathrm{B}^b_1(1a) = (\overline{\Gamma}_8,\overline{K}_6,\overline{M}_5)$ for a different choice of orbital). In this section we restrict ourselves to the coupling of the split EBR of the honeycomb lattice, i.e.~$\mathrm{B}_{1}+\mathrm{B}_{2}$ with $\mathrm{B}_{1}=\mathrm{B}_{1}^{a/b}(2b)$ and $\mathrm{B}_{2}=\mathrm{B}_{2}^{a/b}(2b)$ (see Fig.~\ref{fig_PWV}), with the EBR for the triangular lattice chosen as $\mathrm{B}_1'=\mathrm{B}_{1}^a(1a)$. Furthermore, we fix the occupied subspace as the two bands of $\mathrm{B}_1$ and the unoccupied subspace as the four bands of $\mathrm{B}_2+\mathrm{B}_1'$. Therefore, depending on the split of the EBR of the honeycomb lattice before the coupling, the qBR of the four-band unoccupied subspace is either compatible or incompatible with the set of IRREPs of a BR$-$that is an EBR or a sum of EBRs$-$of the symmetry class AII+L77. We write $\mathrm{B}\sim \mathrm{BR}~(\mathrm{EBR}) $ when the quasiband representation B has a set of IRREPs that is compatible with a BR (EBR), and $\mathrm{B}\not\sim \mathrm{BR}$ when it is not. In the following examples we have $\mathrm{B}_{2}+\mathrm{B}_{1}'=\mathrm{B}_{2}^{a}(2b)+\mathrm{B}_{1}^a(1a) \sim\mathrm{EBR}$ and $\mathrm{B}_{2}+\mathrm{B}_{1}'=\mathrm{B}_{2}^{b}(2b)+\mathrm{B}_{1}^a(1a) \not\sim \mathrm{BR}$.

\subsection{Numerical results}\label{num_results}

\begin{figure}[b]
\centering
\begin{tabular}{c|c} 
	\includegraphics[width=0.5\linewidth]{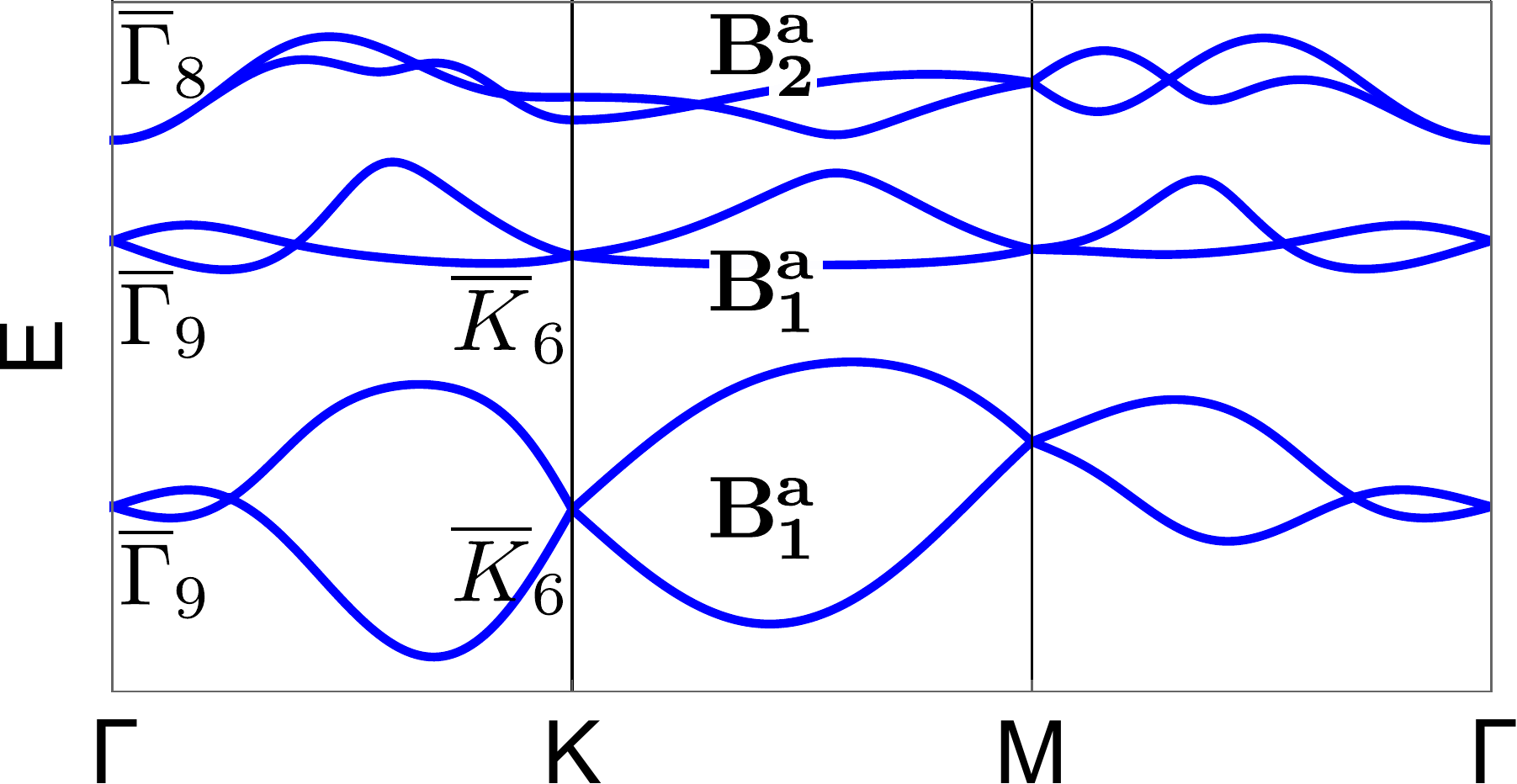} &
	\includegraphics[width=0.5\linewidth]{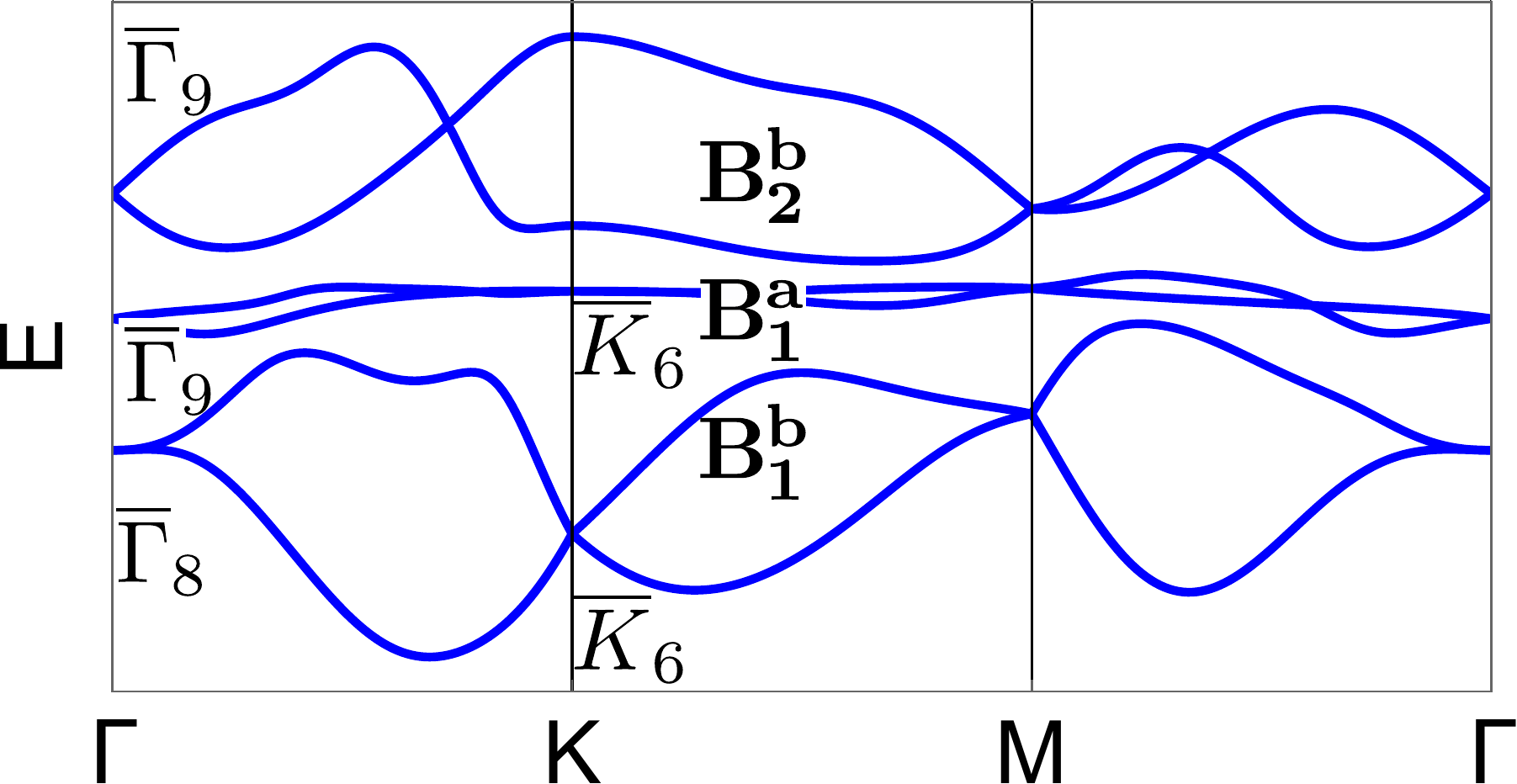} \\
	\includegraphics[width=0.5\linewidth]{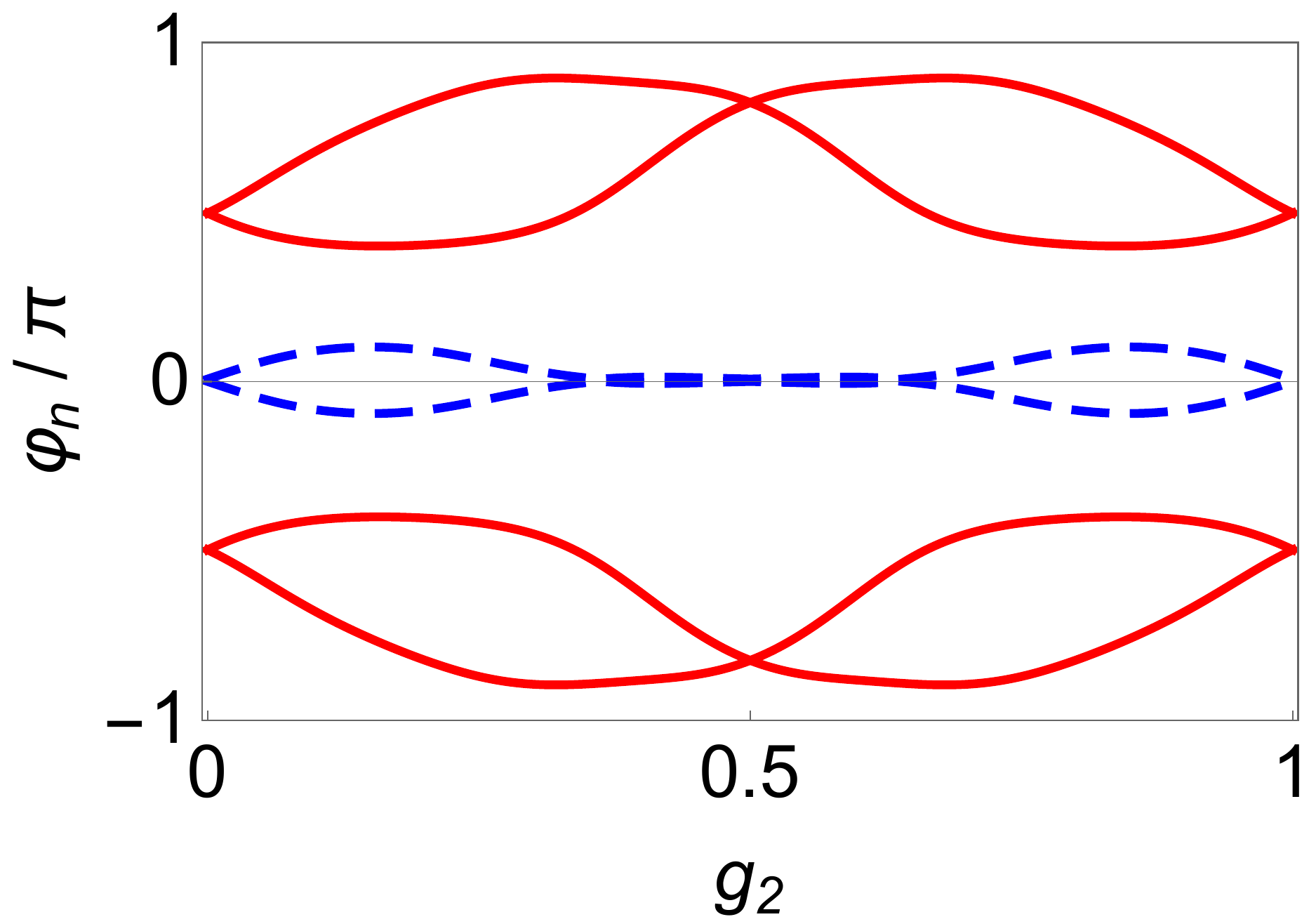} &
	\includegraphics[width=0.5\linewidth]{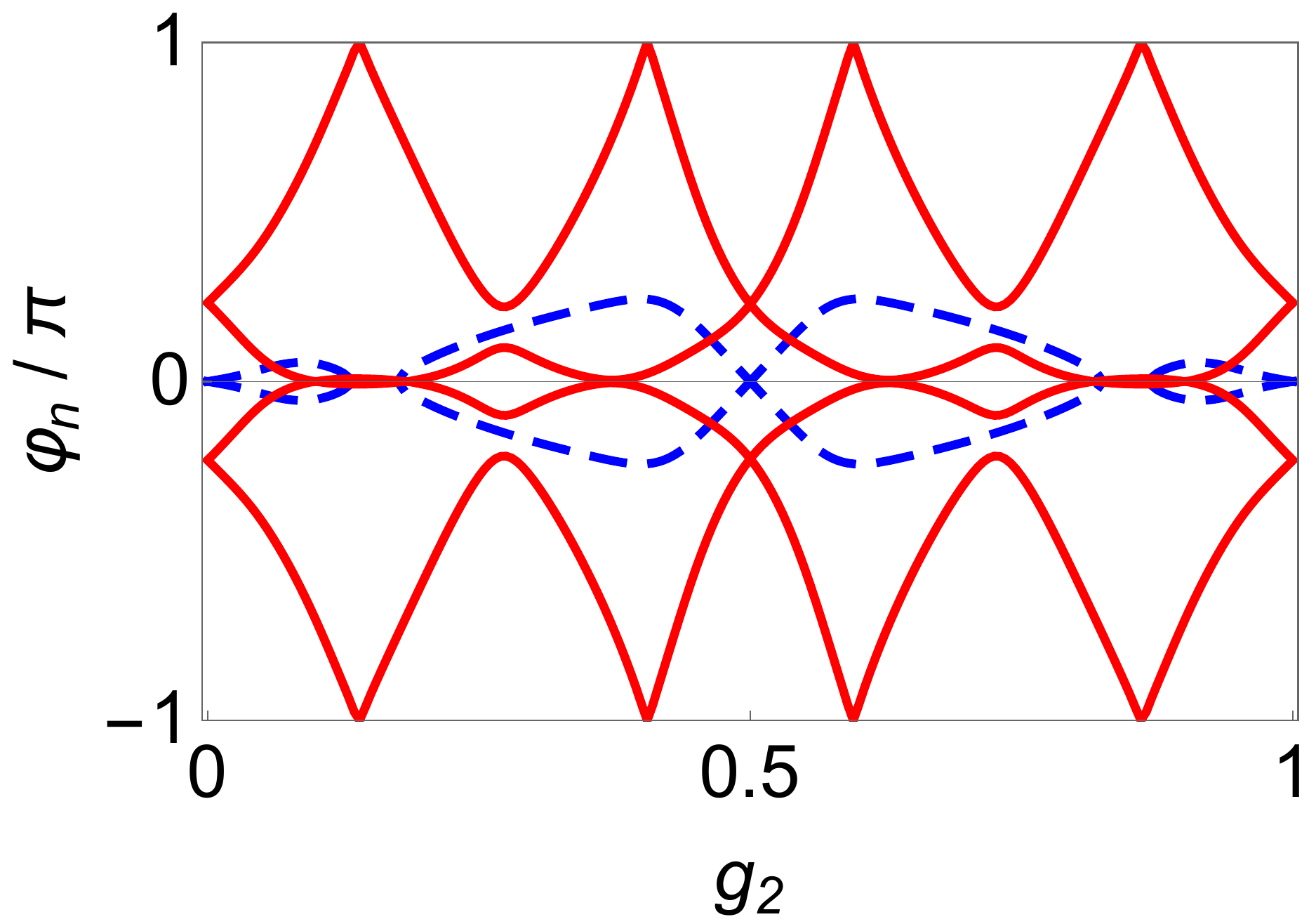} \\
	(a) $W^{\mathrm{B}_1^a}_{II} =0$,  $\nu_{\mathrm{FKM}}=0$, & (b) $W^{\mathrm{B}_1^b}_{II} =0$, $\nu_{\mathrm{FKM}}=0$, \\
	$\nu_{21'} = 0$ & $\nu_{21'} = 0$
\end{tabular}
\caption{\label{fig_PWV_6} Similar to Fig.~\ref{KM_1_899}, but here the six-band topological phases are obtained from the four-band phases of Fig.~\ref{fig_PWV} where inversion symmetry is broken non-adiabatically through Rashba SOC. (a) Split EBR $\mathrm{B}^a_1+\mathrm{B}^a_2$, corresponding to the case discussed in Ref.~\onlinecite{Ft1}. (b) Split EBR $\mathrm{B}^b_1+\mathrm{B}^b_2$.
}
\end{figure}

Figures~\ref{KM_1_899} and Fig.~\ref{fig_PWV_6} show six-band topological insulating phases generated through the coupling of the honeycomb lattice sites (Wyckoff's position $2b$ of L77) with the triangular lattice sites (Wyckoff's position $1a$ of L77). In Fig.~\ref{KM_1_899} and Fig.~\ref{fig_PWV_6}(b) the split EBRs of the honeycomb lattice before coupling were all chosen as $\mathrm{B}_1^b+\mathrm{B}_2^b$. Since the results generated from the other choice of split EBRs, i.e.~$\mathrm{B}_1^a+\mathrm{B}_2^a$, lead to qualitatively identical conclusions we given them in Appendix \ref{num_results_ap} (Fig.~\ref{KM_1_998}). Fig.~\ref{fig_PWV_6}(a) is the analogue to Fig.~\ref{fig_PWV_6}(b) for the split EBR $\mathrm{B}_1^a+\mathrm{B}_2^a$. 

The six-band phases of Fig.~\ref{KM_1_899} are obtained column by column (a-d) from the respective four-band topological phases of Fig.~\ref{adiab_SOC} where inversion symmetry was broken adiabatically. In Fig.~\ref{fig_PWV_6} six-band phases are obtained from the four-band topological phases of Fig.~\ref{fig_PWV} where Rashba SOC had broken inversion symmetry non-adiabatically. The first row in Fig.~\ref{KM_1_899} and Fig.~\ref{fig_PWV_6} has band structures with the same energy ordering of the IRREPs as shown in the first column. The second row in Fig.~\ref{KM_1_899} and Fig.~\ref{fig_PWV_6} gives the flow of Wilsonian phases over the whole BZ for the two-band occupied subspace (blue dashed line) and the four-band unoccupied subspace (red full line). 
In each case we give the Wilson loop winding $W_{II}$ (Eq.~(\ref{windings})) and the Fu-Kane-Mele invariant $\nu_{\mathrm{FKM}}$ of the occupied subspace. 

Analyzing the results we find, on the one hand, when comparing Fig.~\ref{KM_1_899} with Fig.~\ref{adiab_SOC}, and Fig.~\ref{fig_PWV_6} with Fig.~\ref{fig_PWV}, that the topology of the two-band occupied subspace is preserved when two extra bands are included in the unoccupied subspace. On the other hand, the nontrivial Wilson loop winding of the two-band unoccupied subspace is sometimes lost when it is imbedded into the four-band Wilson loop of the six-band system. In particular, we observe that every crossing between Wilsonian branches is avoided away from the phases $0$ and $\pm \pi$ where the degeneracies are protected by $C_{2z}\mathcal{T}$-symmetry, and away from the loops $l_{\Gamma\text{M}}$ ($g_2 = 0,1$) and $l_{\text{M}\text{M}}$ ($g_2=0.5$) where the degeneracies are protected by TRS. In total, we count three distinct Wilson loop patterns of the four-band subspace: (i) each Wilsonian phase winds by $\pm\pi$ over the whole BZ (Fig.~\ref{KM_1_899}(a, d)), (ii) each Wilsonian phase winds by $\pm2\pi$ over the whole BZ (Fig.~\ref{KM_1_899}(b)), (iii) zero winding of each Wilsonian phase (Fig.~\ref{KM_1_899}(c), Fig.~\ref{fig_PWV_6}(a, b)). This suggests a $\nu_{21'}\in\mathbb{Z}_3$ classification of the four-band subspaces protected by TRS and $C_{2z}\mathcal{T}$-symmetry, where the new index $\nu_{21'}$ refers to the magnetic point group $21' = C_{2}\times \{E,\mathcal{T}\}$ that contains TRS and $C_{2z}\mathcal{T}$-symmetry. 

For completeness we show in Fig.~\ref{fig_6by6_BR2_occ}(a) of Appendix \ref{num_results_ap} one example with $\mathrm{B}^a_2$ as the two-band occupied subspace with $W^{\mathrm{B}^a_2}_{II}=4$. The result in this case is analogue to Fig.~\ref{KM_1_899}(c) with a non-zero Wilson loop winding of the occupied subspace and a vanishing Wilson loop winding of the four-band unoccupied subspace. We also show as a reference in Fig.~\ref{fig_6by6_BR2_occ}(b) an example with $\mathrm{B}_1'= \mathrm{B}^a_1(1a)$ ($W^{\mathrm{B}_1'}_{II}=0$) chosen as the occupied subspace. This case has no winding of the Wilson loops. 

Importantly, in all the six-band results of Fig.~\ref{KM_1_899}, Fig.~\ref{fig_PWV_6}, and Fig.~\ref{fig_6by6_BR2_occ}, we have included the two extra trivial bands $\mathrm{B}_1'$ from above in energy such that they had to cross the two bands $B_2$ non-adiabatically, i.e. closing the band gap. However, we have checked that in all the cases the two-band Wilson loop winding of $B_2$ and the zero winding of $B_1'$ are conserved separately whenever there is an energy gap between them. Therefore, while there is no $\mathbb{Z}$-type Wilson loop winding of the four-band subspaces, the Wilson loop winding of two-band subspaces is robust as long as the two bands are separated from the other bands by an energy gap above and below. 

\section{Discussion}\label{discussion}

Based on the six-band numerical results of the previous section we discuss here in detail the topology of two-band and four-band subspaces for the symmetry class AII+L77. We introduce the distinction between stable, unstable, and fragile topologies within vector bundle theory, and relate these to the topology of split EBRs. We then conclude that in the many-band limit the stable Wilson loop winding is determined only by the Fu-Kane-Mele $\mathbb{Z}_2$ invariant implying that further stable topological phases must belong to the class of higher-order topological insulators.\cite{SOT1, SOT2}  For completeness we also briefly discuss the effect of breaking $C_{2z}\mathcal{T}$-symmetry, and TRS, from which Chern insulators with high Chern numbers can be formed.

\subsection{Stable, unstable, and stably trivial or ``fragile" topology} 

The topological classification of band structures, i.e. of vector bundles, can be recast in terms of homotopy groups of the classifying spaces of the Hamiltonian, i.e. the complex Grassmannian $\mathrm{Gr}_m(\mathbb{C}^n)$ where $m$ is the number of occupied bands and $n$ is the total number of bands.\cite{Kitaev,SchnyderClass,DeNittisGomi_1,DeNittisGomi_2} The homotopy groups of a space $X$ generally depends on the dimension of $X$. The classical example is given by the homotopy groups of the $d$-dimensional spheres, i.e. $\pi_{k+d}(\mathbb{S}^d)$. When the dimension $d$ is large enough ($d\geq k+2$) the homotopy groups are independent of $d$, i.e. one says that the homotopy groups are \textit{stable}.\cite{Hatcher_1} On the contrary for small $d$, the homotopy groups strongly depend on $d$ and are called \textit{unstable}. For instance, $\pi_3 (\mathbb{S}^2) = \mathbb{Z}$ (unstable), while $\mathbb{Z}_2= \pi_4(\mathbb{S}^3) = \pi_5(\mathbb{S}^4) \dots$ (stable).\cite{Hatcher_1}

The prototypical example of stable topology in the physics literature are the Chern insulators, i.e. their topology is unaffected by adding a trivial band to the occupied or the unoccupied subspaces, indeed we have $\pi_2(\mathrm{Gr}_1(\mathbb{C}^2)) = \pi_2(\mathrm{Gr}_m(\mathbb{C}^n)) = \mathbb{Z}$. An example of unstable topology are three-dimensional two-band systems, called Hopf insulators, i.e. the classifying space is $\mathrm{Gr}_1(\mathbb{C}^2) \cong \mathbb{S}^2$ and $\pi_3 (\mathbb{S}^2) = \mathbb{Z}$.\cite{Hopf_1,Hopf_2} By adding one band to a Hopf insulator, the dimension of the classifying space is increased and the Hopf topology is lost, as $\pi_3(\mathrm{Gr}_{1,2}(\mathbb{C}^3)) = 0$. Further physical example of unstable topologies are discussed in Ref.~\onlinecite{BzduSigristRobust}. 

The example of the tangent bundle ($T\mathbb{S}^d$) and the normal bundle ($N\mathbb{S}^d$) of a $d$-sphere give yet another possibility. Whenever $d \neq 1,3,7$, the tangent bundle is nontrivial, while the normal bundle is always trivial.\cite{Hatcher_2} It turns out that the ``addition'' (the direct sum) of the tangent and the normal bundles gives a trivial bundle, i.e. $T\mathbb{S}^d\oplus N\mathbb{S}^d \cong \mathbb{S}^d \times \mathbb{R}^{d+1} $. Therefore, a non-trivial bundle ($T\mathbb{S}^d$, $d\neq1,3,7$) can be combined with a trivial one ($N\mathbb{S}^d$) through a direct sum, resulting in a trivial bundle. In this case, one says that the tangent bundle is \textit{stably trivial}, i.e. the (vector) bundle is trivialized by adding one trivial bundle.\cite{Hatcher_2} The above terminology can be directly transferred to the characterization of the topology of band structures, where the direct sum of two vector bundles is interpreted as the grouping of two band subspaces, i.e. by discarding the energy gap between the two set of bands and treating them as a single higher-dimensional subspace. Importantly, we call the topology of a band subspace ``fragile'' when it is stably trivial in the sense of the above example from vector bundle theory.

In the following we distinguish between three kinds of stability. We call it stable topology of an $N$-band occupied subspace when it is robust under the addition of extra trivial bands in the occupied or unoccupied subspace (e.g.~Chern insulators). We call it unstable topology of a system when it is lost under the addition of extra trivial bands either in the occupied or in the unoccupied subspace (e.g.~Hopf insulators). We call it fragile topology of an $N$-band subspace when it is conserved under the inclusion of extra trivial bands separated from the $N$-band subspace by a band gap but lost if the extra bands are added nonadiabatically (i.e. closing the band gap) with respect to the $N$-band subspace. We argue in this section that, based on the numerical results of Section \ref{num_results}, the two-band topology of split EBRs of the honeycomb lattice is fragile for most of the topological sectors identified in the single EBR. In particular, we show that the topology of four-band subspaces of the symmetry class AII+L77($2b+1a$) is independent of whether or not their set of IRREPs is compatible with the set of IRREPs of an EBR or a sum of EBRs of AII+L77($2b+1a$), while two-band subspaces originating from a split EBR conserve their nonzero Wilson winding.

\subsection{Two-band subspace}\label{two_band}

We start with a discussion of two-band subspaces, i.e. two bands separated from all the other bands by an energy gap above and below. Let us first characterize the bands of the triangular lattice alone. Its band structure is made of the single EBR $\mathrm{B}^a_1(1a) = (\overline{\Gamma}_9,\overline{K}_6,\overline{M}_5)$ that cannot be split because of the Kramer's degeneracies at the TRIMPs due to TRS. A basis of smooth Bloch functions that spans the two-band Hilbert space is readily given by $\{\vert \varphi_C, \uparrow , \boldsymbol{k}\rangle ,\vert \varphi_C, \downarrow , \boldsymbol{k}\rangle \}$ which is the Fourier trivialization of the total Bloch bundle.\cite{reference_trivialization} Hence the Berry phase of each component is vanishing over any chosen loop and the two-band Wilson loop has zero winding.\footnote{\unexpanded{It is important to note that $\mathrm{B}^a_1(1a)$ contains simple point nodes at the TRIMPs that are characterized by nontrivial Berry phases. Considering a two-band model for $\mathrm{B}^a_1(1a)$ (i.e.~the triangular lattice sites $C$ with spins) at half-filling, the single occupied band has the Bloch eigenstate $\vert \psi_{occ}, \boldsymbol{k}\rangle \propto -e^{i \theta(\boldsymbol{k})} \vert \varphi_C, \uparrow , \boldsymbol{k}\rangle + \vert \varphi_C, \downarrow , \boldsymbol{k}\rangle$, where $\theta(\boldsymbol{k}) = \sqrt{h_{C\uparrow,C\downarrow}(\boldsymbol{k})/h_{C\uparrow,C\downarrow}(\boldsymbol{k})^*}$ with $h_{C\uparrow,C\downarrow}$ the off-diagonal matrix element of the two-by-two tight-binding Hamiltonian. We readily find that the phase $\theta(\boldsymbol{k})$ imposes a nontrivial Berry phase over any loop that encircles one the simple point nodes at the TRIMPs.}} Similarly, the triviality of any unsplit EBR is obvious as it is for any band structure where the number of occupied bands equals the total number of bands. 

Let us now characterize the two-band occupied subspace of the six-band case introduced in Section \ref{six_band}. First of all, in the case when the occupied subspace is spanned by $\mathrm{B}_2$ we readily find a classification of the Wilson loop windings that matches with Eq.~(\ref{spinful_inversion_homotopy}) as is the case from Eq.~(\ref{BR2}). Indeed, as explained below Eq.~(\ref{BR2}), $\mathrm{B}_2$ has $\tilde{R}^{\text{K}}_{\bar{3}} = -\mathbb{I}$ which readily leads to a symmetry protected quantization of the Wilson loop spectrum\cite{HolAlex_Bloch_Oscillations} over $l_{\Gamma\text{K}}$ given by Eq.~(\ref{BR2}) and leading to the classification of Wilson loop windings of Section \ref{non_adiabatic}, i.e.~$W^{\mathrm{B}_2}_{II} \in \pm\boldsymbol{1}+3\mathbb{Z}$.

When the occupied subspace is spanned by $\mathrm{B}_1$, as chosen in Section \ref{num_results} in Fig.~\ref{KM_1_899} and Fig.~\ref{fig_PWV_6}, we have to generalize the discussion of Section \ref{non_adiabatic} that was based on the spin locking of the doublets at $\Gamma$ and K of a smooth, periodic and rotation-symmetric frame spanning the two-band occupied subspace of the four-band system and over base loops of the BZ. In the six-band case, this approach is straightforwardly generalized by considering the pseudo-spin locking of the doublets at $\Gamma$ and K. Indeed the axis of rotation-symmetries at these HSPs gives a favored quantization axis of the pseudo-spin degrees of freedom, i.e.~only in that spin basis is the matrix representation of rotations diagonal. Actually at $\Gamma$, as a result of the fact that we only consider a single orbital per site, all spin-flip terms still vanish leading to a pure spin decomposition of the doublets $\overline{\Gamma}_{8(9)}$. At K however, the six-band tight-binding model has open spin-flip channels leading to a pseudo-spin structure of the doublets $\overline{K}_6$. For completeness, we give in the Appendix \ref{six_band_ap} the exact expression of the Bloch eigenstates of the six-band model at $\Gamma$ and K.

Assuming the existence of a smooth, periodic, and rotation-symmetric frame of (cell-periodic) Bloch functions that span the two-band occupied subspace $\mathrm{B}_1$ over base loops (see Section \ref{non_adiabatic}), we obtain the same classification as in the four-band model derived in Section \ref{non_adiabatic}. The difference now is that the smooth frame connects the spin components of the doublet  $\overline{\Gamma}_{8(9)}$ at $\Gamma$ to the pseudo-spin components of the doublets $\overline{K}_{6}$ at K, while the dichotomy of $\Gamma$-K spin to pseudo-spin aligned or $\Gamma$-K spin to pseudo-spin flipped still holds. Again using the parallel transport method of Ref.~\onlinecite{Vanderbilt_smooth_gauge}, we have verified that such a smooth and rotation-symmetric frame always exists for any two-band subspace of the six-band model. We then conclude that the symmetry protected quantization of two-band Wilson loop spectrum over $l_{\Gamma\text{K}}$ derived in Section \ref{non_adiabatic} is preserved in the six-band case and thus leads to the same classification of the occupied subspace $\mathrm{B}_1$, i.e.~$W^{\mathrm{B}_1}_{II} \in \left\{ \pm\boldsymbol{1}+3\mathbb{Z},\boldsymbol{0}\right\}$. 

At this point it is reasonable to postulate the existence of a smooth, periodic and rotation-symmetric frame for every two-band occupied subspace of the symmetry class AII+L77 and over base loops of the BZ, thus leading to the same and unique classification independently of the number of unoccupied bands. The only difference in the most general case is that spin-flip terms can also be non-zero at $\Gamma$ (i.e.~if each lattice site is the host of multiple electronic orbitals with opposite parities) thus leading to a pseudo-spin structure of the Bloch eigenstates at $\Gamma$ as well. This however does not change the dichotomy between the phases with a pseudo-spin aligned between $\Gamma$ and K and the phases where the pseudo-spin is flipped between $\Gamma$ and K. The same result applies for any two-band subspace that originates from a split EBR as long as it remains separated from the other bands by an energy gap above and below. We leave the formal proof of the existence of a smooth, periodic and rotation-symmetric frame of any two-band subspace over a base loop for the future. 

We conclude that the obstructed $\mathbb{Z}$-type topological classification, i.e.~$W_{II}\in\left\{ \pm\boldsymbol{1}+3\mathbb{Z},\boldsymbol{0}\right\}$, of any two-band subspace of the symmetry class AII+L77 is stable under the addition of arbitrary many bands in the complement band  subspace and as long as a band gap around the two-band subspace is preserved. This is schematically represented in Fig.~\ref{EBR_stability} showing that two-band occupied subspaces, with zero Wilson loop winding marked in white and non-zero Wilson loop windings marked in green, have their topology unchanged after the coupling with extra bands separated by a band gap. This is true for all six-band models computed in this work and shown in Fig.~\ref{KM_1_899} and Fig.~\ref{fig_PWV_6} in Section \ref{num_results} and Fig.~\ref{KM_1_998} in Appendix \ref{num_results_ap}.

\subsection{Four-band subspace}

We now turn to the topology of the four-band subspaces in the six-band case. Starting from a split EBR of the honeycomb lattice $\mathrm{B}_1+\mathrm{B}_2$ with a known topology, we couple it with the trivial EBR of a triangular lattice $\mathrm{B}_1'$ and characterize the topology of the four-band unoccupied subspace $\mathrm{B}_{2/1}+\mathrm{B}_1'$ where we either choose $\mathrm{B}_{1}$ or $\mathrm{B}_{2}$ as the occupied subspace. Depending on the way the EBR of the honeycomb lattice splits, the four-band unoccupied subspace has a set of IRREPs that either is compatible with an BR (i.e.~an EBR or a sum of EBRs) of AII+L77, or it is not. The combinatorial possibilities for the four-band subspaces of the six-band case are 
\begin{equation}
\begin{aligned}
	\mathrm{B}^{a}_{2}(2b)+\mathrm{B}^a_1(1a) &= (\overline{\Gamma}_9+\overline{\Gamma}_8,\overline{K}_4+\overline{K}_5+\overline{K}_6,2\overline{M}_5) \\
	&\sim \mathrm{EBR} = \mathrm{B}^{a(b)}_{1}(2b)+\mathrm{B}^{a(b)}_2(2b)	\;,\\
	\mathrm{B}^{b}_{2}(2b)+\mathrm{B}^a_1(1a) &= (2\overline{\Gamma}_9,\overline{K}_4+\overline{K}_5+\overline{K}_6,2\overline{M}_5)\\
	& \not\sim \mathrm{BR}	\;,\\
	\mathrm{B}^{a}_{1}(2b) +\mathrm{B}^a_1(1a) &= (2\overline{\Gamma}_9,2\overline{K}_6,2\overline{M}_5) \\
	&\sim  \mathrm{BR}=\mathrm{B}^a_1(1a)+\mathrm{B}^a_1(1a) 	\;,\\
	\mathrm{B}^{b}_{1}(2b) +\mathrm{B}^a_1(1a) &= (\overline{\Gamma}_9+\overline{\Gamma}_8,2\overline{K}_6,2\overline{M}_5) \\
	&\sim \mathrm{BR} =  \mathrm{B}^a_1(1a)+\mathrm{B}^b_1(1a)	\;.
\end{aligned}
\end{equation}
The conclusions of this section are the same if we choose the extra bands $\mathrm{B}_1' $ as $\mathrm{B}^{b}_{1}(1a) =  (\overline{\Gamma}_8,\overline{K}_6,\overline{M}_5)$ instead of $\mathrm{B}^{a}_{1}(1a) =  (\overline{\Gamma}_9,\overline{K}_6,\overline{M}_5)$.

We then ask how the topology of the four-band subspace, given in terms of the winding of the four-band Wilson loop over the BZ, is related to the known topology of split EBRs (Section \ref{non_adiabatic}). From the numerical results of Section \ref{num_results} in Fig.~\ref{KM_1_899} and Fig.~\ref{fig_PWV_6}, and in Fig.~\ref{KM_1_998} of Appendix \ref{num_results_ap}) we conclude that the two-band topology is in general fragile when imbedded into a four-band subspace irrespectively of whether or not the set of IRREPs (of the four-band subspace) is compatible with a BR. 

\begin{figure}[t!] 
\centering
\begin{tabular}{c} 
	\includegraphics[width=1.\linewidth]{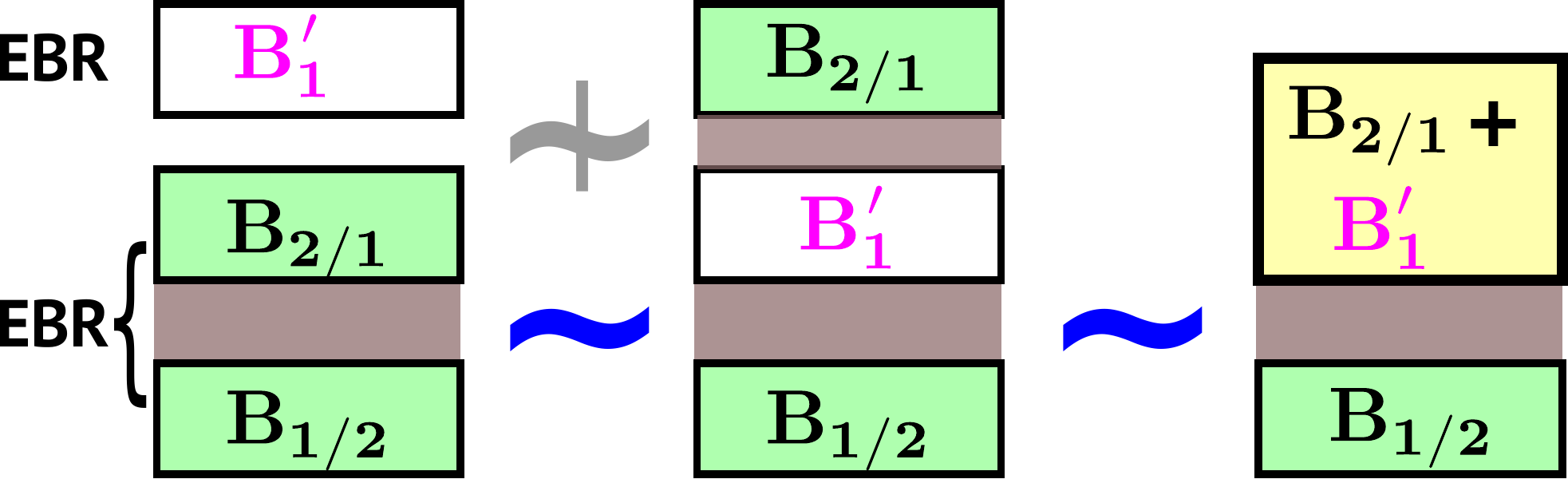} \\
	(a) \\
	\includegraphics[width=1.\linewidth]{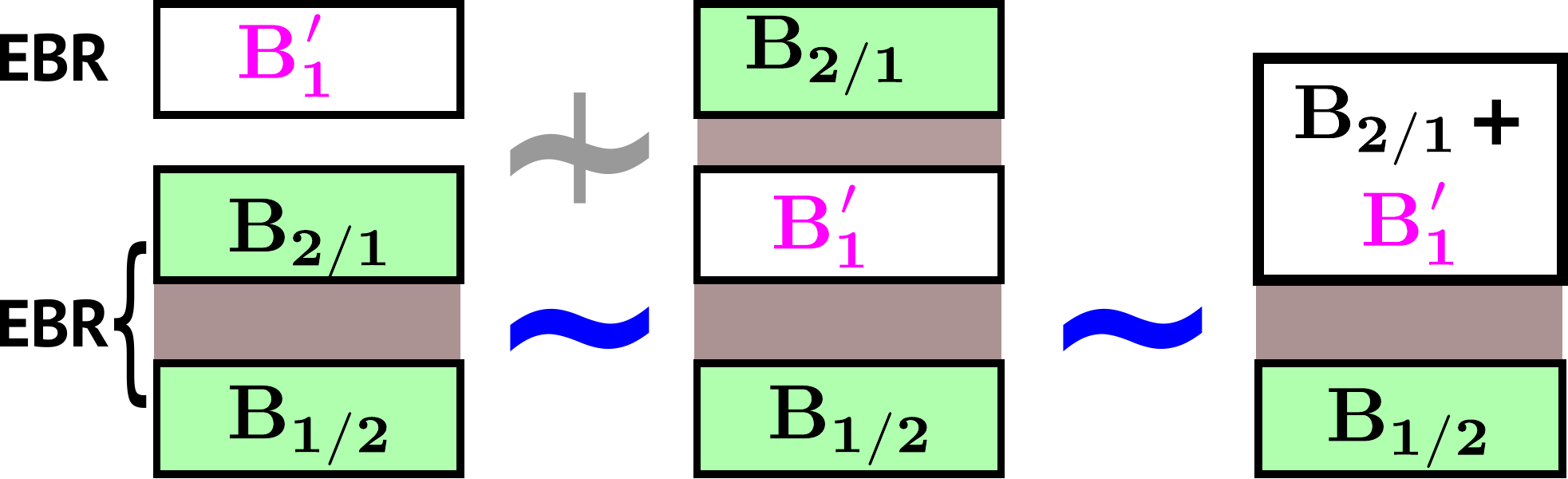} \\
	(b) \\
	 \includegraphics[width=1.\linewidth]{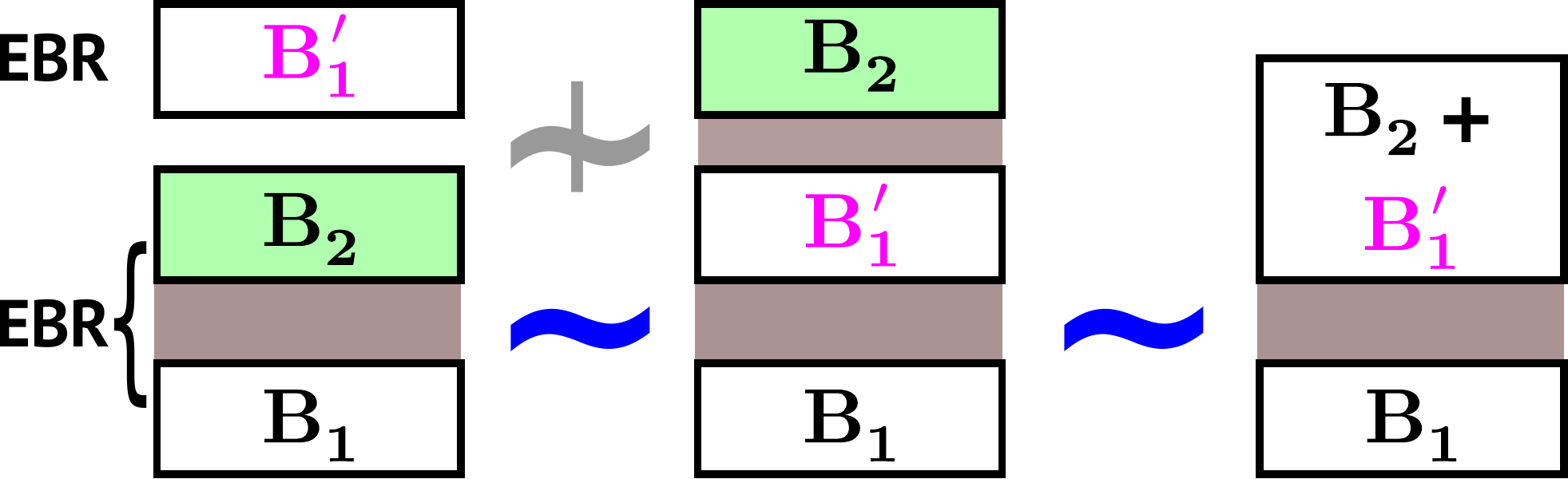} \\
	 (c)
\end{tabular}
\caption{\label{EBR_stability} Fragile topology of the split EBR of the honeycomb lattice $\mathrm{B}_1+\mathrm{B}_2$, with $\mathrm{B}_1=\mathrm{B}_1^{a/b}(2b)$ and $\mathrm{B}_2=\mathrm{B}_2^{a/b}(2b)$, under the coupling with the trivial EBR of the triangular lattice, $\mathrm{B}_1'= \mathrm{B}_1^{a/b}(1a)$. Two-band (four-band) subspaces with non-zero Wilson loop winding are colored in green (yellow), otherwise in white. In (a) and (b) we choose either $\mathrm{B}_1$ or $\mathrm{B}_2$ as the two-band occupied subspace of the six-band insulating phases, in (c) it can only be $\mathrm{B}_1$. The band gaps are drawn in brown. The blue $\boldsymbol{\sim}$ marks that the coupling is adiabatic, i.e.~without closing the band gap, while the barred gray marks a non-adiabatic coupling (i.e. closing the band gap). (a) reveals a stable topology of the split EBR (i.e. here for a nonzero Fu-Kane-Mele invariant).  (b) and (c) reveal a fragile topology of the split EBR (i.e. nonzero two-band Wilson loop winding but with a zero Fu-Kane-Mele invariant). (c) corresponds to the example given in Ref. \onlinecite{Ft1}, and here in Fig.~\ref{fig_PWV_6}.
}
\end{figure}

For clarity, let us summarize our findings in Fig.~\ref{EBR_stability}, where non-trivial two-band (four-band) Wilson loop windings are colored in green (yellow). In the left column, we show the schematic EBRs and their topologies before the coupling. In the middle column, we show the topology of the coupled EBRs, after that the EBR B$_1'$ has crossed from above and non-adiabatically\footnote{Any inversion in energy of two qBRs with distinct IRREPs must close the energy gap between them.} the upper half of the split EBR B$_1+$B$_2$. In the right column we show the topology when there is no gap between B$_1'$ and the upper half of B$_1+$B$_2$, hence characterizing the intrinsic four-band topology of the unoccupied subspace. 

Two possibilities are encountered after coupling a split EBR with an extra EBR, as shown in Fig.~\ref{EBR_stability}: (a) both occupied and unoccupied subspaces have stable topology, (b) and (c) both occupied and unoccupied subspaces have fragile topology (where (c) corresponds to the special case found by Ref.~\onlinecite{Ft1}). Only for special values of the two-band Wilson loop winding before the coupling do we find a topological four-band unoccupied subspace, e.g.~Fig.~\ref{KM_1_899}(a,b,d) and Fig.~\ref{KM_1_998}(a,b,d). Therefore, the obstructed $\mathbb{Z}$-type classification of two-band Wilson loop winding is lost in general in four-band Wilson loops. We hence conclude that the topology of split EBRs is generically fragile when imbedded in many-band structures.  

We further show that even when the unoccupied subspace has a set of IRREPs incompatible with EBRs its topology can be fragile with a vanishing four-band Wilson loop winding. Let us first consider for instance the four-band unoccupied subspace of Fig.~\ref{KM_1_899}(c) that has a zero Wilson loop winding. Indeed, the crossings at $0$ have been lifted and the crossings at $\pm\pi$ can be removed two-by-two through an adiabatic deformation of the Wilson loop branches. This case corresponds to Fig.~\ref{EBR_stability}(b) for the initial split EBR $\mathrm{B}^b_1+\mathrm{B}^b_2$. Similarly, the four-band unoccupied subspace of Fig.~\ref{fig_PWV_6}(b) has a zero Wilson loop winding (here the crossings at $0$ and at $\pm \pi$ can both be removed two-by-two through an adiabatic deformation of the Wilson loop branches). This case corresponds to Fig.~\ref{EBR_stability}(c) for the same initial split EBR. Reversely, we also show that when the unoccupied subspace has a set of IRREPs that is compatible with EBRs its topology can be nontrivial, e.g.~Fig.~\ref{KM_1_998}(a,d,b) which have a non-zero Wilson loop winding. These cases correspond to Fig.~\ref{EBR_stability}(a) for the initial split EBR $\mathrm{B}^a_1+\mathrm{B}^a_2$. These results greatly generalizes the recent observation of a ``fragile topology'' in Ref.~\onlinecite{Ft1}.

We have checked numerically that the four-band Wilson loop spectra in the six-band case are not quantized over $l_{\Gamma\text{K}}$ contrary to the two-band Wilson loop spectra. Furthermore, we have also checked numerically using the Soluyanov-Vanderbilt's smooth gauge construction\cite{Vanderbilt_smooth_gauge} that the four-band subspaces do not support a smooth and periodic frame that, at the same time, also satisfies the rotation-symmetric gauge (making rotation matrix representations diagonal). Therefore, the Wilson loop spectrum of four-band subspaces are generically not quantized by symmetry and we do not expect the $\mathbb{Z}$-type Wilson loop winding of the two-band subspaces to be stable. This confirms the homotopy classification of Wilson loop flows of Section \ref{WLW}. We give a more detailed discussion in the Appendix \ref{four_band_smooth_gauge}. 

We conclude with the characterization of the preserved four-band topologies. We observed in Section \ref{num_results} that the crossings of Wilsonian branches that remain are those at $0$ and $\pm \pi$, as they are protected by $C_{2z}\mathcal{T}$-symmetry, and the Kramer's degeneracies over $l_{\Gamma\text{M}}$ ($g_2=0,1$) and $l_{\text{M}\text{M}}$ ($g_2=0.5$), as they are protected by TRS. This readily leads to a $\nu_{21'}\in\{0,1,2\} \cong \mathbb{Z}_3$ classification of the four-band subspaces with the index $\nu_{21'}$ being protected by the magnetic point group $21' = C_{2}\times \{E,\mathcal{T}\}$ that contains TRS and $C_{2z}\mathcal{T}$-symmetry. In particular, we find that the two-band Wilson loop winding before the coupling directly determines the four-band index $\nu_{21'}$ according to
\begin{equation}
\begin{aligned}
	W^{\mathrm{B}_{2/1}}_{II}~\mathrm{mod}~2=1
	&\Rightarrow \nu_{21'} = 1	\;,\\
	W^{\mathrm{B}_{2/1}}_{II}~\mathrm{mod}~2=0 ~\&~ W^{\mathrm{B}_{2/1}}_{II}~\mathrm{mod}~4=0 &\Rightarrow  \nu_{21'} = 0 \;,\\
	W^{\mathrm{B}_{2/1}}_{II}~\mathrm{mod}~2=0 ~\&~ W^{\mathrm{B}_{2/1}}_{II}~\mathrm{mod}~4=2 &\Rightarrow  \nu_{21'} = 2 \;.
\end{aligned}
\end{equation}
The Fu-Kane-Mele $\mathbb{Z}_2$ invariant is then readily given through $\nu_{\mathrm{FKM}} = \nu_{21'} \mod 2$. 

\subsection{Towards the many-band limit}

If an extra trivial two-band subspace $\mathrm{B}^{a/b}_1$ is added to the occupied subspace, the topological features of the four-band occupied subspace is similar to the four-band unoccupied subspace described above. If instead the extra trivial two-band subspace is added to the four-band unoccupied subspace, only two types of Wilson loop flow are allowed for the six-band subspace, namely either a Wilson loop flow with zero winding (gapped Wilson bands) or with a complete winding (no gap in the Wilson bands). Hence $\nu_{21'}\in\mathbb{Z}_3$ should be understood again as a fragile topology. We also conclude that only a nontrivial Fu-Kane-Mele $\mathbb{Z}_2$ invariant guarantees a complete winding of the Wilson loop in the many-band limit. As a consequence, the further stable topological phases that have been predicted heuristically in Ref.~\onlinecite{2017arXiv171104769K} must be characterized by topological invariants that are not directly related to the Wilson loop winding and fall in the class of the higher-order topological insulators.\cite{SOT1, SOT2}  This falls beyond the scope of the present work and it will be explored in detail elsewhere.

\subsection{Breaking $C_{2z}\mathcal{T}$-symmetry, and TRS}

It is now straightforward to address the breaking of the anti-unitary symmetries. When $C_{2z}\mathcal{T}$-symmetry is broken while conserving TRS the degeneracies of the Wilson loop at $0$ and $\pm \pi$ are lifted and only two types of Wilson loop patterns survive which correspond to the two values of the Fu-Kane-Mele $\mathbb{Z}_2$ invariant. 
When TRS is broken adiabatically the Kramer's degeneracies of the Wilson loop over $l_{\Gamma\text{M}}$ and $l_{\text{M}\text{M}}$ are also lifted and there is no winding of the Wilson loop left. This can be easily understood from the fact that any two-band or four-band subspaces can be decomposed into pair(s) of Chern bands with opposite Chern numbers, \cite{FuKaneZ2} which cancel each other if the number of occupied bands is unchanged when breaking TRS. 
If TRS symmetry is broken non-adiabatically non-zero Chern numbers can be generated. In particular, let us start with a two-band occupied subspace with a high winding of its Wilson loop. Then, by breaking TRS and keeping a single occupied band, we are guaranteed to generate a Chern insulator with a correspondingly high Chern number. Therefore, the two-band Wilson loop winding directly determines the one-band Chern number under the breaking of TRS.

\section{Concluding remarks}

In this work we have studied the relation between Wilson loop flows and symmetry protected topology of band structures. Apart from elucidating nodal points by generalizing Berry curvature arguments to include space group symmetries, thereby outlining a unifying and comprehensive approach, we in particular utilize this procedure to characterize time reversal symmetric crystalline topological insulating phases. Namely, the presented framework can be employed to quantify and elucidate topological invariants that are stable with regard to the addition of extra bands, usually described in terms of K-theories, as well as finer characterizations of band topology coined fragile topology. Our setup allows us to naturally lift spinless notions to spinful ones, extending both homotopy and K-theory. Amongst other things, we find that there exists an obstructed $\mathbb{Z}$-type classification protected by crystalline symmetries and TRS beyond the standard Fu-Kane-Mele $\mathbb{Z}_2$ classification. We have shown that this accounts for all nontrivial Wilson loop windings of a split elementary band representation (EBR). More importantly, this viewpoint can then be adopted to evaluate the proposed scheme of topological quantum chemistry\cite{Clas5,EBR_1,EBR_2} (TQC) by examining the EBR content. In particular, we have shown that while Wilson loop winding of split EBRs can unwind when embedded in higher-dimensional band spaces, two-band subspaces that remain separated by a band gap from the other bands conserve their Wilson loop winding, thus pointing to the fragile topology of split EBRs. This unifying perspective finds natural use in analyzing the origin of the fragile topology found in Ref.~\onlinecite{Ft1}. We then conclude with the observation that the stable Wilson loop winding, i.e. in the many-band limit, is only determined by the Fu-Kane-Mele $\mathbb{Z}_2$ invariant implying that further topological phases predicted in Ref.~\onlinecite{2017arXiv171104769K} must belong to the class of higher-order topological insulators (i.e. characterized through the nested Wilson loop).\cite{SOT1, SOT2} 

On the most general level, our work thus sets the basis of a systematic classification of topological crystalline phases with TRS by rigorously positioning three main classes; 1) unstable topologies that are only defined in few-band models and are rooted in homotopy classes of finite-dimensional Grassmannians, 2) stable topologies that persist upon the addition of an arbitrary amount of trivial bands and are described by K-theories, 3) fragile topological phases bridging the two former classes and described within the formalism of vector bundle theory. It is expected that fragile topology leads to phase sensitive observables\cite{HolAlex_Bloch_Oscillations,EBR_2,Barry_fragile} (e.g. Berry phase measurements) as opposed to the thermodynamical manifestations of stable topology (e.g. robust edge and surface states). Furthermore, our work has important implications for the characterization of obstructed atomic limits\cite{EBR_1} which will be discussed elsewhere. The general applicability and effectiveness of our work is further underlined by potential impact on yet other recently discovered phases. Indeed, due to the similarity of co-dimension 1 and 2 defects,\cite{Codefects1, Codefects2, UnifiedBBc} we anticipate that this approach can shed new light on the recently proposed higher order topological insulators \cite{SOT1, SOT2}. In fact, we already posses some insights on this subject, which we will report elsewhere. The above results should therefore be considered in the general context as an effective illustration of the versatility of the proposed framework.

\textit{Note added in proof}. Recently, we became aware of Ref.~\onlinecite{AlexHol_nogo} that rigorously proves that for any one-band ground state with symmorphic magnetic space group G the vanishing of the first Chern class is equivalent to being a band representation. This is done through the analytical construction of a smooth, periodic and symmetric one-band section of the Bloch bundle. In our work we combine homotopy arguments and the numerical construction of multi-band smooth, periodic and rotation-symmetric Bloch frames (sections) with TRS in order to characterize the topology of a split EBR in class AII leading to the complete classification of two-band subspaces in the layer groups L80 and L77. We furthermore discuss the effect of adding EBRs thus revealing the fragile stability of the topology of split EBRs.

We also became aware of Ref.~\onlinecite{Barry_fragile} that appeared three months after our work and has some overlap with our work.

\section*{Acknowledgements}

We are grateful to Andrei Bernevig for stimulating discussions, and Barry Bradlyn for bringing our attention to their Ref.~\onlinecite{Barry_fragile}, Aris Alexandradinata and Judith H\"oller for insightful comments about their works Ref.~\onlinecite{HolAlex_Bloch_Oscillations,AlexHol_nogo}, and Tom\'{a}\v{s} Bzdu{\v s}ek for a critical reading of an early version of the manuscript. AB and ABS acknowledge financial support from the Swedish Research Council (Vetenskapsr\aa det) Grant No.~621-2014-3721 and the Knut and Alice Wallenberg Foundation.

\appendix

\section{Two-band spinless topology}

\subsection{Tight-binding Hamiltonian and symmetries}\label{spinless_ap}

We give here the generic structure of the spinless tight-binding Hamiltonian based on the layer group L77 with the point group $C_{6v}$. The Bloch basis is built from a complete set of localized Wannier functions through 
\begin{equation}
\label{Bloch_basis}
\begin{aligned}
	\vert \varphi_A , \boldsymbol{k} \rangle &=  \dfrac{1}{\sqrt{N}} \sum_{\boldsymbol{R}_{\nu} } e^{i \boldsymbol{k}\cdot (\boldsymbol{R}_{\nu}+\boldsymbol{r}_A)} \vert w_A , \boldsymbol{R}_{\nu} \rangle \;,\\
	\vert \varphi_B , \boldsymbol{k} \rangle &=  \dfrac{1}{\sqrt{N}} \sum_{\boldsymbol{R}_{\nu} } e^{i \boldsymbol{k}\cdot (\boldsymbol{R}_{\nu}+\boldsymbol{r}_B)} \vert w_B , \boldsymbol{R}_{\nu} \rangle \;,
\end{aligned}
\end{equation}
from which we obtain the tight-binding Hamiltonian in the sub-lattice site basis, 
\begin{equation}
\label{spinless_site}
	\mathcal{H} = \sum_{\boldsymbol{k}} \left(\begin{array}{c} \vert \varphi_A , \boldsymbol{k} \rangle \\
	\vert \varphi_B , \boldsymbol{k} \rangle\end{array} \right)^T 
	\left( \begin{array}{cc} h_{AA}(\boldsymbol{k}) & h_{AB}(\boldsymbol{k}) \\
	h_{BA}(\boldsymbol{k}) & h_{BB}(\boldsymbol{k}) \end{array}\right)
	\left(\begin{array}{c}   \langle \varphi_A , \boldsymbol{k} \vert \\
	 \langle \varphi_B , \boldsymbol{k} \vert \end{array}\right)\;.
\end{equation}
We define the symmetry-Bloch basis\cite{Wi2,BBS_nodal_lines} through
\begin{align}
	\left(\vert \phi_{\Gamma_1} , \boldsymbol{k} \rangle~\vert \phi_{\Gamma_4} , \boldsymbol{k} \rangle\right)  = 
	\left(\vert \varphi_A , \boldsymbol{k} \rangle ~ \vert \varphi_B , \boldsymbol{k} \rangle \right) \hat{U}_S \;,
\end{align}
with $\hat{U}_S = \dfrac{1}{\sqrt{2}} \left(\begin{array}{cc} 1& 1\\ 1& -1 \end{array}\right)$. We then have for the Hamiltonian written in the Bloch-symmetry basis
\begin{align}
	\mathcal{H} &= \sum_{\boldsymbol{k}} \vert \boldsymbol{\phi} , \boldsymbol{k} \rangle
	H_S(\boldsymbol{k})
	 \langle \boldsymbol{\phi} , \boldsymbol{k} \vert  \;,\nonumber\\
	H_S(\boldsymbol{k}) &= 
	\left( \begin{array}{cc} h^{(\Gamma_1)}_{11}(\boldsymbol{k}) & h^{(\Gamma_4)}_{14}(\boldsymbol{k}) \\
	h^{(\Gamma_4)}_{41}(\boldsymbol{k}) & h^{(\Gamma_1)}_{44}(\boldsymbol{k}) \end{array}\right) \;,\\
	 \vert \boldsymbol{\phi} , \boldsymbol{k} \rangle &= \left( \vert \phi_{\Gamma_1} , \boldsymbol{k} \rangle ~\vert \phi_{\Gamma_4} , \boldsymbol{k} \rangle\right) \;,\nonumber
\end{align}
where the upper-script of matrix element $h^{(\Gamma_j)}_{\mu\nu}(\boldsymbol{k})$ marks that it behaves as a basis function of the IRREP $\Gamma_j$ of $C_{6v}$.

Real space crystal symmetries act on the symmetry-Bloch basis as
\begin{equation}
	^{\{g\vert0\}}\vert \phi_{\Gamma_j} , \boldsymbol{k} \rangle =    \vert \phi_{\Gamma_j} , g\boldsymbol{k} \rangle  \chi^{(\Gamma_j)}(g)\;,
\end{equation}
which leads to the following symmetry constraint of the Hamiltonian
\begin{equation}
	\hat{U}_g \cdot H_S( \boldsymbol{k} ) \cdot \hat{U}^{\dagger}_g = H_S( g\boldsymbol{k} )\;,
\end{equation}
where 
\begin{equation}
	\hat{U}_{g} = \bigoplus_{j} \chi^{(\Gamma_j)}(g)\;.
\end{equation}
Under a translation by a reciprocal lattice vector the Bloch functions satisfy
\begin{align}
	\vert \phi_{\Gamma_j} , \boldsymbol{k}+\boldsymbol{K} \rangle &= \vert \phi_{\Gamma_j} , \boldsymbol{k} \rangle \cdot \hat{T}(\boldsymbol{K}) \;,\nonumber\\
	\hat{T}(\boldsymbol{K}) &= \hat{U}^{\dagger}_S \cdot \left(\begin{array}{cc} e^{i \boldsymbol{K} \cdot \boldsymbol{r}_A } & 0 \\
	 0 & e^{i \boldsymbol{K} \cdot \boldsymbol{r}_B } \end{array}\right) \cdot  \hat{U}_S \;,\\
	  H_S( \boldsymbol{k} )&=\hat{T}(\boldsymbol{K}) \cdot H_S( \boldsymbol{k} + \boldsymbol{K}) \cdot \hat{T}^{\dagger}(\boldsymbol{K})\;.\nonumber
\end{align}

Spinless TRS takes a particularly simple form
\begin{equation}
	H^*_S( \boldsymbol{k} ) = H_S(- \boldsymbol{k} ) \;.
\end{equation}

We define the Bloch eigenstates through
\begin{equation}
\begin{aligned}
	\vert \boldsymbol{\psi} , \boldsymbol{k} \rangle &= \vert \boldsymbol{\phi} , \boldsymbol{k} \rangle \cdot \breve{U}(\boldsymbol{k}) 	\;,\\
	\vert \psi_n , \boldsymbol{k} \rangle &= \vert \boldsymbol{\phi} , \boldsymbol{k} \rangle \cdot \breve{U}_n(\boldsymbol{k}) \;,
\end{aligned}
\end{equation}
where $\breve{U}(\boldsymbol{k}) = (\breve{U}_{1}(\boldsymbol{k}) ~\breve{U}_{2}(\boldsymbol{k})) $ is the matrix formed by the eigenvectors $\breve{U}_{n}(\boldsymbol{k})$ of $H_S(\boldsymbol{k})$. In the following we write them as $[\breve{U}_n(\boldsymbol{k})] \equiv \vert u_n, \boldsymbol{k}\rangle $.

\subsubsection{General symmetry properties of the Bloch eigenstates}\label{periodic_sym}

The symmetry properties of the Bloch eigenstates can readily be derived from those of the symmetry-Bloch basis introduced above, i.e.
\begin{equation}
	^{\{g\vert0\}}\vert \psi_n , \boldsymbol{k} \rangle =    \vert \psi_m , g\boldsymbol{k} \rangle  \langle u_m , g\boldsymbol{k} \vert \hat{U}_g \vert u_n , \boldsymbol{k} \rangle\;.
\end{equation}
We define the periodic gauge for the eigenstates by the constraint
\begin{equation}
\label{periodic_gauge}
		\vert u^p_n, \boldsymbol{k}+\boldsymbol{K}\rangle = \hat{T}^{\dagger}(\boldsymbol{K})\vert u^p_n, \boldsymbol{k}\rangle \;.
\end{equation}
It readily follows for a high-symmetry momentum point that contains $g$ in its little co-group,\cite{BradCrack} i.e.~$g\bar{\boldsymbol{k}} = \bar{\boldsymbol{k}} + \boldsymbol{K}_g$,
\begin{equation}
	^{\{g\vert0\}}\vert \psi^p_n ,\bar{\boldsymbol{k}} \rangle =    \vert \psi^p_m , \bar{\boldsymbol{k}} \rangle  \langle u^p_m , \boldsymbol{k} \vert \hat{T}(\boldsymbol{K}_g) \hat{U}_g \vert u^p_n , \bar{\boldsymbol{k}} \rangle\;, 
\end{equation}
and the elements of a matrix representation of $g$ with the IRREP $\Gamma_j$ of the space group is then given by
\begin{align}
	R^{\bar{\boldsymbol{k}}}_{\Gamma_j,mn} &= \langle \psi^p_m(\Gamma_j) , g\bar{\boldsymbol{k}} \vert^{\{g\vert0\}}\vert \psi^p_n(\Gamma_j) , \bar{\boldsymbol{k}} \rangle\;, \nonumber\\
	&= \langle u^p_m(\Gamma_j) , \bar{\boldsymbol{k}} \vert \hat{T}(\boldsymbol{K}_g) \hat{U}_g \vert u^p_n(\Gamma_j) , \bar{\boldsymbol{k}} \rangle\;,
\end{align}
where $\vert \psi^p_n(\Gamma_j) , \bar{\boldsymbol{k}} \rangle$ is a basis function of the IRREP $\Gamma_j$ of the little group at $ \bar{\boldsymbol{k}}$. Also the cell-periodic Bloch function transforms as
\begin{equation}
\label{cell_periodic_sym}
	\vert u^p_n(\Gamma_j) , g \bar{\boldsymbol{k}} \rangle  = \hat{U}_g \vert u^p_m(\Gamma_j) , \bar{\boldsymbol{k}} \rangle R^{\bar{\boldsymbol{k}}\dagger}_{\Gamma_j,mn} \;.
\end{equation}

\subsection{Analytical ansatz of the occupied eigenstate}\label{ansatz}

The tight-binding Hamiltonian written in the symmetry-Bloch basis can be expanded with Pauli matrices as 
\begin{equation}
\label{H_spinless}
	H_S(\boldsymbol{k}) = h^{(\Gamma_1)}_0(\boldsymbol{k}) \hat{1} + h^{(\Gamma_1)}_z(\boldsymbol{k}) \hat{\sigma}_z + h^{(\Gamma_4)}_y(\boldsymbol{k}) \hat{\sigma}_y\;,
\end{equation}
with the component 
\begin{equation}
\begin{aligned}
	h^{(\Gamma_1)}_0(\boldsymbol{k}) &= h_{AA}(\boldsymbol{k}) = h_{BB}(\boldsymbol{k}) \;,\\
	h^{(\Gamma_1)}_z(\boldsymbol{k}) & = \mathrm{Re} [h_{AB}(\boldsymbol{k})] \;,\\
	h^{(\Gamma_4)}_y(\boldsymbol{k}) & = - \mathrm{Im} [h_{AB}(\boldsymbol{k})] \;.
\end{aligned}  
\end{equation}
Crystalline symmetries impose $h^{(\Gamma_1)}_0(\boldsymbol{k})$ and $h^{(\Gamma_1)}_z(\boldsymbol{k})$ to be real and even under inversion, while $h^{(\Gamma_4)}_y(\boldsymbol{k})$ must be real and odd under inversion. It is a special feature of the two-band model, symmetric under the point group $C_{6v}$, that inversion symmetry ($\mathcal{I}$) and spinless TRS ($\mathcal{T}=\mathcal{K}$, $\mathcal{T}^2$=+1) are also effectively satisfied, leading to the effective point group $D_{6h}$ as the symmetry of the system (i.e.~L80) and with the effective AZ class AI. In other words, whenever $C_{6v}$ is assumed the two-band model is too restrictive to allow inversion symmetry and TRS breaking terms. In the following we drop the sup-script and write $h_0$, $h_z$, and $h_y$.  

Thanks to the simplicity of the Hamiltonian, the eigenvalues and eigenvectors take a simple form,
\begin{equation}
\begin{aligned}
	E_1(\boldsymbol{k}) &= h_0(\boldsymbol{k}) - \epsilon_0(\boldsymbol{k}) \\
	E_2(\boldsymbol{k}) &= h_0(\boldsymbol{k}) + \epsilon_0(\boldsymbol{k}) 
\end{aligned}
\end{equation}
with $\epsilon_0(\boldsymbol{k}) = \sqrt{ h_z(\boldsymbol{k})^2 + h_y(\boldsymbol{k})^2}$, and 
\begin{align}
	 \vert u_{1} , \boldsymbol{k}  \rangle = \dfrac{1}{2} \left(\begin{array}{c} 1 - \dfrac{\epsilon_0(\boldsymbol{k})}{h_z(\boldsymbol{k}) - i h_y(\boldsymbol{k}) } \\
	-1 - \dfrac{\epsilon_0(\boldsymbol{k})}{h_z(\boldsymbol{k}) - i h_y(\boldsymbol{k}) } \end{array}\right)\;, \\
	 \vert u_{2}, \boldsymbol{k}  \rangle = \dfrac{1}{2} \left(\begin{array}{c} 1 + \dfrac{\epsilon_0(\boldsymbol{k})}{h_z(\boldsymbol{k}) - i h_y(\boldsymbol{k}) } \\
	-1 + \dfrac{\epsilon_0(\boldsymbol{k})}{h_z(\boldsymbol{k}) - i h_y(\boldsymbol{k}) } \end{array}\right) \;.
\end{align}
It is important to note that in this form the wave function is single-valued and smooth over almost all of the Brillouin zone. These properties are necessary conditions for the noncyclic Berry phase to be meaningful, see below in Appendix \ref{noncyclic_BP}. Hence the above ansatz implicitly defines a smooth reference gauge. We discuss in Appendix \ref{gauge_dependence} the effect of small and large gauge transformations. 

In the following we distinguish between the occupied eigenstate with energy $E_1(\boldsymbol{k})$ and the unoccupied eigenstate with energy $E_2(\boldsymbol{k}) $, and $ E_1(\boldsymbol{k})< E_2(\boldsymbol{k})$.

Since by symmetry $h_y(\Gamma) = 0$, the occupied eigenstate at $\Gamma$ is
\begin{equation}
\label{analitical_ansatz}
	  \vert u_{1} , \Gamma \rangle = \dfrac{1}{2} \left( \begin{array}{c} 1+ s_{\Gamma} \\ -1+s_{\Gamma} \end{array}\right)	\;,
\end{equation}
with $s_{\Gamma} = - \mathrm{sign}\{ h_z(\Gamma)\}$. At $M_1$ (see Fig.~\ref{fig_BZ}), it is 
\begin{equation}
	 \vert u_{1}, M_1 \rangle = \dfrac{1}{2} \left( \begin{array}{c} 1- \dfrac{\epsilon_0(M_1)}{h_z(M_1)-i h_y(M_1)} \\  
	 -1- \dfrac{\epsilon_0(M_1)}{h_z(M_1)-i h_y(M_1)}\end{array}\right)	\;.
\end{equation}
By symmetry $h^{(\Gamma_4)}_y(g\boldsymbol{k}) = - h^{(\Gamma_4)}_y(\boldsymbol{k})$, therefore the eigenvector at $M_2$ is
\begin{equation}
	 \vert u_{1}, M_2 \rangle =  \vert u_{1}, M_1 \rangle^* =  \dfrac{1}{2} \left( \begin{array}{c} 1- \dfrac{\epsilon_0(M_1)}{h_z(M_1)+i h_y(M_1)} \\  
	 -1- \dfrac{\epsilon_0(M_1)}{h_z(M_1)+i h_y(M_1)}\end{array}\right)	\;.
\end{equation}

The energy eigenstate must also be an eigenstate of the symmetries of the system, therefore
\begin{align}
	 ^{\{C_{2z}\vert 0\}}\vert \psi_{1} , M_1\rangle 
	 &=  \vert  \psi_{1} ,M_1 \rangle  \nonumber\\
	 & \quad\langle  u_{1},M_1\vert  \hat{T}(-\boldsymbol{b}_1-\boldsymbol{b}_2)  \hat{U}_{C_{2z}} \vert  u_{1},M_1 \rangle \;,\nonumber\\
	 &= \vert  \psi_{1}, M_1 \rangle \dfrac{h_z(\text{M}_1)- \sqrt{3} h_y(\text{M}_1)}{2 \sqrt{ h_z(\text{M}_1)^2+ h_y(\text{M}_1)^2 }} \;,\nonumber\\
 	&\stackrel{!}{=}  \vert  \psi_{1} , \text{M}_1\rangle s_{\text{M}}\;,
 \end{align}
where $s_{\text{M}} = \pm 1$ determines the IRREP at M. This leads to the equation 
\begin{align}
	 \dfrac{1 - \sqrt{3} t }{2 \sqrt{ 1+ t^2 }} = s_{\text{M}} \dfrac{\sqrt{h_z(\text{M}_1)^2}}{h_z(\text{M}_1)}
\end{align} 
with $t=h_y(\text{M}_1)/h_z(\text{M}_1)$. Since the lefthand side cannot be $-1$, we set $s_{\text{M}} = \mathrm{sign}\{h_z(\text{M}_1)\}$, and we are left with the equation 
\begin{align}
	 \dfrac{1 - \sqrt{3} t }{2 \sqrt{ 1+ t^2 }} = 1\;,
\end{align} 
from which we find $t=h_y(\text{M}_1)/h_z(\text{M}_1) = -\sqrt{3}$. Hence we can rewrite the occupied eigenvector at $M_1$ as,
\begin{align}
	 \vert u_1  ,M_1 \rangle = \dfrac{1}{2} \left( \begin{array}{c} 1-  \dfrac{2 s_{\text{M}}}{1+ i \sqrt{3}} \\  
	 -1-\dfrac{2 s_{\text{M}}}{1+ i \sqrt{3}} \end{array}\right)	\;.
\end{align}

We readily have the $C_{2z}$-symmetry eigenvalues at $\Gamma$ and $M_1$ of the occupied eigenstate are given by
\begin{equation}
\begin{aligned}
	\xi^{\Gamma}_2 &= s_{\Gamma}	=  - \mathrm{sign}\{ h_z(\Gamma)\}\;,\\
	\xi^{\text{M}}_2 &= s_{\text{M}} = \mathrm{sign}\{h_z(\text{M}_1)\}	\;.
\end{aligned}
\end{equation}

\subsection{Real homotopy and topological invariant}

Performing the change of basis corresponding to $(\sigma_y,\sigma_z) \rightarrow (\sigma_z$, $\sigma_x)$, the Hamiltonian becomes real. Terms with the third Pauli matrix must vanish as a consequence of the $C_2\mathcal{T}$ symmetry (indeed $C_2\mathcal{T}$ with $(C_2\mathcal{T})^2=+1$ can be represented as the complex conjugation). The classifying space of the two-band Hamiltonian is then $\mathrm{Gr}_1(\mathbb{R}^2) \cong \mathbb{RP}^1\cong \mathbb{S}^1$.\cite{BzduConversion} Therefore, $\pi_1(\mathrm{Gr}_1(\mathbb{R}^2)) = \pi_1(\mathbb{S}^1) = \mathbb{Z}$, i.e. the topology over any arbitrary one-dimensional base loop is captured by an integer topological invariant. 

The topological invariant for $\pi_1(\mathrm{Gr}_1(\mathbb{R}^2)) =  \mathbb{Z}$ can be computed as the winding number of the non-cyclic Berry phase over a closed loop when the eigenvectors are defined in a gauge that is smooth almost everywhere and is periodic over the Brillouin zone (see also Section \ref{gauge_dependence} on the gauge invariance). Each nodal point of the band structure induces a vortex structure in the eigenvectors so that the winding of the non-cyclic Berry phase captures the \textit{vorticity}\cite{BzduConversion} of the nodal point. The above $\mathbb{Z}$ classification over base loops indicates that the vorticities of multiple nodal points are additive for the base loop that encircles them all. 

While this local result is not affected by the presence of additional crystalline symmetries, we derive below that the hexagonal symmetries act as an obstruction on the winding of the non-cyclic Berry phase. We do this by choosing a symmetric base loop that crosses high-symmetry points and encircle on half of the Brillouin zone, hence revealing the global topology of the crystalline spinless semimetal.

\subsection{Symmetry constrained noncyclic Berry phase}\label{noncyclic_BP}

Below, we derive the consequences for the non-cyclic Berry phase over a symmetric base loop that crosses high-symmetry points, hence revealing the global topology of the crystalline spinless semimetal. 

An oriented base loop in momentum space can be parameterized as $l(t) : [0,1] \rightarrow  \mathbb{S}^1,~ t \mapsto l(t) $ ($l\cong \mathbb{S}^1$). Assuming that the occupied eigenstate $\vert u_1,\boldsymbol{k} \rangle$ is defined smoothly all along an oriented loop $l(t)$, the noncyclic Berry phase is defined as 
\begin{equation}
	\gamma[l(t)] = i \int_{l(t)} d\boldsymbol{k}(t) \cdot \boldsymbol{\mathcal{A}}(\boldsymbol{k}(t))\;,
\end{equation}
with the $U(1)$-Berry connection $\mathcal{A}_{\mu}(\boldsymbol{k}) = \langle u_1,\boldsymbol{k}\vert \partial_{k^{\mu}} u_1,\boldsymbol{k} \rangle$ with $\mu = x,y$.

The geometric phase factor of the occupied state over the closed loop $l_{\beta}$ encircling $K_1$ (see Fig.~\ref{fig_BZ}) is then given by
\begin{equation}
	e^{i \gamma[l_{\beta}]} = \left(e^{i \gamma[l_{\beta}']}\right)^3	\;,
\end{equation}
where the loop segment $l_{\beta}' = l_{\text{M}_1\Gamma} \circ l_{\Gamma\text{M}_2} $ is one third of the total loop, starting at $M_2$, crossing $\Gamma$ and ending at $M_1$ (see Fig.~\ref{fig_BZ}). Using the Wilson loop formalism,\cite{WindingKMZ2, InvTIBernevig, PointGroupsTI, Wi1, Wi3, Wi2, BBS_nodal_lines} we write
\begin{align}
	e^{i \gamma[l_{\beta}'] } =  \langle u_1  ,M_1 \vert \hat{W}_{M_1M_2}  \vert u_1  ,M_2 \rangle \;.
\end{align}
with the Wilson loop operator $\hat{W}_{k_2k_1} = \stackrel{k_2 \leftarrow k_1}{\prod_{\boldsymbol{k}}} \vert u_1 (\boldsymbol{k}) \rangle \langle u_1 (\boldsymbol{k}) \vert $. We now project the occupied eigenvector at every $\boldsymbol{k}$ on the eigenvector at $\Gamma$, i.e.~
\begin{equation}
	\vert \tilde{u}_1, \boldsymbol{k} \rangle = \vert u_1, \boldsymbol{k} \rangle \langle u_1 , \boldsymbol{k} \vert u_1, \Gamma \rangle  \;,
\end{equation}
(note that $\vert\langle u_1 , \boldsymbol{k} \vert u_1, \Gamma \rangle\vert =1$), and rewrite  
\begin{align} 
	e^{i \gamma[l_{\beta}'] } &=   \dfrac{\langle u_1 ,  \Gamma \vert  u_1 , M_2 \rangle }{\langle u_1 , \Gamma \vert  u_1 , M_1 \rangle} ~ \langle \tilde{u}_1 , M_1 \vert \widetilde{\hat{W}}_{M_1M_2} \vert \tilde{u}_1 , M_2 \rangle \;,\nonumber\\
	&=   \dfrac{\langle u_1 ,  \Gamma \vert  u_1 , M_2 \rangle }{\langle u_1 , \Gamma \vert  u_1 , M_1 \rangle} ~ \exp\left\{ - \displaystyle  \int_{l_{\beta}'} d\boldsymbol{k} \cdot \widetilde{\boldsymbol{\mathcal{A}}}(\boldsymbol{k}) \right\} \;.
\end{align}
Now, since $\vert \tilde{u}_1 , \Gamma \rangle$ is real, after projecting the eigenvectors on it, we get the simplification $\widetilde{\boldsymbol{\mathcal{A}}}(\boldsymbol{k}) = \langle \tilde{u}_1 , \boldsymbol{k} \vert \partial_{\boldsymbol{k}} \tilde{u}_1 , \boldsymbol{k}  \rangle = \mathrm{Im}~ \left( \langle \tilde{u}_1 , \boldsymbol{k} \vert \partial_{\boldsymbol{k}} \tilde{u}_1 , \boldsymbol{k} \rangle\right) = 0$. Therefore, 
\begin{align}
	e^{i \gamma[l_{\beta}'] } =   \dfrac{\langle u_1 , \Gamma \vert  u_1 , M_2 \rangle }{\langle u_1 , \Gamma \vert  u_1 , M_1 \rangle}\;.
\end{align}

Substituting with the expressions of the eigenvectors derived above, we eventually find 
\begin{align}
\label{beta_prime}
	e^{i \gamma[l_{\beta}'] } &= \dfrac{2 - (1+i \sqrt{3}) \xi^{\Gamma}_2 \xi^{\text{M}}_2 }{2 - (1-i \sqrt{3})  \xi^{\Gamma}_2 \xi^{\text{M}}_2 } \;. 
\end{align}

\subsection{Gauge dependence}\label{gauge_dependence}

Eq.~(\ref{berry_phase_lg}) of the main text is derived from Eq.~(\ref{beta_prime}). Eq.~(\ref{beta_prime}) has been derived from the analytical expression of the occupied eigenstate Eq.~(\ref{analitical_ansatz}) that fixes the gauge (also Eq.~(\ref{analitical_ansatz}) is defined in terms of a set of Bloch basis functions Eq.~(\ref{Bloch_basis}) that trivializes the total Bloch bundle\cite{reference_trivialization}). It is then relevant to ask whether the result Eq.~(\ref{berry_phase_lg}) is invariant under a gauge transformation $\vert u_1(\boldsymbol{k}) \rangle \rightarrow \vert u_1(\boldsymbol{k}) \rangle e^{i \theta(\boldsymbol{k})}$, $\boldsymbol{k} \in l_{\beta}$, that leads to the change of the Berry phase $\gamma^{(0)}[l] \rightarrow \gamma^{(\theta)}[l]$. Such a gauge transformation can originate for instance from a different choice of trivialization of the total Bloch bundle, e.g.~for distinct choices of gauge of the set of Bloch (Fourier) basis functions Eq.~(\ref{Bloch_basis}),\cite{reference_trivialization} or by using a different ansatz of the eigenstate Eq.~(\ref{analitical_ansatz}). 

Keeping track of the gauge phase as we travel along the base loop $l_{\beta}$, which we parametrize as $t\in [0,1] \mapsto l(t) $ with $l(0) = l(1)$, the single-valuedness of the wave function requires $\left.\theta(\boldsymbol{k}(t))\right\vert_{t=1} = \left.\theta(\boldsymbol{k}(t))\right\vert_{t=0} + n 2\pi$. Let us decompose the gauge phase into the part that winds, $\theta_n$, and the part that goes back to its original value after we run one time through the loop, i.e.~$\theta(t) = \theta_0(t) +\theta_n(t)$ with $\theta_0(t=1) = \theta_0(t=0)$, $\theta_n(t=0) = 0$ and $\theta_n(t=1) = 2n\pi$. Then the non-winding part of the gauge $\theta_0(t)$ can be smoothly mapped to zero without changing the Berry phase, i.e.~$\gamma^{(\theta_0)}[l] = \gamma^{(0)}[l]$. On the contrary, the winding part of the gauge $\theta_n$ defines a large gauge transformation that shifts the Berry phase as $\gamma^{(\theta_n)}[l] = \gamma^{(0)}[l] + n 2\pi$. While large gauge transformations are allowed within equivalence classes defined up to bundle isomorphisms, they lead to distinct homotopy equivalence classes.\cite{Gauge_dependence} Therefore, Eq.~(\ref{berry_phase_lg}) (obtained from Eq.~(\ref{beta_prime})) makes only sense under the assumption that large gauge transformations are excluded which is practically always possible by (i) always using the same reference trivialization of the total Bloch bundle (Eq.~(\ref{Bloch_basis})),\cite{reference_trivialization} and (ii) using the same smooth reference gauge of the eigenstate (Eq.~(\ref{analitical_ansatz})).

\section{Four-band spinful case with inversion symmetry}\label{spinful_inv_ap}

Let us now consider the extension of tight-binding model Eq.~(\ref{spinless_site}) when the spin degrees of the freedom are taken into account. The basis functions can then be taken as $(\vert \phi_{\Gamma_1} , \sigma, \boldsymbol{k} \rangle, \vert \phi_{\Gamma_4} , \sigma, \boldsymbol{k} \rangle)$ with $\sigma= \uparrow, \downarrow$. 
Assuming TRS and inversion symmetry ($\mathcal{I}$), there is no spin-flip (Rashba) SOC and we can split the four-band spinful Hamiltonian into spin-up and spin-down sectors as 
\begin{equation}
\label{spin_sector}
	\mathcal{H}_{\sigma} = \left(\vert \phi_{\Gamma_1,\sigma}, \boldsymbol{k} \rangle ~	\vert \phi_{\Gamma_4,\sigma}, \boldsymbol{k} \rangle \right) H_{S,\sigma}(\boldsymbol{k})	
	\left(\begin{array}{c} \langle \phi_{\Gamma_1,\sigma}, \boldsymbol{k} \vert \\
	\langle \phi_{\Gamma_4,\sigma}, \boldsymbol{k} \vert \end{array}\right)\;,
\end{equation}
where
\begin{multline}
	H_{\sigma}(\boldsymbol{k}) = h^{(\Gamma_1)}_0(\boldsymbol{k}) \hat{1} + h^{(\Gamma_1)}_z(\boldsymbol{k}) \hat{\sigma}_z + \\
	h^{(\Gamma_4)}_y(\boldsymbol{k}) \hat{\sigma}_y + s_{\sigma} h^{(\Gamma_3)}_{x}(\boldsymbol{k}) \hat{\sigma}_x\;,
\end{multline}
with $\sigma=\uparrow,\downarrow$ and $s_{\uparrow} = +1$, $s_{\downarrow} = -1$. The term $h^{(\Gamma_3)}_{x}(\boldsymbol{k})$ contains all the (non-Rashba) SOC terms. Comparing with the Hamiltonian written in the sub-lattice site basis, we have $h^{(\Gamma_3)}_{x}(\boldsymbol{k}) = (h_{AA,\uparrow}(\boldsymbol{k})-h_{BB,\uparrow}(\boldsymbol{k}))/2$.

Each spin-sector of the Hamiltonian $H_{\sigma}(\boldsymbol{k})$ closely resembles the two-band spinless Hamiltonian Eq.~(\ref{H_spinless}). However now the term $h^{(\Gamma_3)}_{x}$ effectively breaks spinless vertical mirror symmetries (the character for $\Gamma_3$ of vertical mirror symmetries are $\chi^{\Gamma_3}_{m_{x,y} } = - \chi^{\Gamma_4}_{m_{x,y} }$) and spinless TRS ($\mathcal{T}=\mathcal{K}$). Therefore, each spin-block only effectively conserves the symmetries of the point group $C_{6h}\subset D_{6h}$, i.e.~L80 (SG191)$\rightarrow$L75 (SG175), and belongs to the AZ class A (instead of AI). The classifying space of the four-band spinful inversion symmetric model is then composed as 
\begin{equation}
\label{TRSI_decomposition}
	\mathcal{H}^{2+2}_{\mathrm{AII+L80}} = \mathcal{H}^{1+1}_{\mathrm{A+L75}} \oplus \mathcal{H}^{1+1}_{\mathrm{A+L75}}\;.
\end{equation} 
Contrary to L77 ($C_{6v}$) and L80 ($D_{6h}$), L75 ($C_{6h}$) has no essential degeneracy at K and the spectrum of each spin-polarized subspace $ \mathcal{H}^{1+1}_{\mathrm{A+L75}}$ can be gapped. Furthermore, since there is only a single orbital per site, the eigenstates of both spin-sectors are either all even or all odd under mirror symmetry with respect to the basal plane $\sigma_h$. As a consequence, the spectrum of each spin subspace must be gapped since no symmetry protected band crossing can be formed over the BZ. It is important to notice that for this reason, $H_{S,\sigma}$ as obtained above is not the most general Hamiltonian of the spinless two-band symmetry class A+L75. See the schematic example of a band structure in Fig.~\ref{fig_BS_IRREPs}(b) where each band is labeled according to their IRREPs at every HSP and away for the HSPs (in this case they are all odd under $\sigma_h$ since we have assumed the orbital $p_z$ at every site). 

\subsection{Analytical ansatz for the spin-polarized eigenstates}

The eigenvalues and eigenstates of one spin-sector Eq.~(\ref{spin_sector}) take again a simple form (we have dropped the superscripts for the components of the Hamiltonian),
\begin{equation}
\begin{aligned}
	E_{1,\sigma}(\boldsymbol{k}) &= h_0(\boldsymbol{k}) - \epsilon_0(\boldsymbol{k}) \\
	E_{2,\sigma}(\boldsymbol{k}) &= h_0(\boldsymbol{k}) + \epsilon_0(\boldsymbol{k}) 
\end{aligned}
\end{equation}
where $\epsilon_0(\boldsymbol{k}) = \sqrt{ h_x(\boldsymbol{k})^2+ h_y(\boldsymbol{k})^2 + h_z(\boldsymbol{k})^2 }$, and 
\begin{align}
	 \vert u_{1} ,\sigma, \boldsymbol{k}  \rangle = \dfrac{1}{\sqrt{2}\mathcal{N}_1} \left(\begin{array}{c} 1 + \dfrac{s_{\sigma} h_x-\epsilon_0(\boldsymbol{k})}{h_z(\boldsymbol{k}) - i h_y(\boldsymbol{k}) } \\
	-1 + \dfrac{s_{\sigma} h_x-\epsilon_0(\boldsymbol{k})}{h_z(\boldsymbol{k}) - i h_y(\boldsymbol{k}) } \end{array}\right)\;, \\
	 \vert u_{2}, \sigma, \boldsymbol{k}  \rangle = \dfrac{1}{\sqrt{2}\mathcal{N}_2} \left(\begin{array}{c} 1 + \dfrac{s_{\sigma} h_x+\epsilon_0(\boldsymbol{k})}{h_z(\boldsymbol{k}) - i h_y(\boldsymbol{k}) } \\
	-1 + \dfrac{s_{\sigma} h_x+\epsilon_0(\boldsymbol{k})}{h_z(\boldsymbol{k}) - i h_y(\boldsymbol{k}) } \end{array}\right) \;,
\end{align}
with the normalization factors $\mathcal{N}_{i}$ defined through the conditions $\langle u_{i} ,\sigma, \boldsymbol{k}  \vert u_{i} ,\sigma, \boldsymbol{k}  \rangle =1$, $i=1,2$.

The above analytical ansatz of the occupied eigenstate, $\vert u_{1} ,\sigma, \boldsymbol{k}  \rangle$, can be used to derive the spin-polarized Chern number in a direct way. This is rather tedious though and we instead use an alternative algebraic approach below. We have checked numerically though that the direct computation of the Chern number with the above ansatz gives the same results as the algebraic results. \\

\subsection{Derivation of Eq.~(\ref{gamma_rho_IRREP})}\label{spinful_Berry_ap}

We have argued in Section \ref{spinful_inversion} that the Chern number is obtained from the flow of Berry phase as we sweep a base loop over the BZ. By symmetry it is enough to compute the contribution to the Chern number from the patch $S_{\rho}$ that covers one sixth of the BZ (Fig.~\ref{fig_BZ}). This is given by $e^{i \gamma_{\rho}} $ ($\gamma_{\rho} \equiv \gamma[l_{\rho}]$). We here use the Wilson loop techniques\cite{WindingKMZ2, InvTIBernevig, PointGroupsTI, Wi1, Wi3, Wi2, BBS_nodal_lines, HolAlex_Bloch_Oscillations} in order to derive the spin-polarized Berry phase (Chern number) algebraically.

We choose the occupied eigenstate as $\vert u_{1} , \boldsymbol{k}  \rangle = \vert u_{1} ,\uparrow, \boldsymbol{k}  \rangle$ (see above). In the following derivation we actually don't need the explicit expression of the eigenstate. We only need the eigenvalues of occupied eigenstate under the rotation $C_{2z}$ at $\Gamma$ and $M_1$, and under $C_{3z}$ at $\Gamma$ and $K_1$. In the following, we neglect the phase factor coming from the rotation of the spin and we use the character table of the single-valued IRREPs. We have argued that each spin-polarized sector can be seen as belonging to the symmetry class A+L75 with point group $C_{6h}$. The eigenvalues under $C_{2z}$ and $C_{3z}$ of the relevant IRREPs are then obtained  from the compatibility relations from the IRREPs of $D_{6h}$ to the IRREPs of $C_{6h}$, these are given in Table \ref{T_comp}.

Let us split the oriented boundary of $S_{\rho}$, $l_{\rho}$ see Fig.~\ref{fig_BZ}, into the successive segments that connect the HSPs $\{\Gamma, M_1,\Gamma',K_1\}$ ($\Gamma' =  \Gamma+\boldsymbol{b}_1+\boldsymbol{b}_2$), i.e.~$l_{\rho} =  l_d\circ l_c\circ l_b\circ l_a$, with $l_d = l_{\Gamma\leftarrow K_1}$, $l_c = l_{K_1\leftarrow \Gamma'}$, $l_b = l_{\Gamma'\leftarrow M_1}$, and $l_a = l_{M_1\leftarrow \Gamma}$. We further choose $l_{\rho}$ such that $l_b = C_{2z} (l^{-1}_a -\boldsymbol{b}_1-\boldsymbol{b}_2)  \sim C_{2z} l^{-1}_a$ and $l_c = C_{3z} (l^{-1}_d-\boldsymbol{b}_2) \sim C_{3z} l^{-1}_d $, where $`\sim'$ means equal up to a translation by a reciprocal lattice vector. We then write the contribution to the Berry phase factor from the segment $l_{a}$ as $e^{i \theta_a} =  \langle u_1, M_1 \vert \hat{W}_{a} \vert u_1, \Gamma \rangle$ where $\hat{W}_{a}$ is the Wilson loop operator along the path $l_{a}$, i.e.~$\hat{W}_{a} = \prod\limits_{\boldsymbol{k}}^{l_{a}} \vert u_1, \boldsymbol{k} \rangle \langle u_1, \boldsymbol{k}  \vert$, and similarly for the other segments $l_b, l_c, l_d$.

\begin{widetext}
The Berry phase factor over the whole loop $l_{\rho}$ is then given by
\begin{align}
	e^{i \gamma_{\rho}} &= e^{i \theta_d}  \langle u_1, K_1 \vert \hat{W}_{c} \vert u_1, \Gamma' \rangle  \langle u_1, \Gamma' \vert \hat{W}_{b} \vert u_1, M_1 \rangle e^{i \theta_a} \;,\nonumber\\
	&= e^{i \theta_d} \langle u_1, C_{3z}(K_1-\boldsymbol{b}_2) \vert \hat{W}_{c} \vert u_1, C_{3z}(\Gamma-\boldsymbol{b}_2) \rangle \langle u_1 , C_{2z}(\Gamma-\boldsymbol{b}_1-\boldsymbol{b}_2) \vert  \hat{W}_{b} \vert u_1, C_{2z}(M_1-\boldsymbol{b}_1-\boldsymbol{b}_2) \rangle e^{i \theta_a}  \;,\nonumber\\
	&= e^{i \theta_d}  R^{\text{K}}_3 \langle u_1, K_1 \vert \hat{T}(-\boldsymbol{b}_2)\hat{C}^{\dagger}_{3z} \hat{W}_{c} \hat{C}_{3z} \hat{T}^{\dagger}(-\boldsymbol{b}_2) \vert u_1, \Gamma \rangle \left( R^{\Gamma}_3 \right)^{\dagger}\nonumber\\
	& \quad\quad\quad\quad R^{\Gamma}_2 \langle u_1 , \Gamma \vert \hat{T}(-\boldsymbol{b}_1-\boldsymbol{b}_2)\hat{C}^{\dagger}_{3
	2z} \hat{W}_{b}\hat{C}_{2z} \hat{T}^{\dagger}(-\boldsymbol{b}_1-\boldsymbol{b}_2) \vert u_1, M_1 \rangle \left( R^{\text{M}}_2 \right)^{\dagger} e^{i \theta_a}  \;,\nonumber\\
	&= e^{i \theta_d} R^{\text{K}}_3 \langle u_1, K_1 \vert \hat{W}^{-1}_{d}\vert u_1, \Gamma \rangle \left( R^{\Gamma}_3 \right)^{\dagger} R^{\Gamma}_2 \langle u_1 , \Gamma \vert \hat{W}^{-1}_{a}  \vert u_1, M_1 \rangle \left( R^{\text{M}}_2 \right)^{\dagger} e^{i \theta_a} \;,\nonumber\\
	&= e^{i \theta_d}~[\xi^{\text{K}}_{3}  e^{-i \theta_d} (\xi^{\Gamma}_{3})^{-1}]~ [\xi^{\Gamma}_{2}  e^{-i \theta_a} (\xi^{\text{M}}_{2})^{-1}]~e^{i \theta_a}  \;,\nonumber \\
	&= \xi^{\text{K}}_{3} \xi^{\Gamma}_{2} (\xi^{\Gamma}_{3} \xi^{\text{M}}_{2})^{-1}\;,
\end{align}
\end{widetext}
where we have use Eq.~(\ref{cell_periodic_sym}) with the eigenvalues of the rotation symmetries (one-dimensional representations) at HSP $\bar{\boldsymbol{k}}$ written as $\xi^{\bar{\boldsymbol{k}}}_{2} $ for $C_{2z}$ and $\xi^{\bar{\boldsymbol{k}}}_{3} $ for $C_{3z}$, and where we have assumed the periodic gauge Eq.~(\ref{periodic_gauge}). Now taking into account all combinatorial ways of ordering the IRREPs of the effective spin-polarized EBR at the HSPs, one example is shown in Fig.~\ref{fig_BS_IRREPs}(b), we eventually arrive at Eq.~(\ref{gamma_rho_IRREP}) in the main text.

\subsection{Split EBR with Wilson loop winding of $W_{II} = +7$}\label{winding_seven}

\begin{figure}[h!]
{\centering
\begin{tabular}{cc}
	\includegraphics[width=0.5\linewidth]{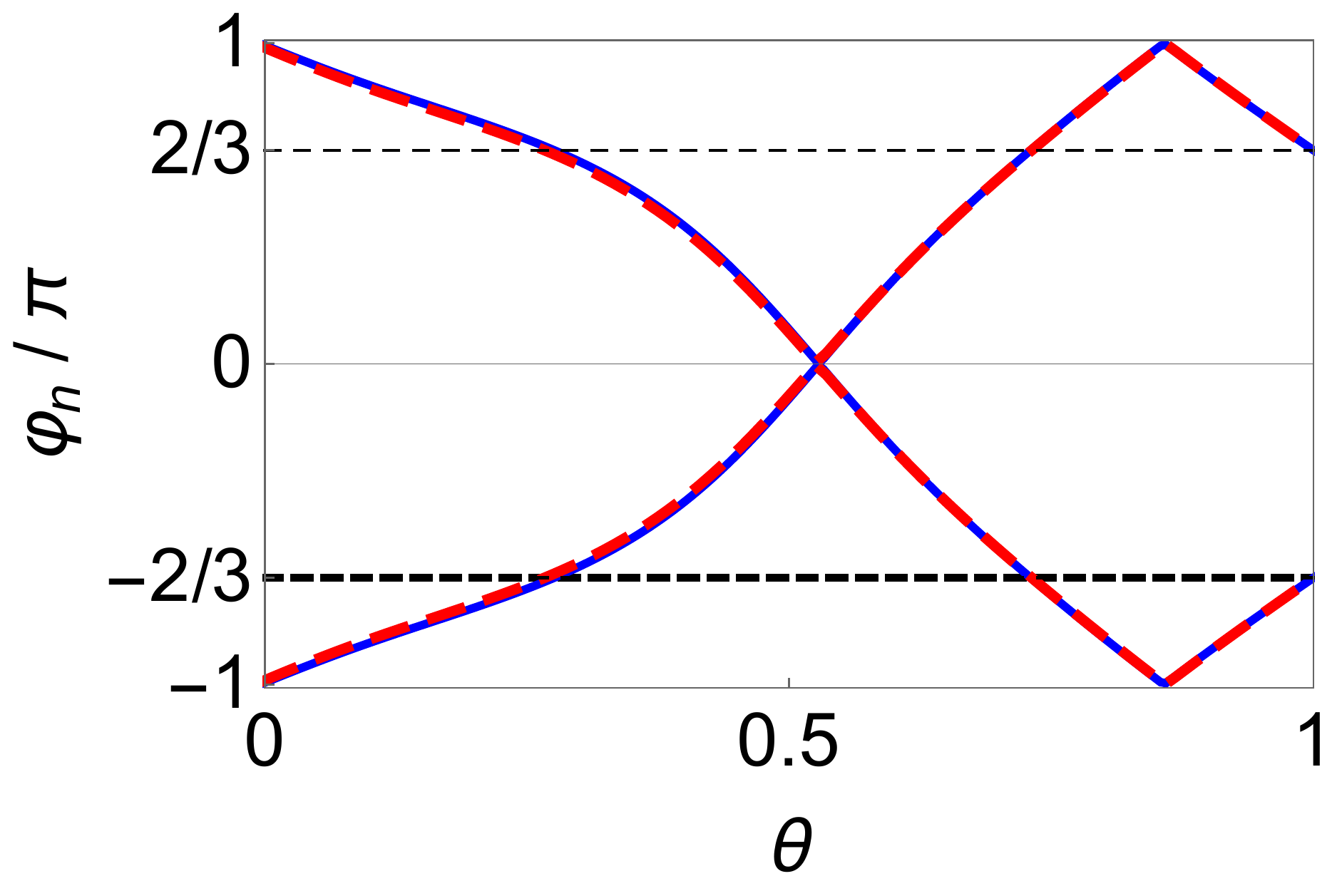} &
	\includegraphics[width=0.5\linewidth]{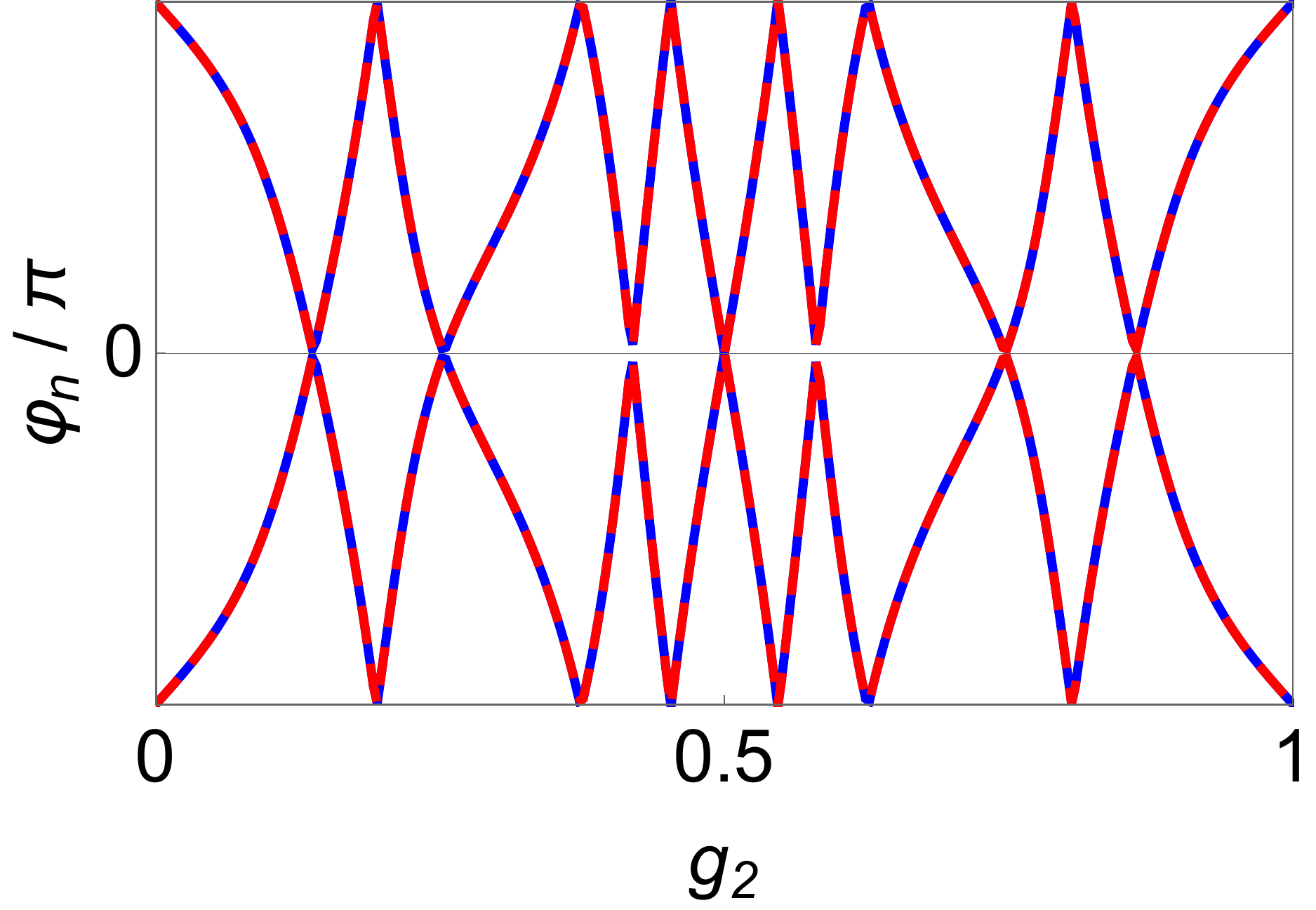} \\
	(a) $W[S_{\rho}] = +7\pi/3$ & (b) $W_{II} = +7$
\end{tabular}
}
\caption{\label{seven_winding} Example of a split EBR with a Wilson loop winging of $W_{II} = +7$. (a) Flow of the Wilsonian phases over the patch $S_{\rho}$. (b) Flow of the Wilsonian phases over the BZ.  
}
\end{figure}

We show in Fig.~\ref{seven_winding} a supplementary example of a four-band spinful topological phase with a Wilson loop winding of $W_{II} = +7$ in each two-band subspaces. This phase was obtained for tight-binding parameters up to the 10th layer of neighbors. This indicates that arbitrary higher Wilson loop windings can be generated by increasing the range of the tight-binding parameters in analogy to topological insulators with high Chern numbers.\cite{high_cern} 

\section{Bloch eigenstates of the six-band case at $\Gamma$ and K}\label{six_band_ap}

After coupling the honeycomb lattice (with the split EBR $\mathrm{B}^a_1+\mathrm{B}^a_2$) to the triangular lattice (with the EBR $\left.\mathrm{B}^{a}_1\right.'$), the doublets for the IRREPs $\overline{\Gamma}_9$ of the six-band model at $\Gamma$ have the form
\begin{equation}
\label{spin_mixture}
\begin{aligned}
	\vert \boldsymbol{\psi}^{\mathrm{B}_1}_{\overline{\Gamma}_9}, \Gamma \rangle &=   \left[\begin{array}{l} 
	\cos\vartheta  \vert \phi_{\Gamma_1} , \uparrow, \Gamma \rangle +  \sin\vartheta \vert \varphi_C , \uparrow, \Gamma \rangle  \\
	\cos\vartheta  \vert \phi_{\Gamma_1} , \downarrow, \Gamma \rangle + \sin\vartheta \vert \varphi_C , \downarrow, \Gamma \rangle 
	   \end{array}\right]^T , \\
	\vert \boldsymbol{\psi}^{\mathrm{B}_1'}_{\overline{\Gamma}_9}, \Gamma \rangle &= \left[\begin{array}{l}  \cos\vartheta \vert \varphi_C , \uparrow, \Gamma \rangle -  \sin\vartheta \vert \phi_{\Gamma_1} , \uparrow, \Gamma \rangle\\
	 \cos\vartheta \vert \varphi_C , \downarrow, \Gamma \rangle - \sin\vartheta \vert \phi_{\Gamma_1} , \downarrow, \Gamma \rangle  \end{array}\right]^T ,
\end{aligned}
\end{equation}
and the doublets for the IRREPs $\overline{K}_6$ at K have now the form
\begin{equation}
\label{spin_mixture}
\begin{aligned}
	\vert \boldsymbol{\psi}^{\mathrm{B}_1}_{\overline{K}_6}, K_1 \rangle &=   \left[\begin{array}{l} 
	\cos\bar{\vartheta}  \vert \varphi_A , \uparrow, K_1 \rangle + \omega \sin\bar{\vartheta} \vert \varphi_C , \downarrow, K_1 \rangle  \\
	\cos\bar{\vartheta}  \vert \varphi_B , \downarrow, K_1 \rangle+ \omega^*\sin\bar{\vartheta} \vert \varphi_C , \uparrow, K_1 \rangle 
	   \end{array}\right]^T , \\
	\vert \boldsymbol{\psi}^{\mathrm{B}_1'}_{\overline{K}_6}, K_1 \rangle &= \left[\begin{array}{l} 
	\cos\bar{\vartheta} \vert \varphi_C , \downarrow, K_1 \rangle -\omega^*\sin\bar{\vartheta} \vert \varphi_A , \uparrow, K_1 \rangle\\
	\cos\bar{\vartheta} \vert \varphi_C , \uparrow, K_1 \rangle -\omega  \sin\bar{\vartheta} \vert \varphi_B , \downarrow, K_1 \rangle 
 \end{array}\right]^T ,
\end{aligned}
\end{equation}
with $\omega = e^{i 2\pi/3}$ and $\vartheta \in [0,\pi/2]$. These forms are readily obtained by inspection of the six-band Hamiltonian at $\Gamma$ and K (we work with the most general tight-binding model allowed by symmetry including up to the 10th layer of neighbors). We have ordered the doublets such that the matrix representations of $C_{3z}^+$ are diagonal with the order $[e^{-i \pi/3},e^{i \pi/3}]$ for $\overline{\Gamma}_9$ with the spin components ordered as $(\uparrow,\downarrow)$, and with $[e^{i \pi/3},e^{-i \pi/3}]$ for $\overline{K}_6$ with the pseudo-spin components ordered as $(\tilde{\uparrow},\tilde{\downarrow})$ for $\mathrm{B}_1$ and as $(\tilde{\downarrow},\tilde{\uparrow})$ for $\mathrm{B}_1'$. The doublets at the inverted momentum -K are then chosen such they are the conjugate partners under TRS, according to Eq.~(\ref{TRS_relation}), with the doublets at K, i.e. 
\begin{equation}
\label{spin_mixture}
\begin{aligned}
	&\vert \boldsymbol{\psi}^{\mathrm{B}_1}_{\overline{K}_6}, -K_1 \rangle =\\
	&\quad  \left[\begin{array}{l} 
	\cos\bar{\vartheta}  \vert \varphi_B , \uparrow, -K_1 \rangle - \omega \sin\bar{\vartheta} \vert \varphi_C , \downarrow, -K_1 \rangle  \\
	\cos\bar{\vartheta}  \vert \varphi_A , \downarrow, -K_1 \rangle- \omega^*\sin\bar{\vartheta} \vert \varphi_C , \uparrow, -K_1 \rangle 
	   \end{array}\right]^T , \\
	&\vert \boldsymbol{\psi}^{\mathrm{B}_1'}_{\overline{K}_6}, -K_1 \rangle = \\
	&\quad\left[\begin{array}{l} 
	\cos\bar{\vartheta} \vert \varphi_C , \downarrow, -K_1 \rangle +\omega^* \sin\bar{\vartheta} \vert \varphi_B , \uparrow, -K_1 \rangle\\
	 \cos\bar{\vartheta} \vert \varphi_C , \uparrow, -K_1 \rangle +\omega  \sin\bar{\vartheta} \vert \varphi_A , \downarrow, -K_1 \rangle
 \end{array}\right]^T \;.
\end{aligned}
\end{equation}
 
\begin{figure*}[t!]
\centering
\begin{tabular}{c|c|c|c} 
	\includegraphics[width=0.25\linewidth]{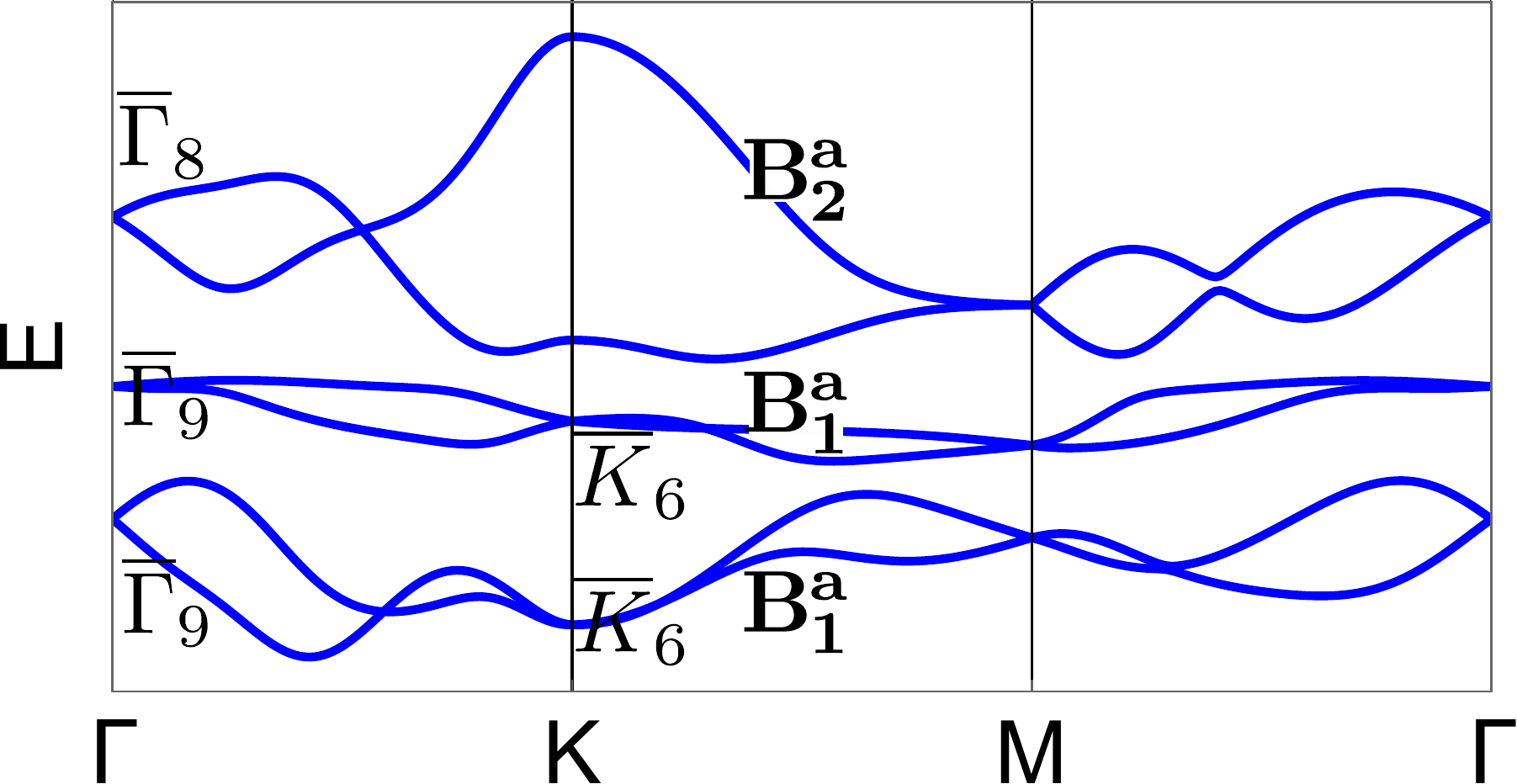} &
	\includegraphics[width=0.25\linewidth]{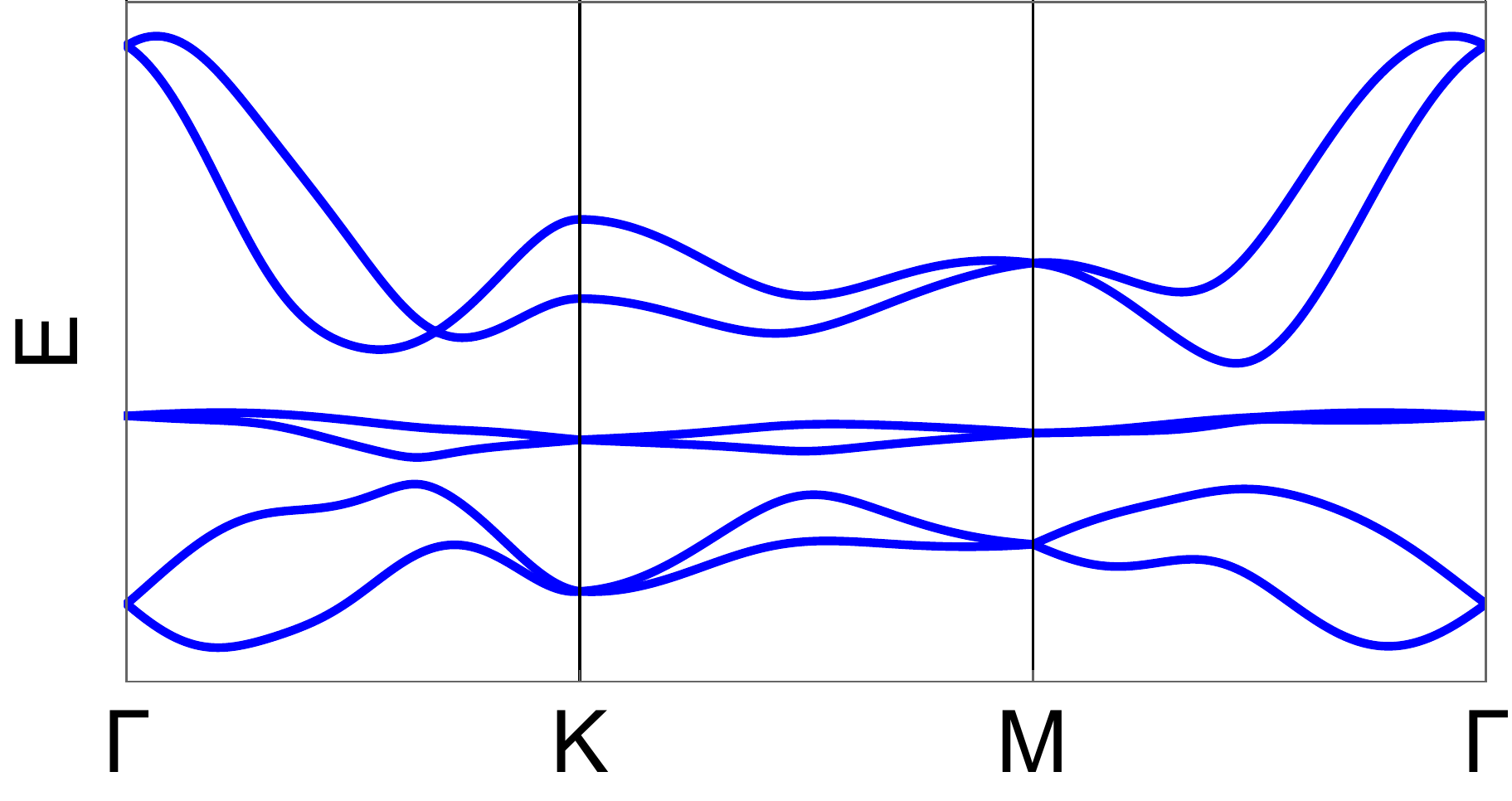} &
	\includegraphics[width=0.25\linewidth]{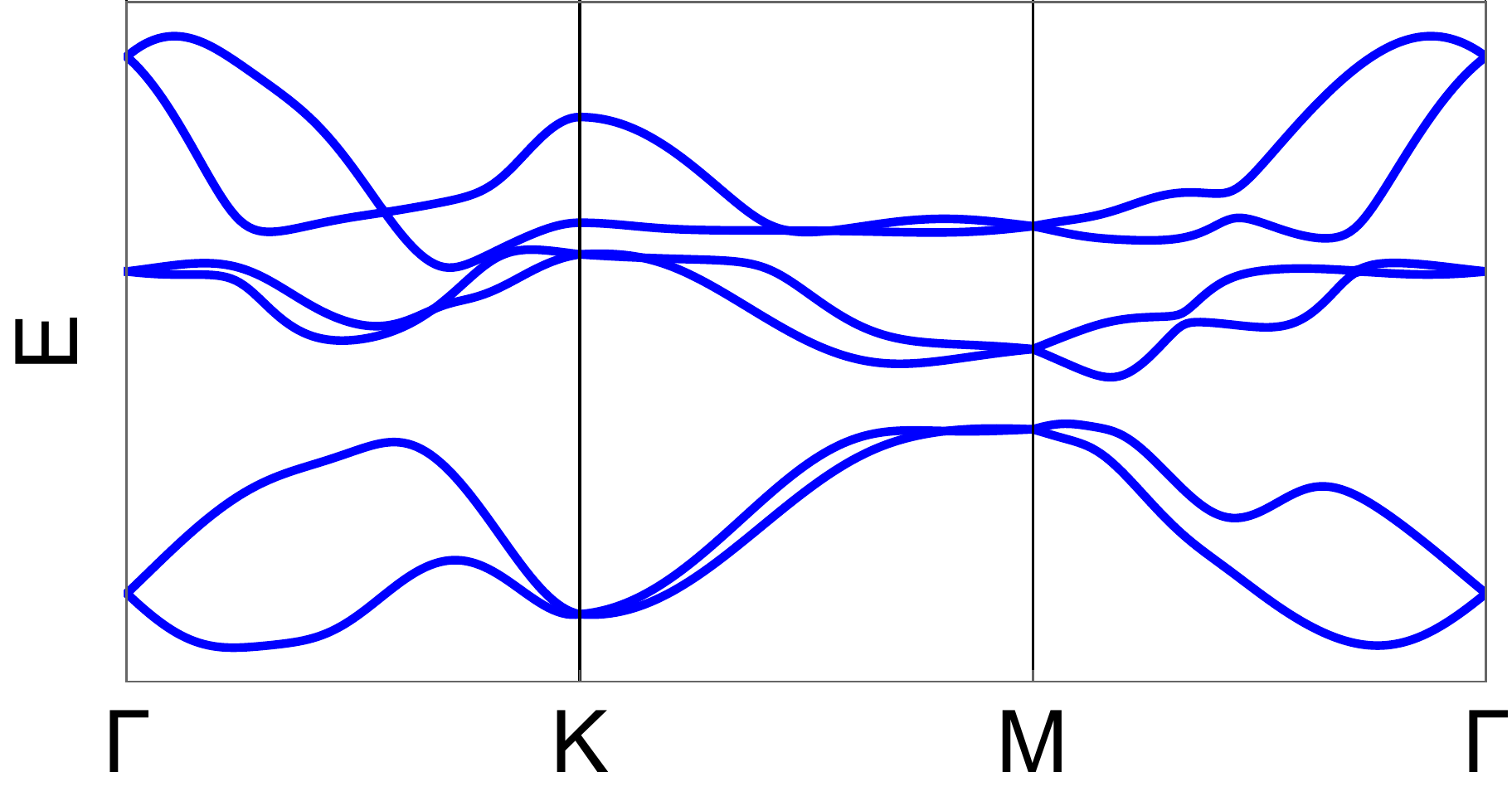} &
	\includegraphics[width=0.25\linewidth]{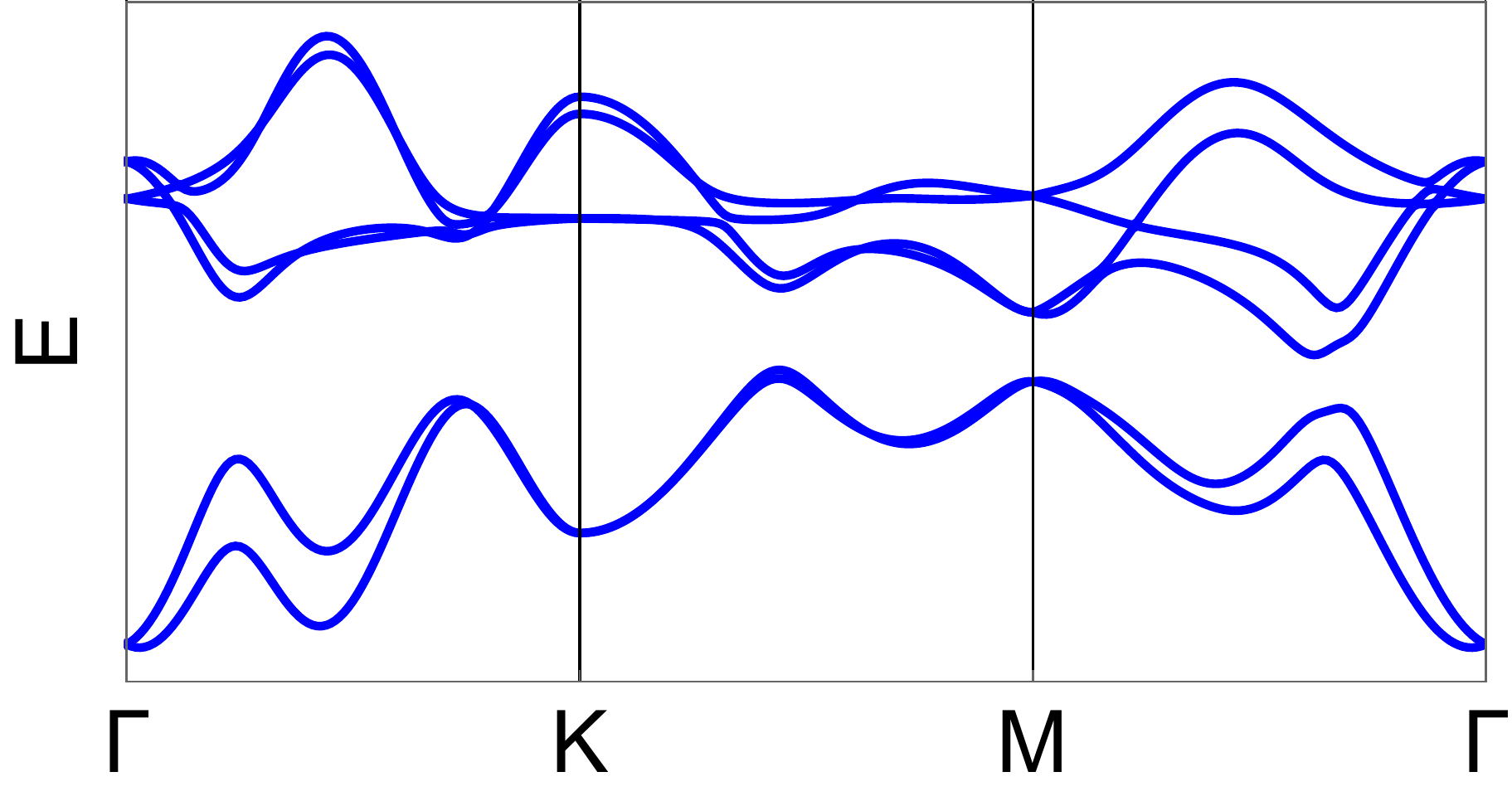} \\
	\includegraphics[width=0.25\linewidth]{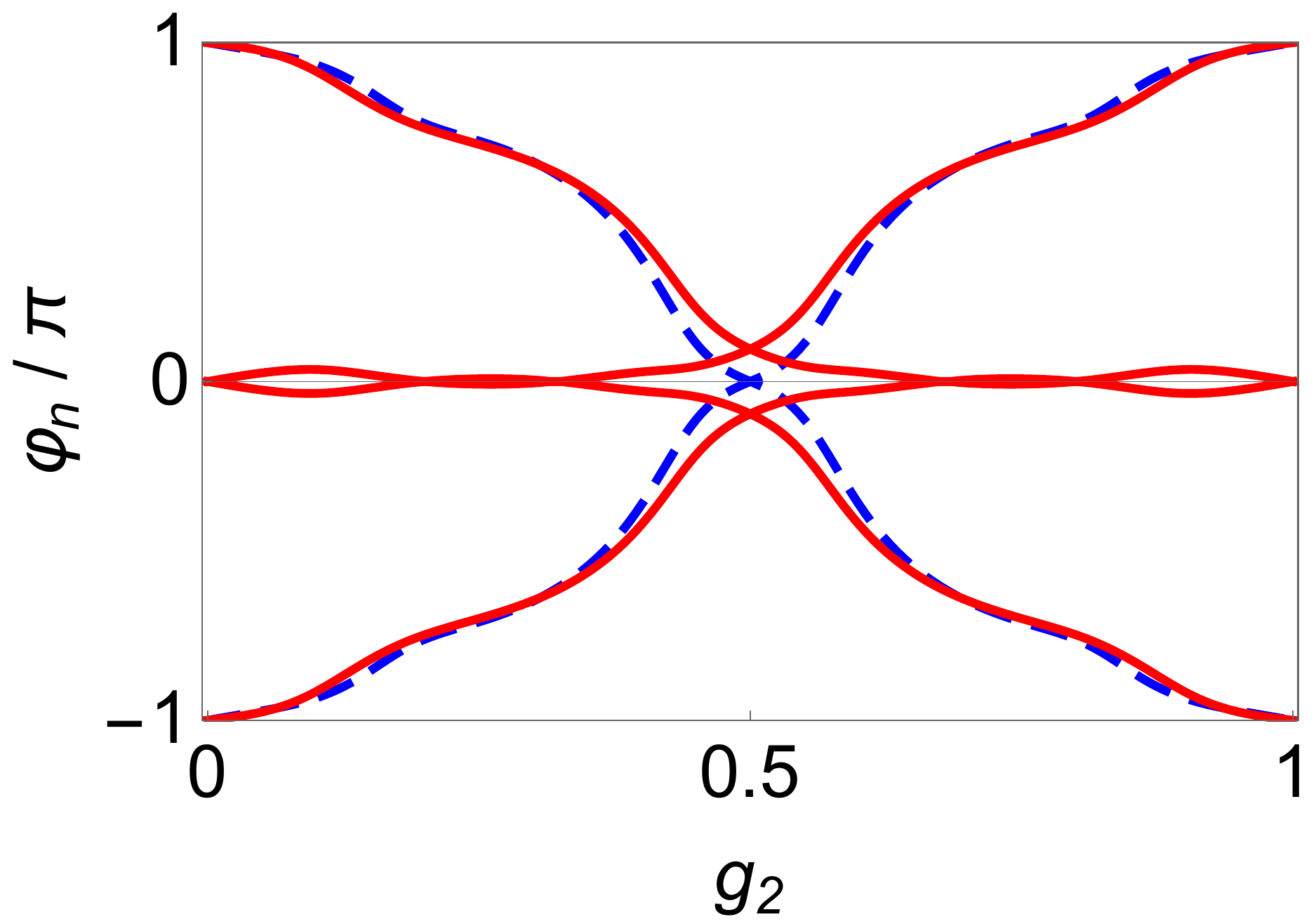} &
	\includegraphics[width=0.25\linewidth]{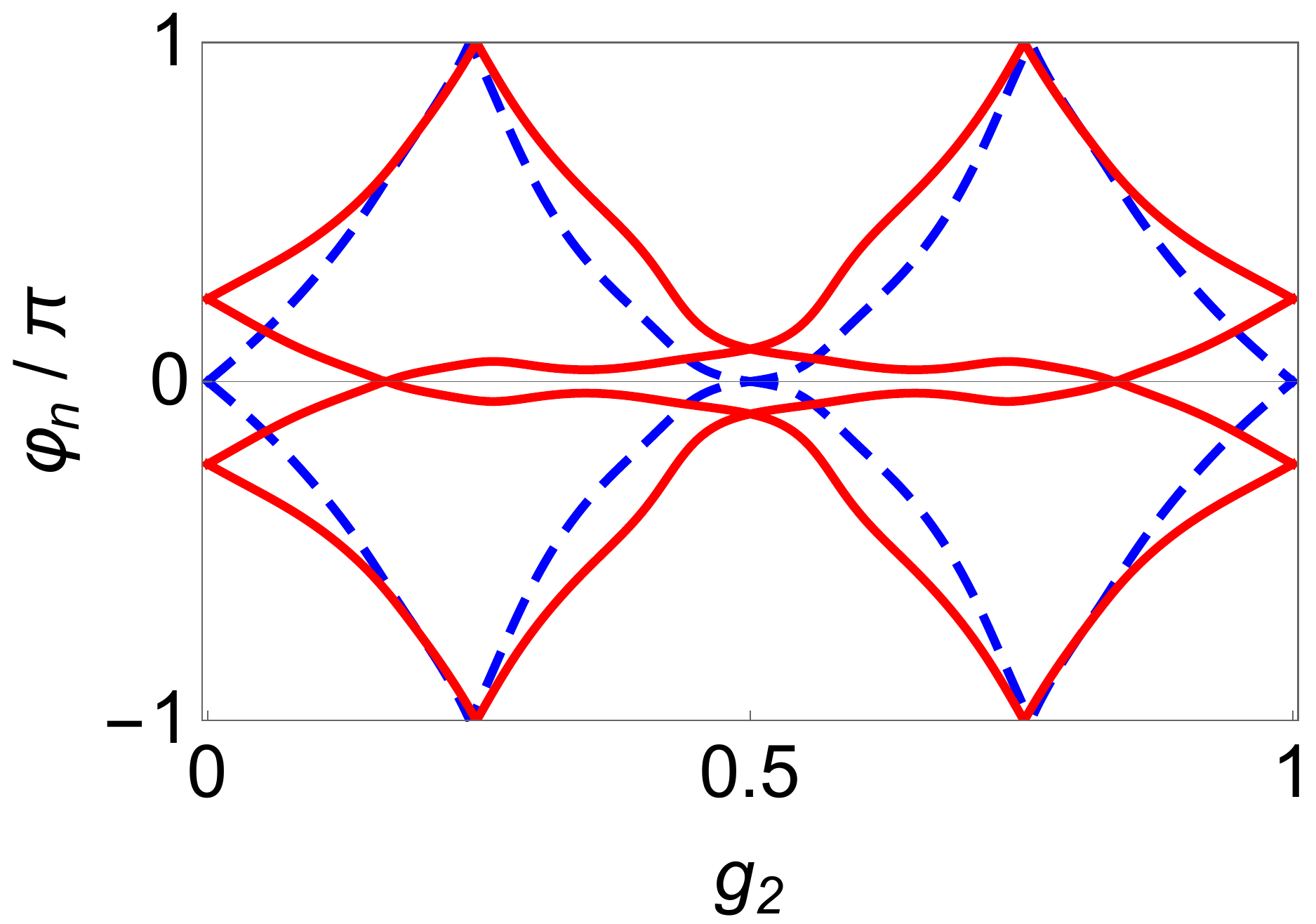} &  
	\includegraphics[width=0.25\linewidth]{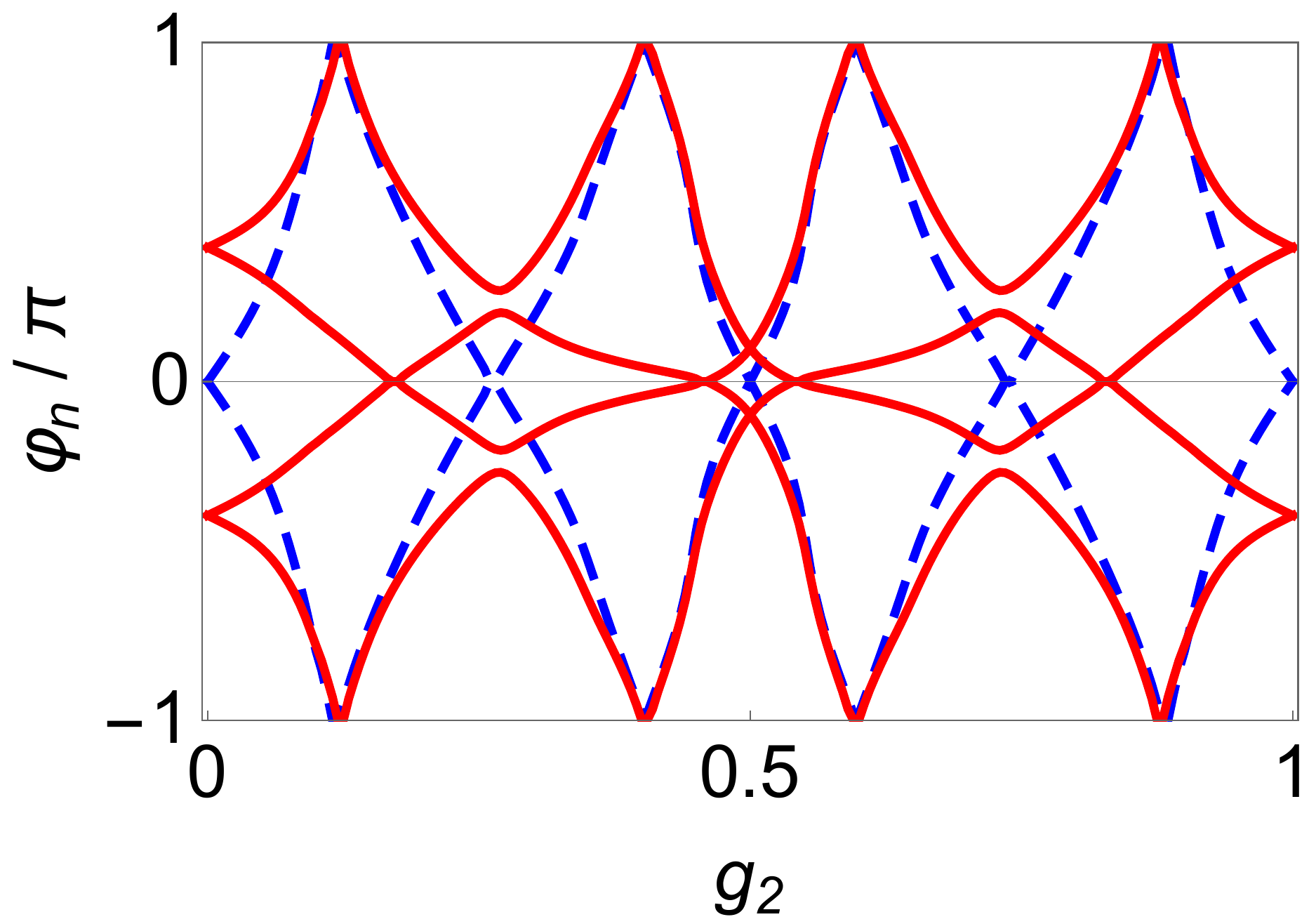} & 
	\includegraphics[width=0.25\linewidth]{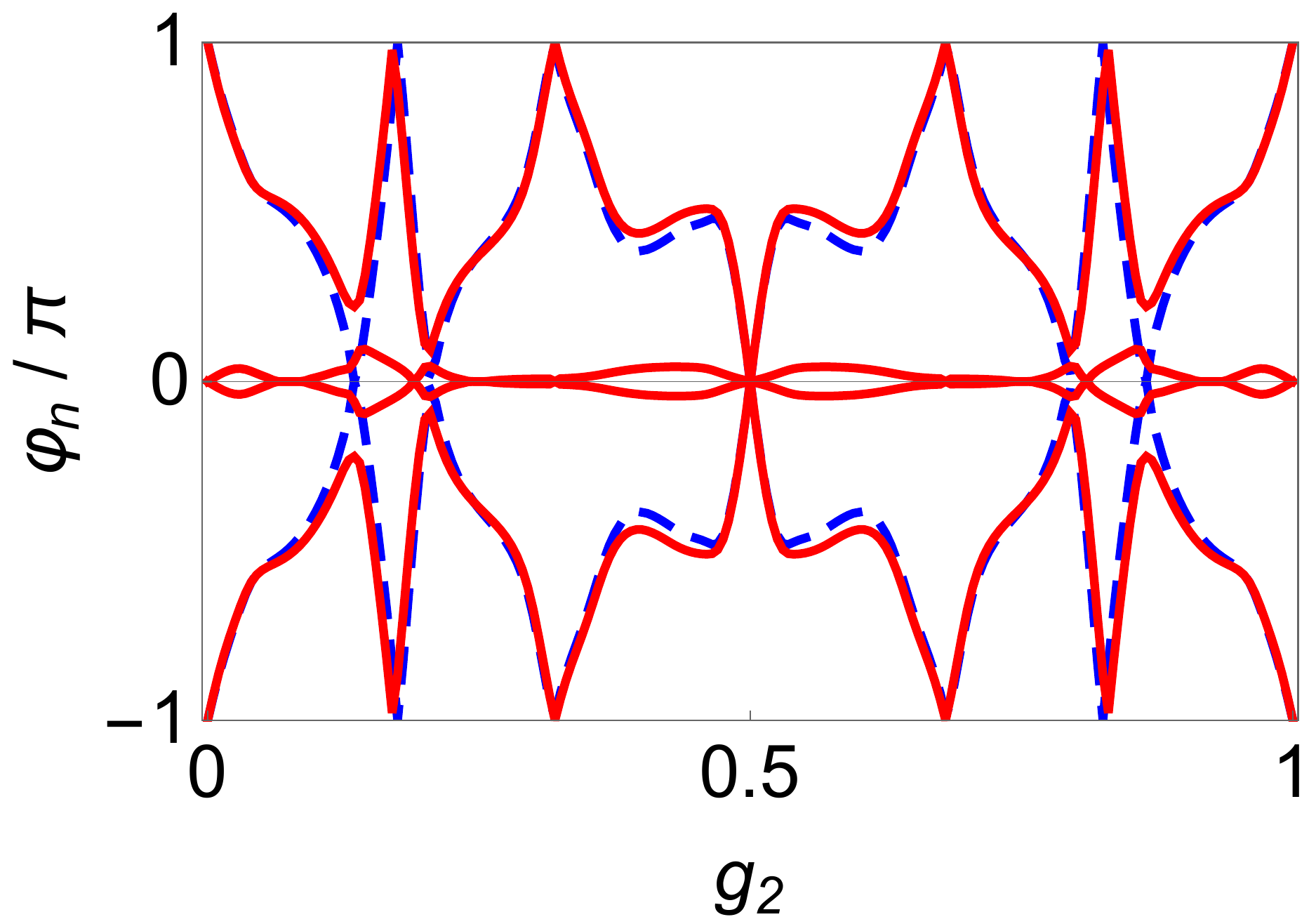} \\
	(a) $W^{\mathrm{B}_{1,occ}^a}_{II} =1$, $\nu_{\mathrm{FKM}} = 1$, &  
	(b) $W^{\mathrm{B}_{1,occ}^a}_{II} =2$, $\nu_{\mathrm{FKM}} = 0$, & 
	 (c) $W^{\mathrm{B}_{1,occ}^a}_{II} = 4$, $\nu_{\mathrm{FKM}} = 0$, & 
	 (d) $W^{\mathrm{B}_{1,occ}^a}_{II} = 5$,  $\nu_{\mathrm{FKM}} = 1$, \\
	 $\nu_{21'} = 1$ & 
	 $\nu_{21'} = 2$ &
	 $\nu_{21'} = 0$ & 
	$\nu_{21'} = 1$
\end{tabular}
\caption{\label{KM_1_998} Six-band topological phases obtained from the coupling of the split EBR $\mathrm{B}^a_1+\mathrm{B}^a_2$ with $\mathrm{B}_1'=\mathrm{B}_1^a$. The rest as for Fig.~\ref{KM_1_899} of the main text. 
}
\end{figure*} 
 
A transformation of the Hamiltonian is here captured through the change of the only remaining free parameter $\vartheta$. It is then straightforward to keep track of the effect at K of the coupling between the four bands $\mathrm{B}_1+\mathrm{B}_2$ (from Wyckoff's position $2b$) with the extra $\mathrm{B}_1'$ (from Wyckoff's position $1a$). Starting with $\vartheta=0$, the triangular lattice ($C$) is completely decoupled from the honeycomb lattice ($A,B$) and the $\overline{K}_6$-doublets both have a pure spin composition with the order $( \uparrow  ,  \downarrow )$ in the basis that makes the matrix representation of $C^+_{3z}$ diagonal (see Section \ref{non_adiabatic}). Switching on the coupling between the two lattices the doublets acquire the mixed spin structure of Eq.~(\ref{spin_mixture}). Then eventually at $\vartheta=\pi/2$, the two lattices are again completely decoupled and the doublets recover their pure spin composition but now in the reversed order $( \downarrow  ,  \uparrow )$ while the ordering of the $C^+_{3z}$-symmetry eigenvalues is kept unchanged.

\section{Complementary numerical six-band results}\label{num_results_ap}

We show in Fig.~\ref{KM_1_998} the six-band results analogue to Fig.~\ref{KM_1_899} in the main text but here generated from the other possible way of splitting the EBR of the honeycomb lattice, i.e.~$\mathrm{B}_1^a(2b)+\mathrm{B}_2^a(2b)$. 

\begin{figure}[t!]
\centering
\begin{tabular}{cc}  
	\includegraphics[width=0.5\linewidth]{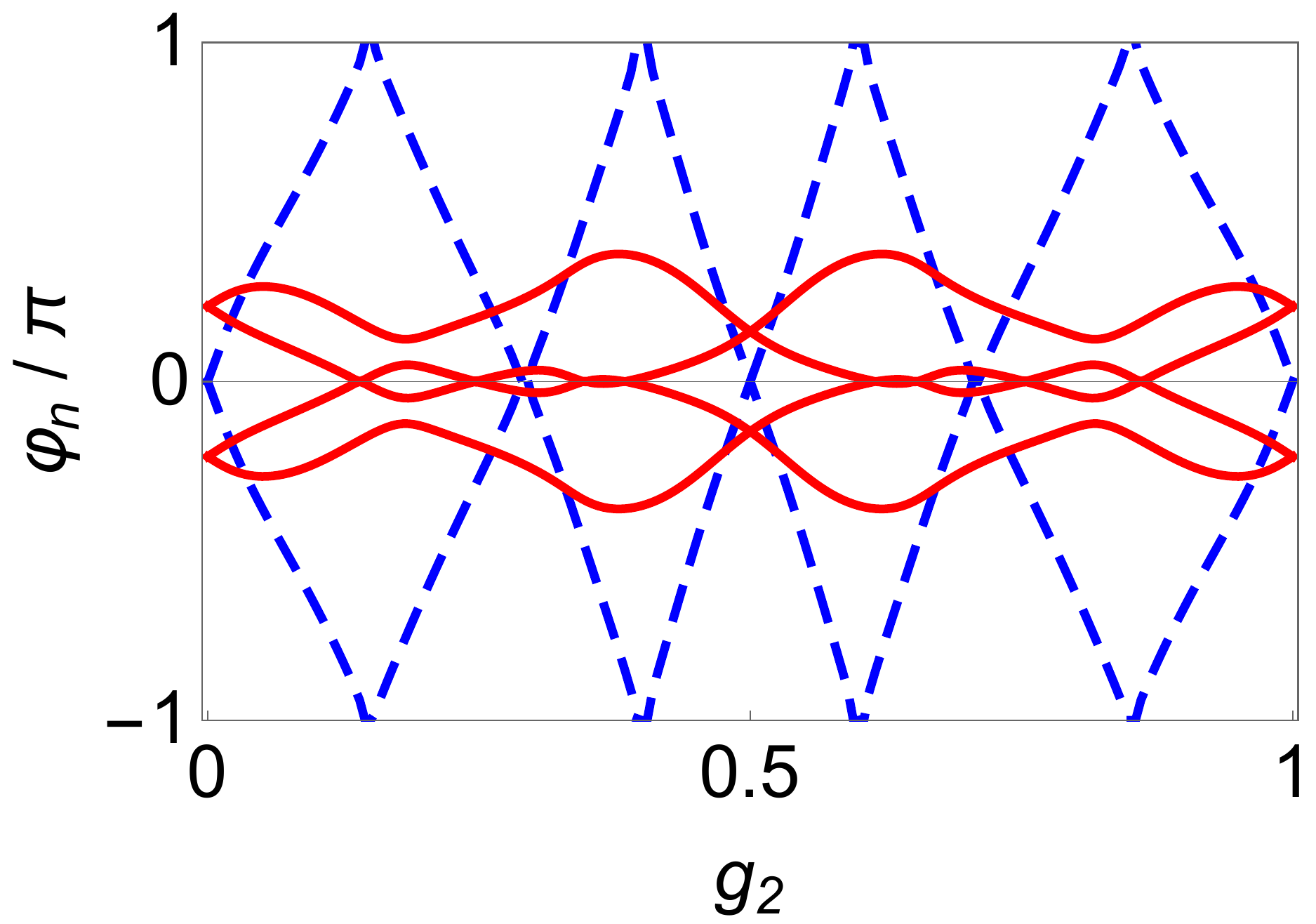} &
	\includegraphics[width=0.5\linewidth]{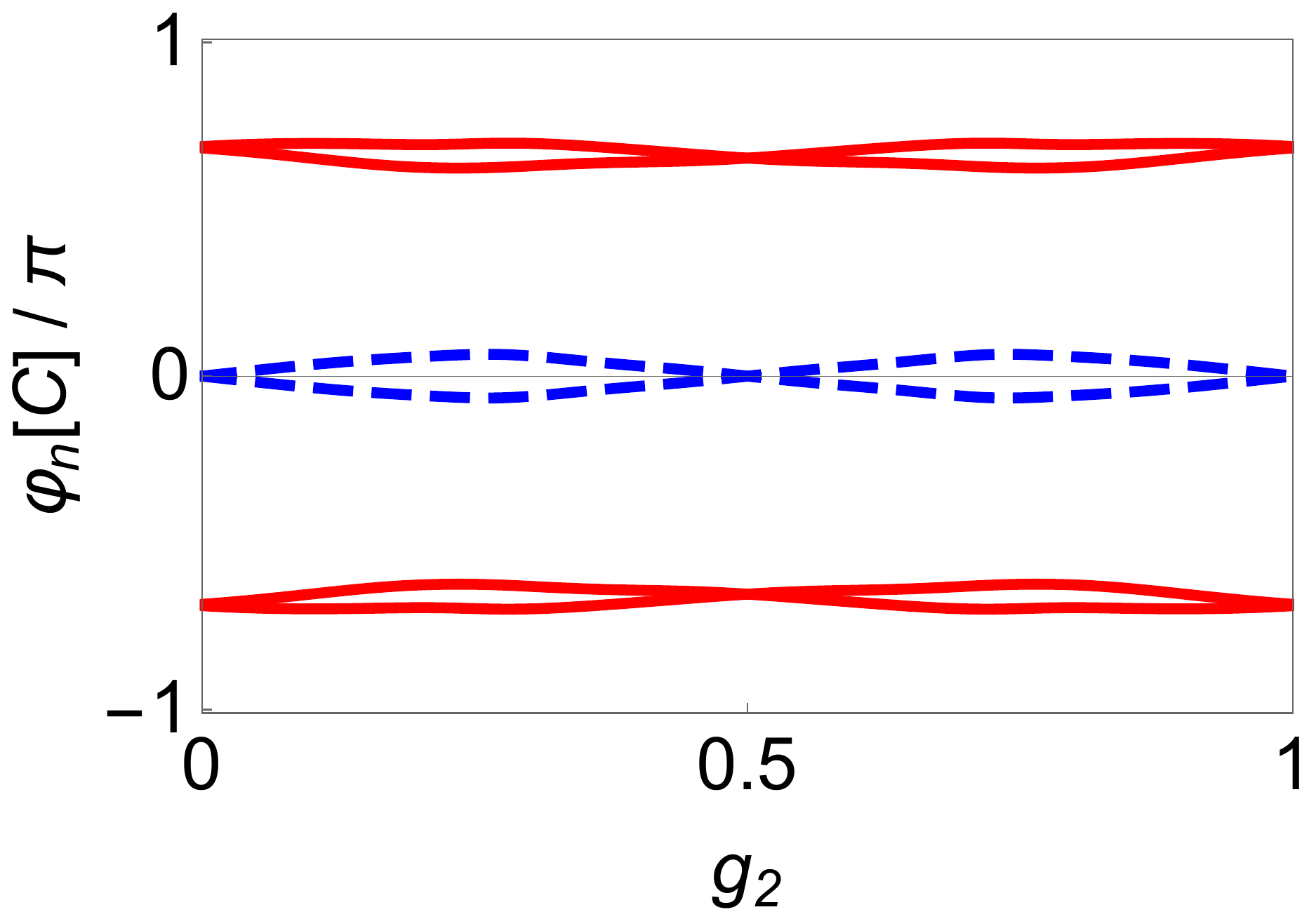} \\
	(a) & (b)
\end{tabular}
\caption{\label{fig_6by6_BR2_occ} Similar to Fig.~\ref{KM_1_998}. (a) Wilson loops for $\mathrm{B}_2=\mathrm{B}_2^a(2b)$ chosen as the two-band occupied subspace before coupling. (b) Wilson loops for $\mathrm{B}_1'=\mathrm{B}_1^a(1a)$ chosen as the two-band occupied subspace before coupling.  
}
\end{figure}

In Fig.~\ref{fig_6by6_BR2_occ}(a) we show the flow of Wilsonian phases over the BZ in the case where $\mathrm{B}_2^a$ is chosen as the two-band occupied subspace before the coupling with the extra bands $\mathrm{B}_1^a$. This configuration is analogue to Fig.~\ref{KM_1_899}(c) and Fig.~\ref{KM_1_998}(c) with a trivialized four-band unoccupied subspace (red line in Fig.~\ref{fig_6by6_BR2_occ}(a)) and a two-band occupied subspace that remains topological with a Wilson loop winding of $W_{II}=\pm4$ (blue dashed line in Fig.~\ref{fig_6by6_BR2_occ}(a)).

In Fig.~\ref{fig_6by6_BR2_occ}(b) we show the flow of Wilsonian phases over the BZ in the case where the EBR of the triangular lattice $\mathrm{B}_1'=\mathrm{B}_1^a(1a)$ has been chosen as the two-band occupied subspace before the coupling. In this case the Wilson loops of the occupied and the unoccupied subspaces have both zero winding indicating a fully trivial topology.

\section{Smooth frame of four-band subspaces}\label{four_band_smooth_gauge}

We give here a more detailed discussion of the stability of Wilson loop windings within a four-band subspace. Let us assume, as it is the case in all our six-band numerical examples (given in Section \ref{six_band} and Appendix \ref{num_results_ap}) that, after the coupling between the split EBR of the honeycomb lattice ($\mathrm{B}_1+\mathrm{B}_2$) and the EBR of the triangular lattice ($\mathrm{B}_1'$), a secondary band gap remains within the unoccupied subspace between $\mathrm{B}_{2/1}$ and $\mathrm{B}_1'$. We can then follow the procedure of Section \ref{non_adiabatic} in order to derive a smooth, periodic and rotation-symmetric frame of each two-band BR composing the four-band subspace, i.e.~taken separately. We have argued in Section \ref{non_adiabatic} and Section \ref{two_band} that such frame can always be found for a two-band subspace. Therefore, the Wilson loop winding of each two-band BR is quantized by symmetry and is classified by $W_{II}\in\{\pm \boldsymbol{1}+3\mathbb{Z},\boldsymbol{0}\}$ (Section \ref{non_adiabatic}). According to Section \ref{two_band}, the two-band Wilson loop winding of each two-band BR is unaffected by the coupling with other bands as long as the band gap separating each two-band BR from the other bands is preserved, i.e.~as long as the coupling is adiabatic with respect to both the main band gap (between occupied and unoccupied subspaces) and the secondary band gap. Therefore, $W^{\mathrm{B}_1'}_{II}=0$, and $W^{\mathrm{B}_{2/1}}_{II}$ keeps the value it had before the coupling. If the two-band topology was stable, we would expect the Wilson loop winding of the four-band subspace $\mathrm{B}_1'+\mathrm{B}_{2/1}$ to be given by $W_{II}^{\mathrm{B}_{2/1}} +W_{II}^{\mathrm{B}_1'}  = W_{II}^{\mathrm{B}_{2/1}} $. We now argue that nothing guarantees \textit{a priori} the stability of the two-band topology once it is imbedded in a four-band subspace. 

In the following we need the orthogonality of the components of a Bloch frame. The smooth frame is obtained through a unitary transformation of the cell-periodic Bloch eigenstates at each $\boldsymbol{k}$, i.e.~$\vert \boldsymbol{v}, \boldsymbol{k} \rangle = (\vert u_1, \boldsymbol{k} \rangle,\vert u_2, \boldsymbol{k} \rangle )^T \mathcal{U}(\boldsymbol{k})$ with $\mathcal{U}(\boldsymbol{k}) \in U(2)$. Then given the orthogonality of the Bloch eigenstates, $\langle u_2, \boldsymbol{k}  \vert u_1, \boldsymbol{k} \rangle=0$, the components of the smooth frame are also orthogonal, i.e.~$\langle v_2, \boldsymbol{k}  \vert v_1, \boldsymbol{k} \rangle=0$, since the columns of $ \mathcal{U}(\boldsymbol{k})$ form themselves an orthonormal basis of $\mathbb{C}^2$.

Let us write the separate smooth, periodic and rotation-symmetric Bloch frames as $\vert \boldsymbol{v}^{\nu} , \boldsymbol{k} \rangle = (\vert v^{\nu}_1 , \boldsymbol{k} \rangle ,\vert v^{\nu}_2 , \boldsymbol{k} \rangle)^T$ with $\nu=1$ for $\mathrm{B}_{2/1}$ and $\nu=2$ for $\mathrm{B}_1'$ (see notations in Section \ref{non_adiabatic}). While each frame is orthogonal they are not necessarily orthogonal to each other, i.e.~$\langle v^{\nu}_i , \boldsymbol{k} \vert v^{\nu}_j , \boldsymbol{k} \rangle = \delta_{ij} $ but $\langle v^1_i , \boldsymbol{k} \vert v^2_j , \boldsymbol{k} \rangle \neq0 $. Therefore, the Wilson loop operator of the four-band subspace written in the separate smooth frames is given through $\widetilde{W}_{l_{\Gamma\text{K}}} = \mathcal{P}^{11} +  \mathcal{P}^{22} + \mathcal{P}^{12}+ \mathcal{P}^{21}$ with the path ordered products of projection operators $\mathcal{P}^{\nu\mu} = \prod\limits_{\boldsymbol{k}}^{l_{\Gamma\text{K}}} \vert \boldsymbol{v}^{\nu} , \boldsymbol{k} \rangle \langle \boldsymbol{v}^{\mu} , \boldsymbol{k} \vert $, i.e.~the cross product terms $\mathcal{P}^{12(21)}$ do not vanish. It then follows that the Wilson loop of the four-band subspace written in the separate smooth frames is in general not diagonal, i.e.~$[\widetilde{\mathcal{W}}_{l_{\Gamma\text{K}}}]^{\nu\mu}_{ij} = \langle v^{\nu}_i , \boldsymbol{k}  \vert \widetilde{W}_{l_{\Gamma\text{K}}}\vert v^{\mu}_j , \boldsymbol{k} \rangle\not\sim \delta_{\nu\mu}\delta_{ij}$. This prevents the symmetry protected quantization of the Wilson loop spectrum. Indeed, the separate two-band smooth frames diagonalize the matrix representation of rotations at $\Gamma$ and K but do not diagonalize the four-band Wilson loop. Reversely if we compute the four-band smooth frame that diagonalizes the four-band Wilson loop it is then not guaranteed to satisfy the rotation-symmetry constraint. 

We have verified this numerically using the Soluyanov-Vanderbilt approach\cite{Vanderbilt_smooth_gauge} and leave a formal proof for a future work. Therefore, we expect that in general the two-band subspace topology is unstable when it is imbedded within a four-band subspace.

\bibliographystyle{apsrev4-1}
\bibliography{Wilsonpaper}
\end{document}